\documentclass[12pt]{article}
\usepackage[dvips]{graphicx}
\usepackage{graphicx}
\usepackage{epsfig}
\usepackage{latexsym,amsmath,amsfonts,amssymb}
\usepackage[latin1]{inputenc}
\usepackage[american]{babel}
\usepackage[dvips]{graphicx}
\usepackage{bbm}
\def\np{Nucl. Phys.}
\def\pl{Phys. Lett.}
\def\prl{Phys. Rev. Lett.}
\def\pr{Phys. Rev.}

\def\im{Invent. Math.}

\def\jhep{J. High Energy Phys.}

\newcommand{\be}{\begin{equation}}
\newcommand{\ee}{\end{equation}}
\newcommand{\beq}{\begin{equation}}
\newcommand{\eeq}{\end{equation}}
\newcommand{\bea}{\begin{eqnarray}}
\newcommand{\eea}{\end{eqnarray}}

\newcommand{\ba}{\begin{eqnarray}}
\newcommand{\ea}{\end{eqnarray}}
\begin{document}
\baselineskip=15.5pt
\pagestyle{plain}
\setcounter{page}{1}


\def\del{{\partial}}
\def\vev#1{\left\langle #1 \right\rangle}
\def\cn{{\cal N}}
\def\co{{\cal O}}
\def\IC{{\mathbb C}}
\def\IR{{\mathbb R}}
\def\IZ{{\mathbb Z}}
\def\RP{{\bf RP}}
\def\CP{{\bf CP}}
\def\Poincare{{Poincar\'e }}
\def\tr{{\rm tr}}
\def\tp{{\tilde \Phi}}

\def\TL{\hfil$\displaystyle{##}$}
\def\TR{$\displaystyle{{}##}$\hfil}
\def\TC{\hfil$\displaystyle{##}$\hfil}
\def\TT{\hbox{##}}
\def\HLINE{\noalign{\vskip1\jot}\hline\noalign{\vskip1\jot}}
\def\seqalign#1#2{\vcenter{\openup1\jot
   \halign{\strut #1\cr #2 \cr}}}
\def\lbldef#1#2{\expandafter\gdef\csname #1\endcsname {#2}}
\def\eqn#1#2{\lbldef{#1}{(\ref{#1})}%
\begin{equation} #2 \label{#1} \end{equation}}
\def\eqalign#1{\vcenter{\openup1\jot
     \halign{\strut\span\TL & \span\TR\cr #1 \cr
    }}}
\def\eno#1{(\ref{#1})}
\def\href#1#2{#2}
\def\half{{1 \over 2}}

\def\ads{{\it AdS}}
\def\adsp{{\it AdS}$_{p+2}$}
\def\cft{{\it CFT}}

\newcommand{\ber}{\begin{eqnarray}}
\newcommand{\eer}{\end{eqnarray}}

\newcommand{\beqar}{\begin{eqnarray}}
\newcommand{\cN}{{\cal N}}
\newcommand{\cO}{{\cal O}}
\newcommand{\cA}{{\cal A}}
\newcommand{\cT}{{\cal T}}
\newcommand{\cF}{{\cal F}}
\newcommand{\cC}{{\cal C}}
\newcommand{\cR}{{\cal R}}
\newcommand{\cW}{{\cal W}}
\newcommand{\eeqar}{\end{eqnarray}}
\newcommand{\tht}{\thteta}
\newcommand{\lm}{\lambda}\newcommand{\Lm}{\Lambda}
\newcommand{\eps}{\epsilon}


\newcommand{\nonu}{\nonumber}
\newcommand{\oh}{\displaystyle{\frac{1}{2}}}
\newcommand{\dsl}
   {\kern.06em\hbox{\raise.15ex\hbox{$/$}\kern-.56em\hbox{$\partial$}}}
\newcommand{\id}{i\!\!\not\!\partial}
\newcommand{\as}{\not\!\! A}
\newcommand{\ps}{\not\! p}
\newcommand{\ks}{\not\! k}
\newcommand{\D}{{\cal{D}}}
\newcommand{\dv}{d^2x}
\newcommand{\Z}{{\cal Z}}
\newcommand{\N}{{\cal N}}
\newcommand{\Dsl}{\not\!\! D}
\newcommand{\Bsl}{\not\!\! B}
\newcommand{\Psl}{\not\!\! P}
\newcommand{\eeqarr}{\end{eqnarray}}
\newcommand{\ZZ}{{\rm \kern 0.275em Z \kern -0.92em Z}\;}


\def\del{{\delta^{\hbox{\sevenrm B}}}} \def\ex{{\hbox{\rm e}}}
\def\azb{A_{\bar z}} \def\az{A_z} \def\bzb{B_{\bar z}} \def\bz{B_z}
\def\czb{C_{\bar z}} \def\cz{C_z} \def\dzb{D_{\bar z}} \def\dz{D_z}
\def\im{{\hbox{\rm Im}}} \def\mod{{\hbox{\rm mod}}} \def\tr{{\hbox{\rm Tr}}}
\def\ch{{\hbox{\rm ch}}} \def\imp{{\hbox{\sevenrm Im}}}
\def\trp{{\hbox{\sevenrm Tr}}} \def\vol{{\hbox{\rm Vol}}}
\def\rl{\Lambda_{\hbox{\sevenrm R}}} \def\wl{\Lambda_{\hbox{\sevenrm W}}}
\def\fc{{\cal F}_{k+\cox}} \def\vev{vacuum expectation value}
\def\nodiv{\mid{\hbox{\hskip-7.8pt/}}}
\def\ie{{\em i.e.}}
\def\ie{\hbox{\it i.e.}}

\def\CC{{\mathchoice
{\rm C\mkern-8mu\vrule height1.45ex depth-.05ex
width.05em\mkern9mu\kern-.05em}
{\rm C\mkern-8mu\vrule height1.45ex depth-.05ex
width.05em\mkern9mu\kern-.05em}
{\rm C\mkern-8mu\vrule height1ex depth-.07ex
width.035em\mkern9mu\kern-.035em}
{\rm C\mkern-8mu\vrule height.65ex depth-.1ex
width.025em\mkern8mu\kern-.025em}}}

\def\RR{{\rm I\kern-1.6pt {\rm R}}}
\def\NN{{\rm I\!N}}
\def\ZZ{{\rm Z}\kern-3.8pt {\rm Z} \kern2pt}
\def\IB{\relax{\rm I\kern-.18em B}}
\def\ID{\relax{\rm I\kern-.18em D}}
\def\II{\relax{\rm I\kern-.18em I}}
\def\IP{\relax{\rm I\kern-.18em P}}
\newcommand{\CS}{{\scriptstyle {\rm CS}}}
\newcommand{\CSs}{{\scriptscriptstyle {\rm CS}}}
\newcommand{\rc}{\nonumber\\}
\newcommand{\bear}{\begin{eqnarray}}
\newcommand{\eear}{\end{eqnarray}}
\newcommand{\W}{{\cal W}}
\newcommand{\F}{{\cal F}}
\newcommand{\x}{{\cal O}}
\newcommand{\LL}{{\cal L}}

\def\mani{{\cal M}}
\def\calo{{\cal O}}
\def\calb{{\cal B}}
\def\calw{{\cal W}}
\def\calz{{\cal Z}}
\def\cald{{\cal D}}
\def\calc{{\cal C}}
\def\to{\rightarrow}
\def\ele{{\hbox{\sevenrm L}}}
\def\ere{{\hbox{\sevenrm R}}}
\def\zb{{\bar z}}
\def\wb{{\bar w}}
\def\nodiv{\mid{\hbox{\hskip-7.8pt/}}}
\def\menos{\hbox{\hskip-2.9pt}}
\def\dr{\dot R_}
\def\drr{\dot r_}
\def\ds{\dot s_}
\def\da{\dot A_}
\def\dga{\dot \gamma_}
\def\ga{\gamma_}
\def\dal{\dot\alpha_}
\def\al{\alpha_}
\def\cl{{closed}}
\def\cls{{closing}}
\def\vev{vacuum expectation value}
\def\tr{{\rm Tr}}
\def\to{\rightarrow}
\def\too{\longrightarrow}

%

\newfont{\namefont}{cmr10}
\newfont{\addfont}{cmti7 scaled 1440}
\newfont{\boldmathfont}{cmbx10}
\newfont{\headfontb}{cmbx10 scaled 1728}
\newcommand{\re}{\,\mathbb{R}\mbox{e}\,}
\newcommand{\hyph}[1]{$#1$\nobreakdash-\hspace{0pt}}
\providecommand{\abs}[1]{\lvert#1\rvert}
\newcommand{\Nugual}[1]{$\mathcal{N}= #1 $}
\newcommand{\sub}[2]{#1_\text{#2}}
\newcommand{\partfrac}[2]{\frac{\partial #1}{\partial #2}}
\newcommand{\bsp}[1]{\begin{equation} \begin{split} #1 \end{split} \end{equation}}
\newcommand{\calF}{\mathcal{F}}
\newcommand{\calO}{\mathcal{O}}
\newcommand{\calM}{\mathcal{M}}
\newcommand{\calV}{\mathcal{V}}
\newcommand{\bbZ}{\mathbb{Z}}
\newcommand{\bbC}{\mathbb{C}}

\numberwithin{equation}{section}

\newcommand{\Tr}{\mbox{Tr}}    


%
\renewcommand{\theequation}{{\rm\thesection.\arabic{equation}}}
\begin{titlepage}
%
\vspace{0.1in}

\begin{center}
\Large \bf String duals of   two-dimensional (4,4) supersymmetric  gauge theories
\end{center}
\vskip 0.2truein
\begin{center}
Daniel Are\'an ${}^{*}$\footnote{arean@fpaxp1.usc.es}, 
Paolo Merlatti${}^{*}$ \footnote{merlatti@fpaxp1.usc.es},
Carlos N\'u\~nez${}^{\dagger}$\footnote{c.nunez@swansea.ac.uk}, 
 and
Alfonso V. Ramallo${}^{*}$\footnote{alfonso@fpaxp1.usc.es}\\
\vspace{0.2in}
${}^{*}$\it{
Departamento de  F\'\i sica de Part\'\i culas, Universidade
de Santiago de
Compostela\\and\\Instituto Galego de F\'\i sica de Altas
Enerx\'\i as (IGFAE)\\E-15782, Santiago de Compostela, Spain
}
\vspace{0.2in}
\vskip 0.1truein
${}^{\dagger}$ \it{Department of Physics\\ University of Swansea, Singleton
Park\\
Swansea SA2 8PP, United Kingdom.}

\vspace{0.2in}
\end{center}
\vspace{0.2in}
\centerline{{\bf Abstract}}
We study duals to field theories in two dimensions
with ${\cal N}=(4,4)$ SUSY. The string backgrounds reproduce certain non-perturbative 
aspects of the dual field theory with a large number of colors $N_c$ and 
a tunable number of flavors $N_f$. Different aspects of the 
two-dimensional field theory are discussed, among them the running of 
gauge couplings, the spectrum of mesons in Higgs and Coulomb branches, 
entanglement entropy and the non-relativistic version of our string 
backgrounds.

\smallskip
\end{titlepage}
\setcounter{footnote}{0}

\tableofcontents


\newpage

\section{Introduction}
Since its inception \cite{'tHooft:1973jz}, the large $N_c$ expansion has 
given many valuable 
insights into QCD dynamics beyond perturbation theory. Indeed, long ago 't 
Hooft showed, by using this expansion,  that two-dimensional models are very good laboratories in which one can learn about various aspects of four-dimensional field theories, like 
the spectrum of bound 
states \cite{'tHooft:1974hx}. Besides, the large $N_c$ expansion sheds light on 
other aspects, like the connection between the Skyrme model and QCD and 
the qualitative understanding of the Zweig rule \cite{Witten:1979kh}.

The number of colors is not the only available parameter in QCD-like 
theories. In the real world, and specially at energies above the masses of the heavy quarks, the ratio 
$N_f/N_c$ between the number of flavors and colors becomes of order one and, then, it seems natural to investigate 
the way in which a large number of flavors modifies the dynamics of the 
large $N_c$ field theory. From a field theory point of view, these ideas 
were introduced in \cite{Veneziano:1976wm}.

In this paper we consider these kind of problems from the perspective 
given by the AdS/CFT correspondence \cite{Maldacena:1997re}, its 
refinements \cite{Gubser:1998bc}, \cite{Witten:1998qj}
and extensions
\cite{Itzhaki:1998dd,MAGOO}. In this approach the addition of matter degrees of freedom can be performed by including extra (flavor) branes in the description \cite{KK} and one has to deal with a system of gravity coupled to brane sources \cite{Erdmenger:2007cm}. When the number of flavors $N_f$ is small compared with the number of colors $N_c$ one can treat the flavor branes as probes in the geometry originated by the color branes. This is the so-called quenched approximation. However, if $N_f\sim N_c$ the backreaction of the flavor branes on the geometry cannot be neglected anymore and one must obtain a new background that incorporates the effect of the fundamental matter.

We will apply these ideas in order to study  the case of supersymmetric field theories in two spacetime dimensions. This  paper represents a first step in this direction as we will concentrate in  the case of field theories with 
${\cal N}=(4,4) $ SUSY. The analysis of  two-dimensional theories with ${\cal N}=(2,2)$ and ${\cal N}=(1,1)$ supersymmetry  is in  preparation and will be reported elsewhere. 

We will proceed by constructing string duals to ${\cal N}=(4,4) $ gauge theories. Our 
backgrounds will be based on D3-branes wrapping two-cycles  of a Calabi-Yau manifold of complex dimension two (leading to a 
2d field theory at low energies) or on D1/D5 systems of fractional branes.
We will explore our backgrounds with ``probe branes'' to learn about 
various aspects of the dual QFT, most notably, the running of couplings 
and  the ``quenched'' spectrum of mesons (in the Higgs or Coulomb branches). 
Then, following the line of research presented in the papers \cite{Casero:2006pt}-\cite{HoyosBadajoz:2008fw}, 
we will add a (large) number of flavor branes to the geometries 
mentioned above, which correspond to flavor 
degrees of freedom in the dual QFT. We will then find string backgrounds dual 
to ${\cal N}=(4,4)$ QFT's in two dimensions with large $N_c, N_f$. Again, we will 
present some checks of the field theory-string theory matching and obtain 
some non-perturbative predictions from our backgrounds.

In the case of backgrounds generated by D3-branes wrapped on a two-cycle, we will adopt an ansatz for the metric and five-form which contains functions depending on two radial coordinates. By imposing the preservation of eight supersymmetries we will find a system of partial differential equations. These equations look rather simple, but  are difficult to solve in general. However, we will be able to find some particular analytic solutions. It turns out that the most interesting of these solutions is provided by the analysis of five-dimensional gauged supergravity, where supersymmetry is realized by means of the so-called topological twist \cite{Maldacena:2000mw}. Indeed, we will verify that this solution encodes non-trivial information on the $d=2$, ${\cal N}=(4,4)$ gauge theory. Moreover, we will be able to modify the BPS equations to include the backreaction of the flavor branes, which in this case are D3-branes extended along the non-compact directions of the normal bundle of the cycle where the color branes are wrapped.  The corresponding solutions, which in this case are numerical, also pass several non-trivial tests.

An alternative string description of the $d=2$, ${\cal N}=(4,4)$ gauge theories is obtained by considering a system of fractional D1-branes on the orbifold 
$\mathbb{R}^{1,5}\times \mathbb{C}^2/\mathbb{Z}_2$. In this case one proceeds by solving directly the second-order equations of motion for the fields of type IIB supergravity, including the ones originating from the twisted sector of the orbifold theory. In this approach the flavor branes are D5-branes that wrap completely  the 
$\mathbb{C}^2/\mathbb{Z}_2$ orbifold. We will verify that this solution also matches the field theory results.

The organization of the rest of this paper is the following. In section \ref{44unflavored} we will introduce our setup of D3-branes wrapped on a two-cycle. The detailed derivation of the system of BPS equations are presented in appendix \ref{BPSequations}. The gauged supergravity solutions of this BPS system are obtained in appendix \ref{gaugedsugra44} and analyzed at the end of section \ref{44unflavored}.  In section \ref{44flavored} we study the addition of flavor to our solution and we obtain the corresponding backreacted geometry. In section \ref{FT} we first review the field content of the $d=2$, ${\cal N}=(4,4)$ gauge theories and the calculation of their one-loop beta functions. By performing a probe calculation we check that our backgrounds match these field theory results. We also analyze in this section the Higgs branch of the theory, based on the study of the supersymmetric embeddings of D3-branes performed in appendix \ref{Higgs-embeddings}. We finish section \ref{FT} by analyzing the entanglement entropy of our model in the UV region, following the recent proposal of refs. \cite{Ryu} and \cite{Klebanov:2007ws}.

Section \ref{Mesons} deals with the study of the mass spectra of mesons for our theory, starting with the case of the Coulomb branch in the quenched approximation. We also discuss the effect of the backreaction, as well as the spectra for the Higgs branch. In section \ref{EH-section} we study additional solutions of the unflavored BPS system.  In particular we show that our equations also admit a solution in which the geometry has an $AdS_3$ factor. We also obtain the background dual to a non-relativistic system, which is generated from our solution by means of a pair of T-dualities, combined with  a suitable  coordinate shift. Section \ref{fractional-section} is devoted to the analysis of the system  of D1/D5 fractional branes. Finally, in section \ref{conclusions} we summarize our results and discuss some of their possible extensions.

\section{The supergravity dual  of the (4,4) gauge theories}
\label{44unflavored}

Let us consider the background of type IIB supergravity created by a stack of $N_c$ D3-branes wrapped on a two-cycle ${\cal C}_2$ of a Calabi-Yau (CY) cone of complex dimension two according to the brane setup:
\begin{center}
\begin{tabular}{|c|c|c|c|c|c|c|c|c|c|c|}
\multicolumn{3}{c}{ }&
\multicolumn{4}{c}{$\overbrace{\phantom{\qquad\qquad\qquad}}^{\text{CY}_2}$}\\
\hline
&\multicolumn{2}{|c|}{$\mathbb{R}^{1,1}$}
&\multicolumn{2}{|c|}{$S^2$}
&\multicolumn{2}{|c|}{$N_2$}
&\multicolumn{4}{|c|}{$\mathbb{R}^{4}$}\\
\hline
D$3$ &$-$&$-$&$\bigcirc$&$\bigcirc$&$\cdot$&$\cdot$&$\cdot$&$\cdot$&$\cdot$&$\cdot$\\
\hline
\end{tabular}
\end{center}
where $S^2$ represent the directions of the two-cycle (which is topologically a two-sphere) and $N_2$ are the directions of the normal bundle to ${\cal C}_2$ . Notice also that the symbols ``$-$" and ``$\cdot"$ represent respectively unwrapped worldvolume directions and transverse directions, while a circle denotes wrapped directions.

 Let us parameterize  ${\cal C}_2$ by means of two angular coordinates $(\theta, \phi)$ and let $\sigma$ be the radial coordinate of the CY cone. Notice that in this setup there is another radial coordinate $\rho$, which represents the distance along $\mathbb{R}^{4}$, the directions orthogonal to both the D3-brane worldvolume and the CY cone. The ansatz for the string frame metric  which we will adopt is the following:
\bear
&&ds_{st}^2\,=\,H^{-{1\over 2}}\,\,\Big[\,dx_{1,1}^2\,+\,{z\over m^2}\,
\Big(\,d\theta^2\,+\,\sin^2\theta\,d\phi^2\,\,\Big)\,\Big]\,+\,\rc\rc
&&\qquad\qquad+\,
H^{{1\over 2}}\,\,\,\Big[\,{1\over z}\,\,\Big(\,
d\sigma^2\,+\,\sigma^2\,
\Big(\,d\psi+\cos\theta d\phi\,\Big)^2\,\Big)
\,+\,
d\rho^2\,+\,\rho^2\,d\Omega_3^2\,\Big]\,\,,
\qquad\qquad
\label{D3metric}
\eear
where $m$ is a constant with units of mass which, for convenience, we will take as:
\beq
{1\over m^2}\,=\,\sqrt{4\pi g_s N_c}\,\,\alpha'\,\,,
\label{m}
\eeq
with $g_s$  and  $\alpha'$ being  respectively the string coupling constant and the Regge slope of superstring theory. In eq. (\ref{D3metric}) $dx_{1,1}^2$ denotes the two-dimensional Minkowski metric for the coordinates $x^0, x^1$ and the range of the variables $\theta$, $\phi$, $\psi$,  $\rho$ and $\sigma$ is the following:
\beq
0\,\le \theta\le \pi\,\,,\qquad
0\,\le \phi, \psi<2\pi\,\,,\qquad
0\,\le\rho\,,\,\sigma<\infty\,\,.
\eeq
Moreover, the function $z$ (which controls the size of the cycle) and   the warp factor  $H$ should be considered as functions of the two holographic variables $(\rho, \sigma)$:
\beq
H\,=\,H(\rho,\sigma)\,\,,\qquad\qquad
z\,=\,z(\rho,\sigma)\,\,,
\eeq
while $\psi$, which is a coordinate of the CY,  is fibered over the cycle  ${\cal C}_2$. Moreover,  $d\Omega_3^2$ is the metric of a unit three-sphere, which we will parameterize in terms of three angles $\alpha_i$ ($i=1,2,3$) as:
\beq
d\Omega_3^2\,=\,d\alpha_1^2\,+\,\sin^2\alpha_1\,(\,d\alpha_2^2\,+\,\sin^2\alpha_2\,d\alpha_3^2\,)\,\,.
\eeq
As in any background created by D3-branes, our solution should be endowed with a 
a self-dual RR five-form $F_5$, that we write as:
\beq
F_5\,=\,{\cal F}_5\,+\,{}^*{\cal F}_5\,\,,
\label{F5}
\eeq
where ${\cal F}_5$ can be represented in terms of a four-form potential ${\cal C}_4$ as ${\cal F}_5\,=\,d{\cal C}_4$. We shall adopt  the following ansatz for ${\cal C}_4$ :
\beq
{\cal C}_4\,=\,g\,\omega_3\,\wedge\,(d\psi+\cos\theta \,d\phi)\,\,,
\qquad g=g(\rho,\sigma)\,\,,
\label{CalC4}
\eeq
where $g(\rho,\sigma)$ is a new function and $\omega_3$ is the volume element of the
three-sphere:
\beq
\omega_3=\sin^2\alpha_1 \sin\alpha_2\, d\alpha_1\wedge d\alpha_2 \wedge d\alpha_3
\,\,.
\label{omega3}
\eeq
To determine the functions $H$, $z$ and $g$ entering our ansatz we will impose that our background preserves eight supersymmetries, which is the right number of SUSYs that the supergravity dual of a $d=2$, ${\cal N}=(4,4)$ gauge theory must leave unbroken. The detailed analysis is performed in appendix \ref{BPSequations}. It turns out that, in order to preserve this required amount of supersymmetry, the three functions $H$, $z$ and $g$ must satisfy the following system of partial differential equations (PDEs):
\bear
m^2\,g\,=\,\rho^3\,z'\,\,,\rc\rc
m^2\,H\,=\,{z\dot z\over \sigma}\,\,,\rc\rc
g'\,=\,-\sigma\,\rho^3\,\dot H\,\,,\rc\rc
\dot g\,=\,\,{\sigma \rho^3\over z}\,\,H'\,-\,{\sigma\over z^2}\,m^2\,g\,H\,\,,
\label{BPSsystem}
\eear
where we have denoted:
\beq
'\,\equiv\,\partial_{\rho}\,\,,\qquad\qquad
\dot{}\,\equiv\,\partial_{\sigma}\,\,.
\eeq

Actually,  the four BPS equations (\ref{BPSsystem}) are not all independent. Indeed, one can check that, for example,  the last equation in the system (\ref{BPSsystem}) is a consequence of the others. Furthermore, one can combine  the first-order BPS equations (\ref{BPSsystem})  to obtain  the following  second-order PDE equation for $z(\rho, \sigma)$:
\beq
\rho\,z\,(\,\dot z\,-\,\sigma\, \ddot z\,)\,=\,\sigma\,(\,\rho\,\dot z^2\,+\,
\rho\,z''\,+\, 3 z'\,)\,\,.
\label{PDE-z-unflavored}
\eeq
Moreover, once the function $z(\rho,\sigma)$ is known, one can obtain, by using the first two equations in (\ref{BPSsystem}), the values of the other two functions of our ansatz, namely  $H(\rho, \sigma)$ and $g(\rho, \sigma)$.

It is easy to verify that, when the system (\ref{BPSsystem}) holds, the field strength $F_5$ satisfies the Bianchi identity $dF_5=0$. Indeed, one can check this fact directly by using the equations in (\ref{BPSsystem}) (see appendix \ref{BPSequations}) or, alternatively, one can verify that $F_5$ can be represented in terms of a four-form potential $C_4$ as $F_5=dC_4$. The actual expression of $C_4$ can be taken as:
\beq
C_4\,=\,g\,\omega_3\,\wedge\, (d\psi+\cos\theta d\phi)\,+\,
dx^0\wedge dx^1\wedge \Big[\,{z \over m^2 H}\,\omega_2\,-\,{\sigma\over z}\,d\sigma\wedge
(d\psi+\cos\theta d\phi)\,\Big]\,\,,
\label{C4}
\eeq
where $\omega_2$ is the volume element of the $(\theta, \phi)$ two-sphere, namely:
\beq
\omega_2=\sin\theta d\theta\wedge d\phi\,\,.
\label{omega2}
\eeq
One can also verify that the system (\ref{BPSsystem}) ensures the fulfillment of the second-order Einstein equations. The details of this verification are given in appendix \ref{BPSequations}. 

\subsection{Integration of the BPS system}

The system (\ref{BPSsystem}) can be easily integrated for $g$ and $z$ when $\sigma=0$ and $\rho$ varies. Indeed, it follows from the last two equations in (\ref{BPSsystem}) that the five-form function $g$ is independent of $\rho$ when $\sigma=0$. Let us write:
\beq
g(\rho, \sigma=0)=g_0\,\,,
\eeq
with $g_0$ being a constant which should be determined by flux quantization  of the RR five-form (see below).  Then, the first equation in  (\ref{BPSsystem}) for $\sigma=0$ can be written as:
\beq
z'(\rho, 0)\,=\,{m^2\,g_0\over \rho^3}\,\,,
\label{zprime-sigmazero}
\eeq
which can be integrated as:
\beq
z(\rho, 0)\,=\,-{ m^2\,g_0\over 2\rho^2}\,+\,{\rm constant}\,\,.
\label{z-sigma0}
\eeq
In order to get an explicit solution of the BPS equation (\ref{BPSsystem}) for all values of the $(\sigma, \rho)$ coordinates, it is interesting to notice that the brane setup analyzed here can be studied in the framework of five-dimensional gauged supergravity. Actually, the corresponding topological twisting was studied in \cite{Maldacena:2000mw} and is summarized in appendix \ref{gaugedsugra44}. One of the advantages of using gauged supergravity is the fact that the non-trivial fibering of the $\psi$ coordinate in the metric comes up naturally in the uplifting  from five to ten dimensions as a consequence of the twisting. Also, the RR five-form is given by the uplifting formulae of ref. \cite{Cvetic:1999xp}. The ansatz of gauged supergravity contains three functions which depend on a single holographic coordinate (see appendix 
\ref{gaugedsugra44}). The corresponding BPS equations are a system of ordinary differential equations which, as shown in appendix  \ref{gaugedsugra44}, can be analytically  integrated. After performing a suitable change of variables the uplifted metric and five-form can be written as in (\ref{D3metric}) and (\ref{F5})-(\ref{CalC4}). This is worked out in detail in 
 appendix  \ref{gaugedsugra44}. Let us summarize here the main results. With this purpose, we define the 
 function $\Gamma(z)$ as follows:
\beq
\Gamma(z)\,\equiv\,z_*\,+\,(z_*\,-\,z)\,\big[\,\log(z_*\,-\,z)\,+\,\kappa\,\big]\,\,,
\label{Gamma(z)}
\eeq
where $z_*$ and $\kappa$ are constants\footnote{In all the numerical calculations performed in this paper we take $\kappa=0$ .}. The function $\Gamma(z)$ defined in (\ref{Gamma(z)}) can be used to determine implicitly $z$ as a function of $(\rho, \sigma)$. Indeed, it is shown in appendix  \ref{gaugedsugra44} that 
$z(\rho, \sigma)$ satisfies:
\beq
\Big[\,\rho^2\,+\,{\sigma^2\over \Gamma(z)}\,\Big]\,(z_*-z)\,=\,{1\over 2 m^2}\,\,,
\label{implicit-sol}
\eeq
where $m$ is the same constant as in (\ref{m}). Notice that, by taking $\sigma=0$  in (\ref{implicit-sol}) one can immediately solve for $z$, with the result:
\beq
z(\rho, \sigma=0)\,=\, z_*\,-\,{1\over 2 m^2\rho^2}\,\,.
\label{z-sigma0-gsugra}
\eeq
This result is consistent with (\ref{z-sigma0}) if the constant $g_0$ is taken to be:
\beq
 g_0\,=\,{1\over m^4}\,=\,4\pi\,g_s\,(\alpha')^2\,N_c\,\,.
 \label{g0}
 \eeq
One can check that, non-trivially, the function $z(\rho,\sigma)$ defined implicitly in (\ref{implicit-sol}) satisfies the second-order PDE  (\ref{PDE-z-unflavored}) (see appendix \ref{gaugedsugra44}).

From the gauged supergravity approach of appendix \ref{gaugedsugra44}, one can also 
find the function $g(\rho, \sigma)$ in terms of $z(\rho,\sigma)$. One obtains:
\beq
g\,=\,{2\over m^2}\,\,{(z_*-z)\,\rho^4\over 
\rho^2+{z\over \Gamma^2(z)}\,\sigma^2}\,\,.
\label{g-rho-sigma}
\eeq
As a check of (\ref{g-rho-sigma}) it is easy to verify using (\ref{z-sigma0-gsugra}) that $g(\rho, \sigma=0)=g_0=m^{-4}$. Moreover,  one can also obtain  the expression for the warp factor $H$, namely:
\beq
H\,=\,{2z  (z_*-z)\over 
m^2\,\Big[\,\rho^2+{z\over \Gamma^2(z)}\,\sigma^2\,\Big]\,\Gamma(z)}\,\,.
\label{H-rho-sigma}
\eeq
 In particular we can now obtain the value of the warp factor $H$ at $\sigma=0$. One obtains:
\beq
H(\rho,\sigma=0)\,=\,{1\over m^4\rho^4}\,\,{2m^2\rho^2 z_*-1
\over 2m^2\rho^2 z_*-\log(2m^2\rho^2)+\kappa}\,\,.
\eeq
\begin{figure}[ht]
\begin{center}
\includegraphics[width=0.65\textwidth]{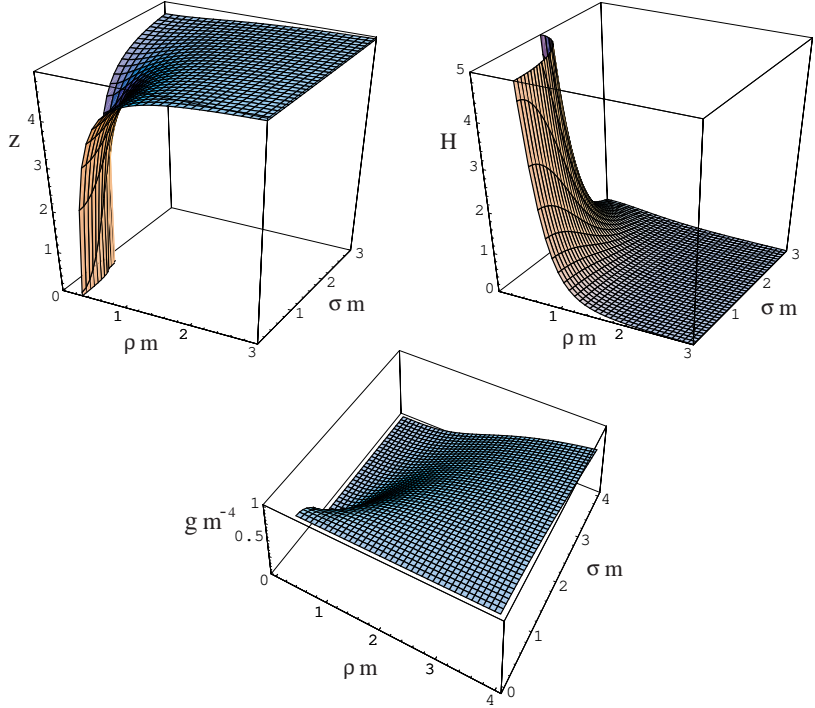}
\end{center}
\caption[Unflavor]{From left to right and from top to bottom plots of $z$, $H$ and
$g\,m^{-4}$ versus the dimensionless variables $\rho\,m$ and
$\sigma\,m$. We are setting $z_*=5$ and $\kappa=0$. } 
\label{Unflavor}
\end{figure}

Let us now study the solution for arbitrary values of $\rho$ and $\sigma$. First of all, it is clear from (\ref{implicit-sol}) that, when either $\rho$ or $\sigma$ are large, the function $z$ approaches the constant value $z=z_*$. Moreover,  by solving numerically the implicit relation (\ref{implicit-sol}) one can obtain the function $z(\rho,\sigma)$ and, then, by using this result in (\ref{H-rho-sigma}) and (\ref{g-rho-sigma}) one can obtain $H$ and $g$.  The result of this numerical analysis has been plotted in figure \ref{Unflavor}.  One important thing that can be observed in this result is that the function $z$ becomes negative  when $\rho$ and $\sigma$ are small enough and, therefore, the supergravity solution is inconsistent in this region. This phenomenon is related to the so-called enhan\c{c}on mechanism \cite{Johnson:1999qt}. Indeed, we will explicitly show in subsection \ref{probe-analysis} that when $z$ vanishes the color brane probes become tensionless, signalling the appearance of new massless degrees of freedom in the IR. One can estimate the scale at which this mechanism starts  by computing the value of $\rho$ at which $z$ vanishes for $\sigma=0$. From (\ref{z-sigma0-gsugra}) we obtain that this value of $\rho$ is given by:
\beq
\rho_*\,=\,{1\over m\sqrt{2 z_*}}\,\,.
\label{rho*}
\eeq
Thus, the asymptotic value of $z$ also controls the IR scale of our solution. 

In spite of the fact that we cannot solve analytically for $z(\rho,\sigma)$ in the implicit equation (\ref{implicit-sol}), one can solve this equation by means of an expansion around the constant asymptotic value $z=z_*$. Notice that, at zeroth order in this expansion, $\Gamma(z_*)\approx z_*$. Using this result to evaluate the left-hand side of  (\ref{implicit-sol}) one finds that, keeping the first non-trivial term, $z(\rho,\sigma)$ can be approximated to be:
\beq
z(\rho,\sigma)\,\approx\,z_*\,\,-\,{z_*\over 2m^2}\,{1\over \sigma^2\,+\,z_*\,\rho^2}\,\,.
\label{zapprox}
\eeq
Notice that this expression gives the exact result for $\sigma=0$ (see eq. (\ref{z-sigma0-gsugra})). Moreover, it gives a rather good approximation for $\sigma>0$. Actually, when one compares the function $z(\rho,\sigma)$ obtained by a numerical solution of (\ref{implicit-sol}) with the one written in (\ref{zapprox}), one realizes that the two functions only differ significantly when we approach the enhan{\c{c}}on point. Moreover, by substituting $z(\rho,\sigma)$ as given by (\ref{zapprox}) in the right-hand side of eqs. (\ref{g-rho-sigma}) and (\ref{H-rho-sigma}) one gets a very good approximation to the functions $g$ and $H$. These expressions greatly simplify if we assume that $\rho$ or $\sigma$ are large and we  keep the leading terms. In this case, one gets:
\beq
H(\rho,\sigma)\,\approx\,{z_*^2\over m^4}\,\,{1\over \Big(\,\sigma^2\,+\,z_*\,\rho^2\,\Big)^2}\,\,,
\qquad\qquad
g(\rho,\sigma)\,\approx\,{z_*^2\over m^4}\,\,{\rho^4\over \Big(\,\sigma^2\,+\,z_*\,\rho^2\,\Big)^2}\,\,,
\label{asymp-expansionHg}
\eeq
which are good estimates far from  the enhan\c{c}on point.  This analysis suggests that in the UV region the particular combination 
$\sigma^2\,+\,z_*\,\rho^2$ of the radial variables plays a relevant role. Accordingly, we define new variables $u$ and $\hat\alpha$ as follows:
\beq
u\,=\,\sqrt{\sigma^2+z_*\rho^2}\,\,,
\qquad\qquad
\tan{\hat\alpha}\,=\,{\sigma\over \sqrt{z_*}\,\rho}\,\,,
\qquad\qquad
0\le\hat\alpha\le {\pi\over 2}\,\,.
\label{u-hatalpha}
\eeq
The functions $H$ and $z$ of the metric will generally depend on both coordinates $u$ and $\hat\alpha$:
\beq
H\,=\,H(u,\hat\alpha)\,\,,
\qquad\qquad
z\,=\,z(u,\hat\alpha)\,\,.
\eeq
The metric  in the new coordinates takes the form:
\bear
&&ds^2\,=\,H^{-{1\over 2}}\,\Big[\,dx^2_{1,1}\,+\,{z\over m^2}\,
(d\theta^2+\sin^2\theta\,d\phi^2\,)\,\Big]\,+
H^{{1\over 2}}\,\Big[\,\Big(\,{\sin^2\hat\alpha\over z}+{\cos^2\hat\alpha\over z_*}\,\Big)\,du^2\,+\rc\rc
&&\qquad\qquad\,+\,u^2\,\Big(\,{\cos^2\hat\alpha\over z}+{\sin^2\hat\alpha\over z_*}\,\Big)\,d\hat\alpha^2\,+\,2u\sin\hat\alpha\cos\hat\alpha\,
\Big(\,{1\over z_*}-{1\over z}\,\Big)\,du\, d\hat\alpha\,+\,\rc\rc
&&\qquad\qquad+\,{u^2\over z}\,\sin^2\hat\alpha\,(d\psi+\cos\theta d\phi)^2\,+\,
{u^2\over z_*}\,\cos^2\hat\alpha\,d\Omega_3^2\,\Big]\,\,.
\eear
It follows from (\ref{zapprox}) and  (\ref{asymp-expansionHg}) that, 
when $u$ is large, the functions $H$ and $z$ become independent of $\hat\alpha$ and are given by:
\beq
z\to z_*\,\,,\qquad\qquad
H\to {z_*^2\over m^4}\,\,{1\over u^4}\,\,.
\eeq
Using these values for $z$ and $H$ in the metric ansatz , we get the following:
\bear
&&ds^2_{UV}\approx {m^2\over z_*}\,u^2\,\Big[\,
dx^2_{1,1}\,+\,{z_*\over m^2}\,(d\theta^2+\sin^2\theta\,d\phi^2\,)\,\Big]\,+\,
{1\over m^2}\,{du^2\over u^2}\,+\,\rc\rc
&&\qquad\qquad+\,{1\over m^2}\,
\Big[\,d\hat\alpha^2\,+\,\sin^2\hat\alpha\,(d\psi+\cos\theta d\phi)^2\,+\,
\cos^2\hat\alpha\,d\Omega_3^2\,\Big]\,\,.
\label{UVmetric}
\eear
The first line in (\ref{UVmetric}) is the metric of an $AdS_5$ space, in which two of its directions are compactified on an $S^2$. The second line of (\ref{UVmetric}) is the line element of an $S^5$ fibered over the $S^2$.  Notice that this result is in agreement with the origin of the solution as wrapped D3-branes in  gauged supergravity, with the fibering being given by the gauge field in this approach. Actually, the variables $u$ and $\hat\alpha$ can be easily identified with the ones used in the gauged sugra approach. In this case the UV region corresponds to the large $\tau$ region in eq. (\ref{gsugrasolution}) of appendix \ref{gaugedsugra44}. From the definition of $\rho$ and $\sigma$ in (\ref{sigma-rho}) one has the following relation, in the UV,  between the coordinates $(u, \hat\alpha)$  and the five-dimensional coordinates $(\tau, \tilde\theta)$:
\beq
u\,=\,{\sqrt{\alpha}\over m}\,e^{\tau}\,\,,\qquad\qquad
\hat\alpha\,=\,\tilde\theta\,\,,
\eeq
where $\alpha$ is the integration constant appearing in (\ref{gsugrasolution}). Notice that, with this identification, the metric (\ref{gaugesugraD3metric}) becomes  for large $\tau$ exactly the one written in (\ref{UVmetric}).

\section{The dual of the (4,4) theory with flavor}
\label{44flavored}

In the previous section we have succeeded in finding the gravity dual of ${\cal N}=(4,4)$ super Yang-Mills theory in two space-time dimensions with a vector multiplet in the adjoint representation of the gauge group. The main purpose of the present section is the extension of this result to include matter hypermultiplets in the fundamental representation.  We will follow the now standard procedure of adding flavor branes to the setup of color branes discussed in section \ref{44unflavored}. These flavor branes add a new open string sector to the theory, which can be interpreted as the gravity dual of the addition of matter hypermultiplets on the gauge theory side. The flavor branes should fill the Minkowski spacetime and should extend along some holographic  direction while wrapping some internal submanifold of the Calabi-Yau manifold. We will show below that the appropriate flavor branes for our case are D3-branes
extended along the Minkowski space-time directions $x^0, x^1$, as well as
the directions $(\sigma, \psi)$ of the  normal bundle of the Calabi-Yau, according to the array:
 \begin{center}
\begin{tabular}{|c|c|c|c|c|c|c|c|c|c|c|}
\multicolumn{3}{c}{ }&
\multicolumn{4}{c}{$\overbrace{\phantom{\qquad\qquad\qquad}}^{\text{CY}_2}$}\\
\hline
&\multicolumn{2}{|c|}{$\mathbb{R}^{1,1}$}
&\multicolumn{2}{|c|}{$S^2$}
&\multicolumn{2}{|c|}{$N_2$}
&\multicolumn{4}{|c|}{$\mathbb{R}^{4}$}\\
\hline
$N_c$\,\,\,D$3$ (color) &$-$&$-$&$\bigcirc$&$\bigcirc$&$\cdot$&$\cdot$&$\cdot$&$\cdot$&$\cdot$&$\cdot$\\
\hline
$N_f$\,\,\,D$3$ (flavor) &$-$&$-$&$\cdot$&$\cdot$&$-$&$-$&$\cdot$&$\cdot$&$\cdot$&$\cdot$\\
\hline
\end{tabular}
\label{44flavored-array}
\end{center}
The $N_f$ D3-branes in this array are located at a fixed point on the $S^2$ sphere parameterized by the angles $(\theta, \phi)$ and at a fixed point in the transverse $\mathbb{R}^{4}$. Actually, the value $\rho_Q$ of the coordinate $\rho$ of the flavor branes represents the distance between the two sets of branes and, according to the radius-energy relation (\ref{radius-energy}), is related to the mass $m_Q$ of the hypermultiplets as follows:
\beq
m_Q={\rho_Q\over 2\pi\alpha'}\,\,.
\eeq
One of the arguments in favor of considering these configurations as the right ones to add flavor to the ${\cal N}=(4,4)$ gauge theory is the fact that this setup preserves the same supersymmetries \footnote{This does not happen if one adds D7-branes as flavor branes.} as that considered in section \ref{44unflavored}. This statement is   verified in appendix \ref{Higgs-embeddings} by using the kappa symmetry of the DBI action. Notice also that, as expected on general grounds,  the flavor D3-branes of our setup are extended along the non-compact direction of the CY cone.  In general, if  $X^{M}$ denote  ten-dimensional coordinates, the D3-brane embedding will be characterized by a set of functions $X^{M} (\xi^a)$, where $\xi^a$ ($a=0,\cdots, 3$) is a system of worldvolume coordinates of the D3-brane.  In what follows we shall choose the 
following set of worldvolume coordinates:
\beq
\xi^{a}\,=\,(x^0, x^1, \psi, \sigma)\,\,,
\label{D3-wvcoordinates}
\eeq
and our embedding will be characterized by having a constant value of the remaining ten-dimensional coordinates, namely:
\beq
\rho=\rho_Q\,\,,\qquad\qquad
\theta\,,\,\phi\,,\,\alpha_i\,=\,{\rm constant}\,\,.
\label{Coulomb-flavor-embedding}
\eeq
Notice the particular ordering we are using for the coordinates $\xi^a$ in (\ref{D3-wvcoordinates}). This ordering determines the orientation of the worldvolume of the flavor branes, which in our case differs from the one induced by the ten-dimensional background.

When $N_f<< N_c$ one can neglect the backreaction of the flavor branes on the geometry and, thus, one can treat them as probes. This is the so-called quenched approximation, which corresponds, in the field theory side, to suppressing quark loops (that contribute with powers of $1/N_c$) in the 't Hooft large $N_c$ expansion.  By analyzing the normalizable fluctuations of these brane probes one can extract the meson spectrum of the model (see \cite{Erdmenger:2007cm} for a review). We will postpone this analysis until subsection \ref{Quenched-mesons}. In the remainder of this section we will study how to incorporate the effect of the backreaction in our brane setup.

\subsection{Including the backreaction}

When the number of flavors $N_f$ is of the same order as the number of colors $N_c$ the backreaction of the flavor branes cannot be neglected anymore and one is led to consider the full coupled gravity plus branes system. Including the backreaction is the analogue, on the field theory side, of considering the effects of quark loops, which are suppressed in the 't Hooft large $N_c$ limit. In what follows in this subsection we will assume that $N_f$ is large and $N_f/N_c$ is fixed.  To construct the corresponding supergravity dual we will follow the approach pioneered in ref. \cite{Casero:2006pt} and we shall consider a suitable continuous distribution of flavor branes (see \cite{noncritical} for a similar analysis in the context of non-critical string theory).  This approach has been successfully applied to study backgrounds with different amount of supersymmetry which are dual to gauge theories in several space-time dimensions \cite{Casero:2007pz}-\cite{HoyosBadajoz:2008fw}. In section \ref{conclusions} we will comment on an interesting subtlety about the smearing and the validity of the DBI+WZ action

The starting point in our analysis is the observation that the embeddings considered at the beginning of this  section are mutually supersymmetric for any value of the coordinates transverse to the flavor D3-brane. Therefore, when $N_f\to\infty$, we can homogeneously distribute the $N_f$ flavor branes in some of their transverse directions. Actually,  we shall locate them at a particular value $\rho=\rho_Q$ of the $\rho$ coordinate, and we will smear them along the angular directions $(\theta, \phi)$ as well as along the three-angles $\alpha_i$ of the three-sphere.   

The action of a stack of $N_f$ D3-branes is given by:
\beq
S_{flavor}\,=\,S_{DBI}+S_{WZ}\,\,,
\eeq
where the DBI and WZ terms are given by:
\beq
S_{DBI}\,=\,-T_3\,\sum_{N_f}\,\int_{{\cal M}_4}\,d^{4}\xi\,\sqrt{-\det\hat G_4}\,\,,
\qquad\qquad
S_{WZ}\,=\,T_3\,\sum_{N_f}\,\int_{{\cal M}_4}\,\hat C_4\,\,,
\label{flavor-DBI-WZ}
\eeq
with $\hat G_4$ being the induced metric on the worldvolume ${\cal M}_4$ and 
the hat over the RR four-form potential denotes its pullback to ${\cal M}_4$.
The smearing procedure\footnote{Notice that we add brane sources via the DBI+WZ action. These are the branes we smear. This is different from what is studied in \cite{Marolf:1999uq} and hence the comments in that paper do not apply to our setups. On the contrary, our approach is closer to \cite{Koerber:2007hd}, where a general analysis of the eqs. of motion was performed and shown to be always consistent. We are grateful to Ingo Kirsch for discussions on this point.} amounts to performing the following substitution in the WZ term 
in (\ref{flavor-DBI-WZ}):
\beq
\sum_{N_f}\,\int_{{\cal M}_4}\,\hat C_4\,\rightarrow\,\int_{{\cal M}_{10}}\,
\Omega\wedge C_4\,\,,
\eeq
where $\Omega$ is a six-form proportional to the volume  form of the  space transverse to the worldvolume of the flavor brane:
\beq
\Omega\,=\,-{N_f\over 8\pi^3}\,\,{\rm Vol}\,({\cal Y}_6)\,\,,
\label{Omega}
\eeq
with ${\rm Vol}\,({\cal Y}_6)$ being:
\beq
{\rm Vol}\,({\cal Y}_6)\,=\,\delta(\rho-\rho_Q)\,d\rho\wedge \omega_3\wedge\omega_2\,\,.
\eeq
The normalization constant in $\Omega$ has been chosen to satisfy the normalization condition:
\beq
\int \Omega\,=\,-N_f\,\,,
\eeq
where the minus sign is due to the orientation of the worldvolume. Therefore, after the smearing, the WZ term of the action of the flavor branes takes the form:
\beq
S_{WZ}\,=\,T_3\,\,\int_{{\cal M}_{10}}\Omega\wedge C_4\,\,.
\label{smearedWZ}
\eeq
In the DBI part of the action  the smearing is performed by means of the substitution:
\beq
\sum_{N_f}\,\int_{{\cal M}_4}\,d^{4}\xi\,\sqrt{-\det\hat G_4}\,
\rightarrow\,\int_{{\cal M}_{10}}\,d^{10} x\,
\sqrt{- \det G}\,\big|\,\Omega\,\big|\,\,,
\eeq
where $\big|\,\Omega\,\big|$ is the modulus of the form $\Omega$, defined as:
\beq
\Big|\,\Omega\,\Big|\,=\,\sqrt{{1\over 6!}\,
\Omega_{M_1\cdots M_6}\,
\Omega_{N_1\cdots N_6}\,\,
\prod_{k=1}^{6}\,G^{M_k N_k}}\,\,.
\label{modulusOmega}
\eeq
The smeared DBI+WZ action of the flavor branes is thus:
\beq
S_{flavor}\,=\,-T_3\,
\int_{{\cal M}_{10}}\,d^{10} x\,
\sqrt{- \det G}\,\big|\,\Omega\,\big|\,+\,
T_3\,\,\int_{{\cal M}_{10}}\Omega\wedge C_4\,\,.
\label{smearedDBI}
\eeq

We will assume that the metric of the backreacted background will be still given by the ansatz (\ref{D3metric}), {\it i.e.} it can be written in terms of the warp factor $H(\rho,\sigma)$ and  the function $z(\rho,\sigma)$. On the contrary,  it is evident from the form of the WZ action in (\ref{smearedWZ}) that the flavor branes act as sources of the RR form $F_5$ and, as a consequence, they will induce a violation of its Bianchi identity in the backreacted geometry. Therefore, to obtain the backreacted background we are looking for, we will have to modify our ansatz (\ref{F5})-(\ref{CalC4}) for $F_5$.  It is clear from (\ref{smearedWZ})  that the modified Bianchi identity in this case is:
\beq
dF_5\,=\,-2\pi\,g_s\,(\alpha')^2\,N_f\,\delta(\rho-\rho_Q)\,d\rho\wedge \omega_3\wedge\omega_2\,\,.
\label{newBianchi}
\eeq
Accordingly,  let us represent $ F_5$ as in (\ref{F5}) with ${\cal F}_5$ being given by:
\beq
{\cal F}_5\,=\,f_5\,-\,2\pi\,g_s\,(\alpha')^2\,N_f\,\Theta(\rho-\rho_Q)\,\omega_3\wedge \omega_2\,\,,
\label{calF5-flavored}
\eeq
with  $f_5$ such that  $df_5=0$.  We shall represent $f_5$ in terms of a potential ${\cal C}_4$ as $f_5=d{\cal C}_4$, where  ${\cal C}_4$ is parameterized by the function $g(\rho,\sigma)$ as in (\ref{CalC4}).

Proceeding as in the unflavored case and substituting our ansatz for $F_5$ in the equations for the SUSY variations of the  dilatino and gravitino (see appendix \ref{BPSequations}), we get the following set of BPS equations:
\bear
m^2\,\big[\,g\,-\,2\pi\,g_s\,(\alpha')^2\,N_f\,\Theta(\rho-\rho_Q)\,\big]\,=\,\rho^3\,z'\,\,,\rc\rc
m^2\,H\,=\,{z\dot z\over \sigma}\,\,,\rc\rc
g'\,=\,-\sigma\,\rho^3\,\dot H\,\,,\rc\rc
\dot g\,=\,{\sigma \rho^3\over z}\,\,H'\,-\,{\sigma\over z^2}\,H\,m^2\,
\big[\,g\,-\,2\pi\,g_s\,(\alpha')^2\,N_f\,\Theta(\rho-\rho_Q)\,\big]
\,\,.
\label{flavored-BPSsystem}
\eear
Clearly, when $N_f=0$  or $\rho<\rho_Q$  the system (\ref{flavored-BPSsystem}) reduces to (\ref{BPSsystem}). Moreover, one can prove that $z(\rho,\sigma)$ satisfies the following PDE:
\beq
\rho\,z\,(\dot z\,-\,\sigma\,\ddot z\,)\,=\,\sigma\,(\, \rho\dot z^2\,+\,\rho z''\,+\,3 z'\,)\,+\,
{N_f\over 2 N_c}\,{\sigma\over m^2\,\rho^2}\,\,\delta (\rho-\rho_Q)\,\,.
\label{flavoredPDE}
\eeq
Notice that the flavors contribute in (\ref{flavoredPDE}) as a source localized at $\rho=\rho_Q$.

The set of projections to be imposed on the Killing spinors in order 
 to arrive at the system (\ref{flavored-BPSsystem}) is just the one written in (\ref{eta-projections}), {\it i.e.} they are  same as in the unflavored case. Therefore, any solution of (\ref{flavored-BPSsystem}) preserves eight supersymmetries. Moreover, as 
in (\ref{BPSsystem}), the last equation in (\ref{flavored-BPSsystem}) can be derived from the first two. On the other hand it can be checked that the equation (\ref{newBianchi}) for $F_5$ is just a consequence of the system (\ref{flavored-BPSsystem}) (see appendix \ref{BPSequations}). We also verify in appendix \ref{BPSequations} that the Einstein equations, including the contribution of the DBI term of the flavor branes action (\ref{smearedDBI}), are satisfied by any solution of (\ref{flavored-BPSsystem}).

As in the unflavored case, we can integrate the function $z$ for $\sigma=0$. Indeed, it follows from (\ref{flavored-BPSsystem}) that $g$ is independent of $\rho$ when the variable $\sigma$ vanishes. Let  $g_0$ denote  this constant value of $g$. Then for $\rho >\rho_Q$, we have:
\beq
z'(\rho, 0)\,=\,m^2\,\Big[\,g_0\,-\,2\pi g_s\, (\alpha')^2\,N_f\,\Big]\,\,{1\over \rho^3}\,\,,
\qquad\qquad (\rho>\rho_Q)\,\,.
\eeq
Using for $g_0$ the same value as in the unflavored case, the above equation becomes:
\beq
z'(\rho, 0)\,=\,4\pi\,m^2\,g_s\,(\alpha')^2\,\Big[\,N_c\,-\,{N_f\over 2}\,\Big]
\,\,{1\over \rho^3}\,\,,
\qquad\qquad (\rho>\rho_Q)\,\,.
\label{zprime-sigmazero-flavored}
\eeq

\begin{figure}[ht]
\begin{center}
\includegraphics[width=0.75\textwidth]{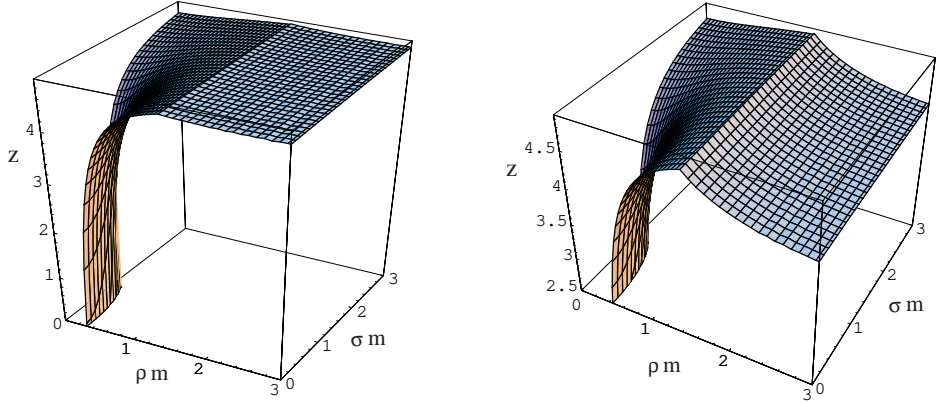}
\end{center}
\caption[Flavored]{Plots of $z(\rho,\sigma)$ for the backreacted case. On the left
we present the solution with $N_f = N_c$ and on the right that
corresponding to $N_f=7N_c$. On the right graph we have zoomed in on the
region of interest so that one can easily see that the slope becomes
negative at $\rho=\rho_Q$. We have set
$z_*=5$ and $\rho_Q=1.5$. } 
\label{Flavored}
\end{figure}
\noindent When $\rho<\rho_Q$ one simply puts $N_f=0$ on the right-hand side of (\ref{zprime-sigmazero-flavored}) and  (\ref{zprime-sigmazero}) is recovered. Notice that $z'(\rho, 0)$ jumps when $\rho$ passes through the point $\rho=\rho_Q$. Actually, the slope of the 
$z(\rho, 0)$ curve becomes negative when $N_f>2N_c$. We will argue below that this is the geometric counterpart of the behavior of the beta function  in  the field theory dual. Let us now integrate (\ref{zprime-sigmazero-flavored}) and impose the condition that the solution  $z(\rho, 0)$ is continuous at $\rho=\rho_Q$ (although its first $\rho$ derivative is not). One gets:
 \beq
 z(\rho, 0)\,=\,z_{*}\,-\,{\pi m^2 g_s\,(\alpha')^2\over \rho_Q^2}\,N_f\,
 \Theta(\rho-\rho_Q)\,-\,
 {2\pi\, m^2\,g_s\,(\alpha')^2\over \rho^2}
 \,\Big[\,N_c\,-\,{N_f\over 2}\, \Theta(\rho-\rho_Q)\,\Big]\,\,,
 \label{z-sigma0-flavored}
 \eeq
where the constant $z_{*}$ is the same as in (\ref{z-sigma0-gsugra}).  In order to evaluate $z$ for arbitrary values of $\rho$ and $\sigma$ we must numerically integrate the flavored system (\ref{flavored-BPSsystem}). The result of this numerical integration in two particular cases is presented in figure \ref{Flavored}. To obtain these results we have assumed that the solution reduces to the unflavored one for $\rho\le\rho_Q$. We also assumed that $g(\rho,\sigma)$ is continuous at $\rho=\rho_Q$ and thus (see the first equation in (\ref{flavored-BPSsystem})) $z'(\rho, \sigma)$ has a discontinuity at $\rho=\rho_Q$ which is independent of $\sigma$ and given by:
 \beq
 z'(\rho_Q+\epsilon, \sigma)\,-\, z'(\rho_Q-\epsilon, \sigma)\,=\,-{1\over m^2\rho_Q^3}\,
 {N_f\over 2N_c}\,\,.
 \eeq
In the plots of figure \ref{Flavored} it is quite evident the qualitative difference between the cases with $N_f<2N_c$ and $N_f>2N_c$. Indeed, when $N_f<2N_c$, the function $z$ continues to grow for $\rho>\rho_Q$ as we move away from the origin in the $(\rho,\sigma)$ plane, whereas for  $N_f>2N_c$ there is a change of behavior at $\rho=\rho_Q$, where $z$ starts to decrease and,  if $\rho_Q$ is not too large\footnote{One can estimate the maximal value of $\rho_Q$ such that $z$ becomes negative when $N_f>2N_c$ from the exact expression (\ref{z-sigma0-flavored}) of $z$ at $\sigma=0$. Indeed, a simple analysis of (\ref{z-sigma0-flavored}) shows that this maximal value of $\rho_Q$ is $\sqrt{{N_f\over 2N_c}}\,\rho_*$, where $\rho_*$ is the IR scale defined in (\ref{rho*}).}, $z$ becomes negative  and the space ends.

 \section{Matching the field theory} 
 \label{FT}
 The two-dimensional gauge theories with ${\cal N}=(4,4)$ supersymmetry can be obtained by dimensional reduction of four-dimensional ${\cal N}=2$ gauge theories \footnote{Alternatively, they can also be obtained by dimensional reduction of ${\cal N}=(1,0)$ supersymmetric theories in six dimensions.}. As in four-dimensional ${\cal N}=2$ theories, the ${\cal N}=(4,4)$ theories in $d=2$ have two massless representations, namely the vector multiplet and the hypermultiplet. 
 
 The field content of the $d=2$, ${\cal N}=(4,4)$ vector multiplet can be obtained by dimensional reduction of the corresponding vector multiplet of the  $d=4$, ${\cal N}=2$  theory. Recall that the latter is composed of a vector field $A_\mu$, a Dirac spinor $\lambda$ and a complex scalar $\phi$.  To obtain the dimensional reduction of the vector field $A_\mu$ one simply distinguishes the components with $\mu=0,1$ from those with $\mu=2,3$.  This results in a two-dimensional vector field and two real scalars. Moreover, the dimensional reduction of the $d=4$ Dirac spinor $\lambda$ gives rise to two complex fermions in $d=2$ whereas, by dropping the dependence on the coordinates $x^2$ and $x^3$ of the $d=4$ complex scalar 
$\phi$, one gets two more real scalars in the two-dimensional theory. Therefore, the total content of the $d=2$, ${\cal N}=(4,4)$ vector multiplet is one vector field, two Dirac fermions and four real scalars (which parameterize the Coulomb branch of the theory). 
 
 The $d=2$, ${\cal N}=(4,4)$ hypermultiplet  is similarly obtained from the $d=4$, ${\cal N}=2$  hypermultiplet. The latter contains a Dirac fermion and two complex scalars which, upon dimensional reduction, give rise to two Dirac fermions and four real scalars. These scalars parameterize the Higgs branch of the theory \footnote{The vector and hyper multiplets of the $d=2$, ${\cal N}=(4,4)$  theories can also be regarded as being  composed of supermultiplets of    $d=2$, ${\cal N}=(2,2)$ supersymmetry. In terms of the ${\cal N}=(2,2)$ superalgebra the vector multiplet contains a chiral multiplet and a twisted chiral multiplet, while the hypermultiplet decomposes into two chiral multiplets.}.  
 
 The R-symmetry of the $(4,4)$ theory is $SU(2)_R \times Spin(4)$; the first part is inherited from the R-symmetry of the six dimensional theory with eight supercharges and the $Spin(4)\sim SO(4)$ is coming from the dimensional reduction on four directions of the initial six. Notice that our background (\ref{D3metric}) realizes the $SO(4)$ explicitly as rotations in the three-sphere whose line element is $d\Omega_3^2$, while it only realizes explicitly a $U(1)$ subgroup of the $SU(2)_R$, represented by traslations in the angle $\psi$.

 Given the field theory content described above, it is possible to find the running coupling constant in perturbation theory for the ${\cal N}=(4,4)$ super Yang-Mills theory \cite{Fradkin}. Here we will follow the presentation of the appendix C of \cite{Divecchia}, for the particular case of two-dimensional theories. The one loop running coupling constant is given by:
\beq
 {1\over g_{YM}^2(\mu)}\,=\, 
 {1\over g^2_{YM}}\,\Big(\,1\,+\,{g^2_{YM}\over 4\pi\mu^2}\,b\,\,\Big)\,\,,
 \eeq
 where $\mu$ is the energy scale and $b$ is a constant which depends on the field content of the theory. Let us consider a theory with $n_s$ scalar fields, $n_v$ vector fields and $n_f$ Dirac fermions. In this case the value of $b$ is the following:
 \beq
 b\,=\,{n_s\over 6}\,c_s\,-\,4\,n_v\,c_v\,+\,{2\over 3}\,n_f\,c_f\,\,,
 \eeq
 where $c_I$ for $I=s,v,f$ is the normalization constant of the generators of the gauge group in the representation of the different fields ($tr (T^aT^b)=c_I\delta ^{ab}$). With the conventions chosen in \cite{Divecchia} one has for the gauge group $SU(N_c)$:
 \beq
 c^{{\rm fundamental}}\,=\,{1\over 2}\,\,,
 \qquad\qquad
 c^{{\rm adjoint}}\,=\,N_c\,\,.
 \eeq
 Therefore,  if all fields are in the adjoint of $SU(N_c)$, one gets:
 \beq
 b^{{\rm adjoint}}\,=\,\Big[\,{n_s\over 6}\,-\,4n_v\,+\,{2\over 3}\,n_f\,\Big]\,N_c\,\,,
 \eeq
 whereas if they are in the fundamental representation the corresponding value of $b$ is:
 \beq
 b^{{\rm fundamental}}\,=\,{n_s\over 12}\,-\,2\,n_v\,+\,{n_f\over 3}\,\,.
 \eeq
 As discussed above,  in the vector multiplet of (4,4) supersymmetry in two dimensions all fields are in the adjoint and $n_s=4$, $n_f=2$ and $n_v=1$. Therefore, one has:
 \beq
 b^{{\rm vector \,\,multiplet}}\,=\,-2N_c\,\,.
 \eeq
On the other hand, for a hypermultiplet in the fundamental representation $n_s=4$ and $n_f=2$ and the value of $b$ in this case is given by:
 \beq
 b^{{\rm hypermultiplet}}\,=\,1\,\,.
 \eeq	
The total value of $b$ in a ${\cal N}=(4,4)$ gauge theory with $N_f$ matter hypermultiplets is thus $-2N_c+N_f$ and the running coupling constant is:
\beq
 {1\over g_{YM}^2(\mu)}\,=\, 
 {1\over g^2_{YM}}\,\Big[\,1\,-\,{g^2_{YM}\over 2\pi\mu^2}\,\Big(\,N_c-{N_f\over 2}\,\Big)
 \,\,\Big]\,\,.
 \label{gYM-QFT}
 \eeq
Eq. (\ref{gYM-QFT}) shows that the (4,4) theory has negative beta function when $N_f<2N_c$, while for $N_f>2N_c$  the beta function changes its sign and becomes positive. In the borderline case $N_f=2N_c$ the one-loop beta function vanishes and the coupling does not run any more in perturbation theory. In the next subsection we will be able to reproduce the running (\ref{gYM-QFT}) from a probe analysis of our gravity solutions, both from the unflavored and backreacted geometries, and, therefore, to characterize geometrically how the different regimes of $N_c$ and $N_f$ are encoded in our solutions.

Some other interesting non-perturbative aspects of 2d ${\cal N}=(4,4)$ theories, mostly related to conformal IR points, were studied in \cite{Diaconescu:1997gu,Aharony:1999dw,Brodie:1997wn}.

 \subsection{Probe analysis}
 \label{probe-analysis}
 In order to extract the information on the gauge theory contained in the gravity dual we have found, let us study the dynamics of a D3-brane probe moving under the influence of the metric and RR form of the background. The action of such a probe will be given by the standard sum of DBI and WZ terms:
  \beq
 S\,=\,-T_3\,\int d^4\xi\,e^{-\Phi}\,\,\sqrt{-\det (\,\hat G_4\,+\,2\pi\alpha'\,F)}\,+\,
 T_3\,\int \hat C_4\,\,,
 \label{actionD3}
 \eeq
 where $\xi^a$ ($a=0,\cdots, 3$) is a set of worldvolume coordinates, $F$ is the field strength of the worldvolume gauge field and the hat over $G_4$ and $C_4$ denotes the pullback over the worldvolume of the D3-brane. The probe we are interested in is 
 a color brane probe extended along the directions 
 $\xi^a=(x^0, x^1, \theta,\phi)$ and located at a fixed  value of the $\sigma$  coordinate at  a constant value of all the remaining coordinates.  We shall first consider the configuration in which the worldvolume gauge field vanishes.  For such a configuration, 
the induced metric on the brane worldvolume is given by:
 \beq
 \hat G^{(4)} _{ab}\,d\xi^a d\xi^b\,=\,H^{-{1\over 2}}\,dx^2_{1,1}\,+\,
 {z H^{-{1\over 2}}\over m^2}\,\,\Big[\,(d\theta)^2\,+\,\sin^2\theta\,
 \Big(1\,+\,\sigma^2\,{m^2 H\over z^2}\,\cot^2\theta\,\Big)\,(d\phi)^2\,\Big]\,\,.
 \eeq
 The determinant of this induced metric is:
 \beq
 \sqrt{-\det \hat G_4}\,=\,{z\sin\theta\over m^2 H}\,\,
 \sqrt{1\,+\,\sigma^2\,{m^2 H\over z^2}\,\cot^2\theta}\,\,.
 \eeq
 Moreover, the pullback of the RR four-form potential  (\ref{C4}) is:
 \beq
 \hat C_4\,=\,{z\sin\theta\over m^2 H}\,\,dx^0\wedge dx^1\wedge d\theta\wedge d\phi
 \,\,.
 \eeq
 If we now add the DBI and WZ contributions as in (\ref{actionD3}), we get:
  \beq
 S_{pot}\,=\,-T_3\,\int d^2 x\,d\theta d\phi\,\,{z\sin\theta\over m^2 H}\,\,
 \Big[\, \sqrt{1\,+\,\sigma^2\,{m^2 H\over z^2}\,\cot^2\theta}\,-\,1\,\Big]\,\,.
 \label{potD3}
 \eeq
 Notice that the right-hand side of (\ref{potD3}) is minus the static potential between the stack of $N_c$ color branes and the additional probe. This potential is in general non-vanishing, which means that there is a non-zero force between these branes. However, the potential does vanish for $\sigma=0$, which should be interpreted as the supersymmetric locus  of the branes inside the Calabi-Yau space. Notice that the branes can be at any point in the $\mathbb{R}^{4}$ directions. Let us parameterize these flat directions in terms of four cartesian coordinates $Z^i$ ($i=1,\cdots, 4$) as follows:
\beq
d\rho^2\,+\,\rho^2\,d\Omega_3^3\,=\,(dZ^i)^2\,\,.
\eeq
Next, let us assume that we are at the no-force point $\sigma=0$ and let us allow the 
 $Z^i$ scalars to depend only on the Minkowski coordinates $x^{\mu}=(x^0, x^1)$. Moreover, let us assume that we switch on  the worldvolume gauge field $F$ in such a way that its only non-vanishing components are those along the unwrapped directions 
$x^{\mu}$.  The lagrangian density of the DBI term of the probe brane action is:
 \beq
 {\cal L}_{DBI}\,=\,-T_3\,{z\over m^2 H}\,\sin\theta\,\Big[\,1\,+\,
 {(2\pi\alpha')^2\over 2}\,H\,F_{\mu\nu}\,F^{\mu\nu}\,+\,
 H\,(\partial_{\mu} Z^i)^2\,\Big]^{{1\over 2}}\,\,,
 \label{DBI-F-phi}
 \eeq
 where all functions are evaluated at $\sigma=0$. 
 Let us also re-express the transverse coordinates $Z^i$ in terms of the scalar fields $\varphi^i$ of the gauge theory living on the color brane as:
 \beq
 Z^i\,=\,2\pi\alpha'\,\varphi^i\,\,.
 \label{Z-varphi}
 \eeq
 In order to get the gravity expression of 
the Yang-Mills coupling we have to expand the lagrangian density (\ref{DBI-F-phi})  (integrated over the angular variables ($\theta, \phi$)) up to quadratic terms  in the fields and remember that passing from the abelian to the non-abelian theory amounts 
to substituting:
 \beq
 F_{\mu\nu}F^{\mu\nu}\to tr[ F_{\mu\nu}F^{\mu\nu}]\,=\,{1\over 2}\,
  F_{\mu\nu}^a F^{\mu\nu, a}\,\,,
 \eeq
 and similarly for the scalars $\varphi^i$. By doing so, we get:
 \beq
 \int\,d\theta d\phi\, {\cal L}_{DBI}\Big |_{quadratic}\,=\,
 -{(2\pi)^3\,(\alpha')^2\,T_3\over m^2}\,\,z\,\Bigg[\,{1\over 2}\,
  tr[ F_{\mu\nu}F^{\mu\nu}]\,+\,tr[\partial_{\mu}\varphi^i\partial^{\mu}\varphi^i]\,\Bigg]\,\,.
  \label{quadraticDBI}
 \eeq
 The canonical kinetic  term for the gauge field is:
 \beq
- {1\over  2g^2_{YM}(\mu)}\,tr[ F_{\mu\nu}F^{\mu\nu}]\,=\,-{1\over 4 g^2_{YM}(\mu)}\,
 F_{\mu\nu}^a F^{\mu\nu, a}\,\,,
 \eeq
 where we have introduced a dependence of the Yang-Mills coupling on a renormalization  energy scale $\mu$. This energy scale should be related to the holographic coordinate $\rho$ (see below). 
 
  By comparing with the expression  (\ref{quadraticDBI}) obtained by expanding the DBI action and taking into account that $(2\pi)^3\,(\alpha')^2\,T_3\,=\,1/g_s$, we arrive at the following gravity value of the Yang-Mills coupling:
 \beq
 {1\over g^2_{YM}}\,=\,{z(\rho, \sigma=0)\over m^2 g_s}\,\,.
 \label{gYM-gravity}
 \eeq
 Taking into account the relation (\ref{Z-varphi})  between the coordinates $Z^i$ and the scalar fields $\varphi^i$,  which do not acquire anomalous dimensions,  it is natural to assume that the energy scale $\mu$ is related to the holographic coordinate $\rho$ as follows:
  \beq
 \rho\,=\,2\pi\,\alpha'\,\mu\,\,.
 \label{radius-energy}
 \eeq
 Let us now substitute in (\ref{gYM-gravity}) the value of $z(\rho, \sigma=0)$ as given by (\ref{z-sigma0-flavored}) for the general flavored solution. For concreteness we will assume that $\rho>\rho_Q$ or, equivalently, that $\mu>m_Q$.  To obtain the corresponding result for $\mu\le m_Q$ one should simply take $N_f=0$ in the equations that follow. After writing $z(\rho, \sigma=0)$  in terms of $\mu$, we get the following YM coupling:
 \beq
 {1\over g_{YM}^2(\mu)}\,=\,{z_{*}\over m^2\, g_s}\,-\,{N_f\over 4\pi m_Q^2}
 \,-\,{1\over  2\pi \mu^2}\,\,\Big(\,N_c\,-\,{N_f\over 2}\,\Big) \,\,.
 \eeq
 Let us next define the bare UV  Yang-Mills  coupling as:
 \beq
 {1\over g^2_{YM}}\,=\,{z_{*}\over m^2\, g_s}\,-\,{N_f\over 4\pi m_Q^2}
 \label{GYM-zinfty}\,\,.
 \eeq
 Then, we get the following running coupling:
 \beq
 {1\over g_{YM}^2(\mu)}\,=\,
 {1\over g^2_{YM}}\,\Big(\,1\,-\,{g^2_{YM}\over 2\pi\mu^2}\,\big(\,N_c\,-\,{N_f\over 2}\,\big)\Big)\,\,.
 \label{44gymrunning-gravity}
 \eeq
 By comparing (\ref{44gymrunning-gravity}) with the result obtained in the previous subsection for the one loop running coupling constant  in perturbation theory (eq. (\ref{gYM-QFT})) we see that they coincide exactly. Moreover, it follows from (\ref{quadraticDBI}) that the action for the scalar fields takes the form:
  \beq
 -{1\over g^2_{YM}(\mu)}\,tr[\,\partial_{\mu}\varphi^{i}\,\partial^{\mu}\varphi^i\,]\,\,.
 \eeq
 This implies that the metric of the moduli space in the Coulomb branch of the theory  is of the form:
 \beq
 ds^2_{{\cal M}}\,=\,{1\over g^2_{YM}(\mu)}\,\,\Big(\,d\mu^2\,+\,\mu^2\,d\Omega_3^2\,\Big) \,\,,
 \eeq
 which is the result also obtained in the field theory \cite{Diaconescu:1997gu}. Thus, we have succeeded in recovering from our gravity solutions the relevant perturbative information of the Coulomb branch of the gauge theory living on the branes.
 
Let us now consider the possibility of placing the brane probe at $\sigma\not=0$. We have already shown that if $\sigma$ is constant and non-vanishing the energy of the vacuum configuration is non-zero, signalling the breaking of supersymmetry of this configuration. One can confirm this conclusion by studying the kappa-symmetric embeddings of probes (see appendix \ref{Higgs-embeddings}).  Actually, one can argue that moving away from the point $\sigma=0$ is dual to switching on a Fayet-Ilioupoulos parameter on the gauge theory side. The rigid displacement of the color brane in the direction of the $\sigma$  coordinate gives rise to a configuration of the probe that breaks the supersymmetry of the $\sigma=0$ point. In order to recover this supersymmetry we must allow some non-trivial profile of the $\sigma$ coordinate which, in some cases, can be interpreted as the recombination of the color and flavor branes. We will explore this possibility in subsection \ref{Higgs} in our description of the string dual to the  Higgs branch of the theory.

\subsection{The Higgs branch}

\label{Higgs}
In this subsection we will present a holographic description of the Higgs branch of the $d=2$,  ${\cal N}=(4,4)$ gauge theory. We begin by recalling that the Higgs branch is a phase of the theory in which the quark fields $Q$ and $\tilde Q$ acquire a non-vanishing expectation value.  It is also interesting to point out that the gauge group that corresponds to our brane setup is, actually, $G=U(N_c)\approx SU(N_c)\times U(1)$.  The presence of the $U(1)$ factor in $G$ allows us to turn on a Fayet-Ilioupoulos (FI) coupling on the lagrangian which, as we will argue below, will induce a non-zero VEV of the quark bilinear  $ \tilde Q Q$ and, thus, will force the system to enter into the Higgs branch\footnote{Though in two-dimensional field theories an operator  cannot have a VEV, we should think about VEVs in this section  in a Born-Oppenheimer semiclassical approximation sense. The solutions describe a semiclassical  state that is almost not changing in time.}. To see how this can happen, let us recall \cite{Antoniadis:1995vb} that the FI term can be recast as an extra term in the superpotential   linear in the adjoint chiral superfield. Let $\Phi$ denote the $U(1)$ component of such a field. Notice that $\Phi$ is naturally coupled to the matter fields $Q$ and $ \tilde Q$ by means of the interaction $\tilde Q\, Q\,\Phi$. Then, the superpotential $W$ can be written as:
\beq
W\,=\, \tilde Q\, Q\,\Phi\,-\,r\,\Phi\,+\,\cdots\,\,,
\eeq
where the dots refer to terms which do not contain $\Phi$,  and 
$r$ is the FI coupling. By extremizing $W$ with respect to $\Phi$, one gets:
\beq
\tilde Q\, Q\,=\,r\,\,,
\label{QQ-VEV}
\eeq
which shows that a non-vanishing value of the FI coupling implies a non-zero value of the
bilinear $\tilde Q\, Q$, as claimed. 

To figure out how this mechanism can be implemented in our approach, it is interesting to recall the analogous Hanany-Witten brane setup in the type IIA theory (see \cite{Giveon:1998sr} for a review). In this setup the $(4,4)$ two-dimensional theories are engineered by considering a stack of $N_c$ D2-branes, with a compact dimension in their worldvolume. Moreover, the D2-branes are suspended between two parallel NS5-branes. The flavor branes in this setup are $N_f$ D4-branes located between the two NS5-branes. If the latter are at different position in their transverse space, a FI term is induced in the gauge theory living in the D2-branes, with the FI coupling being the relative displacement between the two NS5-branes. When the two NS5-branes are misaligned in this way, the only possibility  to preserve SUSY is by making the D2-brane end on the D4-branes, in such a way that both types of brane are recombined. Thus, we learn from this analysis that the Higgs mechanism is described as the reconnection of color and flavor branes which, in our $d=2$, ${\cal N}=(4,4)$ system, is induced by a non-zero FI term.

Coming back to our branes plus gravity setup, it is clear that, in order to find a holographic description of the Higgs branch, we should look for an embedding of a D3-brane which could be interpreted as representing a recombination of flavor and color branes. This is precisely the purpose of appendix \ref{Higgs-embeddings}, where we find a general class of D3-brane embeddings that preserve all the supersymmetries of the background.  In these embeddings the D3-brane is extended along a holomorphic curve inside the $CY_2$ manifold. We will argue that one of these embeddings has the right properties to be considered as the string dual of the Higgs branch.  Moreover, in subsection \ref{mesons-Higgs} we will analyze the meson spectrum in the Higgs branch in the probe approximation. Similar analysis in other brane setups can be found in ref. \cite{Arean:2007nh}.

We want to find D3-brane embeddings that preserve all the supersymmetries of the background and that interpolate between color and flavor branes. Recall that both types of branes are extended along two different two-cycles of the $CY_2$ cone. Indeed, the color branes are extended along $(\theta, \phi)$ (at $\sigma=0$), whereas the flavor branes of the Coulomb branch extend along $(\psi, \sigma)$, at fixed values of $\theta$ and $\phi$. To generalize these configurations it is natural to consider a flavor brane in which $\theta$ and $\phi$ are no longer constant.  Let us thus consider an embedding for the flavor D3-brane in which the worldvolume coordinates are the same as in section \ref{44flavored}, namely $(x^0, x^1, \psi,\sigma)$,  and the scalars $\theta$ and $\phi$ have a non-trivial dependence on the other coordinates of the $CY_2$, namely:
\beq
\theta\,=\,\theta(\psi,\sigma)\,\,,
\qquad\qquad
\phi\,=\,\phi(\psi,\sigma)\,\,.
\label{embedding-ansatz}
\eeq
In order to determine the D3-brane embeddings of the form (\ref{embedding-ansatz}) which preserve the supersymmetries of the background one has to study the kappa symmetry of the brane probe. This analysis is performed in detail in appendix \ref{Higgs-embeddings}. The final result found in this appendix can be nicely recast in terms of the following two complex coordinates of the $CY_2$: 
\beq
\zeta_1\,\equiv\,\tan \Big( {\theta\over 2}\Big)\,e^{i\phi}\,\,,\qquad\qquad
\zeta_2\,\equiv\, \sigma \sin \theta\,e^{-i\psi}\,\,.
\label{def-zetas}
\eeq
It turns out that any holomorphic embedding of the type $\zeta_1=f(\zeta_2)$ solves the kappa symmetry equations and, thus, preserves the supersymmetry of the background.
In principle any holomorphic function $f$  will solve the kappa symmetry conditions. However, in order to make contact with the field theory analysis, it is rather natural to restrict ourselves to embeddings characterized by a polynomial equation of the type:
\beq
\zeta_1^{p_1}\,\,\zeta_2^{p_2}\,=\,{\rm constant}\,\,,
\label{polynomial-embedding}
\eeq
where the exponents $p_1$ and  $p_2$ are constant integers. Notice that, in the Coulomb branch, color branes are extended along $\zeta_1$ (at $\zeta_2=0$), whereas the flavor branes span $\zeta_2$ at fixed $\zeta_1$.  For this reason, when $p_1=p_2=1$ in (\ref{polynomial-embedding}) the corresponding embedding is symmetric with respect to the directions of the color and flavor branes that are recombined in the Higgs branch. Notice also the similarity between the embedding equation in this case, namely $\zeta_1\,\zeta_2\,={\rm constant}$, and the F-term equation (\ref{QQ-VEV}) of the field theory. Indeed, the directions corresponding to the four scalars in the matter hypermultiplet can be identified, in our gravity dual, with those of the $CY_2$, and then  one can regard (\ref{polynomial-embedding}) for $p_1=p_2=1$ as the geometric realization of (\ref{QQ-VEV}) in our holographic setup.

In the next subsection we will proceed to explore another interesting observable in a quantum field theory: the entanglement entropy.

\subsection{Entanglement entropy}

In quantum field theory the entanglement entropy between two complementary spatial regions $A$ and $B$ is defined as the entropy seen by an observer in $A$ who does not have access to the degrees of freedom of $B$. It can be  calculated from the reduced density matrix obtained after taking the trace over the degrees of freedom of $B$. In ref. \cite{Ryu} the authors proposed a simple geometric method to compute the entanglement entropy in quantum field theories which admit a gravity dual (see also \cite{Klebanov:2007ws}). This method consists in finding the eight-dimensional surface $\Sigma$ with minimal area such that its boundary coincides with the boundary of $A$. Then, the entanglement entropy between $A$ and its complementary region $B$ is given by the integral:
\beq
S\,=\,{1\over 4 G_{10}}\,\int_{\Sigma}\,d^8\xi\, e^{-2\phi}\,
\sqrt{\det\hat G_8}\,\,,
\eeq
where $G_{10}$ is the ten-dimensional Newton constant, given by $G_{10}=8\pi^6\alpha'^4 g_s^2$ and $\hat G_8$ is the induced metric on $\Sigma$. We will consider a constant time surface $\Sigma$, obtained by minimizing $S$ over all surfaces that approach the boundary of $A$ at the boundary of the ten-dimensional  bulk manifold and that are extended along the remaining spatial directions. 

For a gravity dual of a two-dimensional field theory the natural choice for the region $A$ is just a line segment in the Minkowski direction $x\equiv x^1$. We will denote by $l$ the length of this segment and we will choose our coordinate system such that $A$ is given by  $\{-l/2\le x\le l/2\}$. Moreover, for simplicity
we will consider a surface $\Sigma$ that does not penetrate much in the IR region of the metric, in such a way that the UV metric (\ref{UVmetric}) can be used. 
Furthermore, we will parameterize this 8d surface by the coordinates:
\beq
\xi^a\,=\,(x, \theta,\phi,\hat\alpha,\psi, \alpha^i)\,\,,
\eeq
and we will assume that the surface is described by the function:
\beq
u\,=\,u(x)\,\,.
\label{entropy-embedding}
\eeq
It can be easily shown that this is a consistent ansatz for the minimal surface in the UV region of large $u$. 
By computing the induced metric $\hat G_8$, we get that $S$ is given by:
\beq
S\,=\,{\pi^4\over m^6\,G_{10}}\,\,\int_{-{l\over 2}}^{{l\over 2}}\,
dx\,u\,\Big[u'\,^2\,+\,{m^4\over z_*}\,u^4\,\Big]^{{1\over 2}}\,\,.
\eeq
Since the function $S$  does not depend on $x$, the Euler-Lagrange equation derived from $S$ can be integrated once and the result can be written as:
\beq
{u^5\over \Big[u'\,^2\,+\,{m^4\over z_*}\,u^4\,\Big]^{{1\over 2}}}\,=\,
{\sqrt{z_*}\over m^2}\,\,u_0^3\,\,,
\label{u-uprime}
\eeq
where $u_0$ is the minimum value of $u$. For consistency we have to assume that 
$u_0$ is also large and we are always in the region in which the UV form of the metric (\ref{UVmetric}) is valid.  
From the expression (\ref{u-uprime}) we can obtain $u'$ as a function of $u$:
\beq
u'\,=\,\pm {m^2\over \sqrt{z_*}}\,u^2\,\sqrt{\Big({u\over u_0}\Big)^6\,-\,1}\,\,.
\label{uprime}
\eeq
From this result we can compute the length $l$ as a function of the turning point $u_0$ of the holographic coordinate. One has:
\beq
l\,=\,2\int_{u_0}^{\infty}\,\,{du\over |u'(u)|}\,=\,
{2\sqrt{z_*}\over m^2}\,\,\int_{u_0}^{\infty}\,\,
{du\over u^2\,\sqrt{\Big({u\over u_0}\Big)^6\,-\,1}}\,\,.
\eeq
It is  now convenient to change the integration variable from $u$ to $\xi=u/u_0$:
\beq
l\,=\,{2\sqrt{z_*}\over m^2 u_0}\,
\int_{1}^{\infty}\,\,
{1\over\xi^2\, \sqrt{\xi^6-1}}\,\,d\xi\,\,.
\eeq
This integral can be performed exactly and one can obtain the relation between $l$ and
$u_0$:
\beq
l\,=\,{2\sqrt{\pi}\sqrt{z_*}\over m^2 }\,\,
{\Gamma\big({2\over 3}\big)\over \Gamma\big({1\over 6}\big)}\,
{1\over u_0}\,\,.
\label{l-u0}
\eeq

We can use (\ref{uprime})  to eliminate $u'$ in the entropy functional $S$.  The resulting integral is divergent if the upper limit is infinite. Let us regulate this divergence by integrating up to some value $u_{\infty}$ of $u$. In terms of the variable $\xi$, one gets:
\beq
S\,=\,{2\pi^4 u_0^2\over m^6\,G_{10}}\,\int_{1}^{{u_{\infty}\over u_0}}\,\,
{\xi^4\over \sqrt{\xi^6-1}}\,\,d\xi\,\,.
\label{S-integral}
\eeq
The integral appearing on the right-hand side of  (\ref{S-integral}) takes the value:
\beq
\int_{1}^{{u_{\infty}\over u_0}}\,\,
{\xi^4\over \sqrt{\xi^6-1}}\,\,d\xi\,=\,{1\over 2}\,\Big({u_{\infty}\over u_0}\Big)^2\,
F\Big(-{1\over 3}, {1\over 2},{2\over 3}, \Big(\,{u_{0}\over u_{\infty}}\Big)^6\,\Big)\,-\,
{\sqrt{\pi}\over 2}\,\,{\Gamma\Big({2\over 3}\Big)\over \Gamma\Big({1\over 6}\Big)}\,\,.
\eeq
Sending $u_{\infty}\to\infty$ we obtain a term that diverges and terms that are finite:
\beq
\int_{1}^{{u_{\infty}\over u_0}}\,\,
{\xi^4\over \sqrt{\xi^6-1}}\,\,d\xi\,\approx\,{1\over 2}\,\Big({u_{\infty}\over u_0}\Big)^2\,-\,
{\sqrt{\pi}\over 2}\,\,{\Gamma\big({2\over 3}\big)\over \Gamma\big({1\over 6}\big)}\,+\,
{\cal O}\Big(\,\big(\,{u_{0}\over u_{\infty}}\big)^4\,\Big)\,\,.
\eeq
Plugging this result into (\ref{S-integral}), we get the following equation for the entropy:
\beq
S\,\approx\,{\pi^4\over m^6\,G_{10}}\,\,u_{\infty}^2\,-\,
{\pi^4 \sqrt{\pi}\over m^6\,G_{10}}\,
{\Gamma\big({2\over 3}\big)\over \Gamma\big({1\over 6}\big)}\,u_0^2\,\,.
\eeq
As pointed out above, the first term in the above expression is divergent. Let us consider the finite part of $S$, obtained by subtracting the divergent part:
\beq
S^{finite}\,=\,S\,-\,{\pi^4\over m^6\,G_{10}}\,\,u_{\infty}^2\,\,.
\eeq
Moreover, 
let us use (\ref{l-u0}) to write  the value of $S^{finite}$ in terms of the lenght $l$. After some calculation we get:
\beq
S^{finite}(l)\,=\,-\,2\sqrt{\pi}\,
\Bigg[\,{\Gamma\big({2\over 3}\big)\over \Gamma\big({1\over 6}\big)}\,\Bigg]^3\,
V_*\,{N_c^2\over l^2}\,\,,
\label{finite-S}
\eeq
where $V_*$ is the volume of the sphere of radius $\sqrt{z_*}/m$ along which our D3-branes are wrapped:
\beq
V_*\,=\,{4\pi  z_*\over m^2}\,\,.
\eeq
Notice that the entropy (\ref{finite-S}) scales as $N_c^2/l^2$. Actually,
 the expression we have just found is the one that corresponds to a 3+1  dimensional gauge theory compactified on a two-sphere (compare with eq. (42) of \cite{Klebanov:2007ws}). Indeed, from the fact that our solution is obtained from a stack of wrapped D3-branes, this UV behavior was to be expected. It would be interesting to explore the IR behavior of the entropy  for our model.  In particular one should  investigate if, as happens in \cite{Klebanov:2007ws}, there is a critical value of $l$ such that the behavior (\ref{finite-S}) is modified. In our case finding the minimal surface in the IR region would require us to generalize the ansatz (\ref{entropy-embedding}) for the embedding (the coordinate $u$ would have to depend on the coordinate $\hat\alpha$ defined in (\ref{u-hatalpha})). We will not attempt to perform this generalization here.

\section{Meson spectrum}
\label{Mesons}
In this section we  discuss general aspects of the computation of the
spectrum of mesons. By now there is a well known technique based on 
considering a (stable) probe brane in a  given background and studying the 
normalizable fluctuations of the fields on the brane. The problem is of 
Sturm-Liouville type and typically admits a discrete set of normalizable 
fluctuations.

The string computation described above 
corresponds to the so called `quenched approximation' in lattice 
field theory. Indeed, what happens  in this case is that the background encodes the 
non-perturbative physics of the field theory with $SU(N_c)$ gauge group, 
and the dynamics of the flavors 
(encoded on the `probe' flavor branes) does not modify the color dynamics. 
This was discussed in many papers and it is known to correspond, in the dual field theory,  to 
ignoring diagrams  where fundamentals propagate inside the loops. The 
justification for this approximation is that this type of diagrams are weighted by a factor of 
$N_f/N_c$ (see figure 3 and section 7 of 
\cite{HoyosBadajoz:2008fw} for a detailed discussion on this subject). It is observed in 
lattice simulations that the results of the quenched and unquenched calculations of the 
masses of mesons do not  differ qualitatively. Nevertheless, in some cases the numerical differences are relevant to match experimental data. This is one of the reasons why lattice people insist on more expensive unquenched computations. Besides, it is only for the meson spectrum that the quenched approximation
works qualitatively well, while for thermodynamical, finite densite or other properties it gives the wrong physics.

In this paper and in other papers along these lines, the backreaction of 
flavors has been considered. Hence it is natural to ask how to compute 
the mass of the mesons in this more realistic scenario.

In principle, we can use what we know about computing the spectrum 
of glueballs. Following the ideas introduced in \cite{Witten:1998zw}
and elaborated on in \cite{Csaki:1998qr}, we need to fluctuate
the fields of type IIB supergravity and find  sub-sets of fluctuations that constitute  a 
consistent truncation. Remind that in general
the glueball states may be linear combinations of the fluctuations of 
metric, dilaton and RR fields that diagonalize the linear fluctuation 
equations.

In the case of backreacted flavors, we should find a 
consistent sub-set of  fluctuations that includes closed and open string fields (the supergravity fields and the fields on the brane). More 
concretely, if hatted fields indicate the background solutions we 
discussed in the first sections of  this paper and $\epsilon$ a small 
parameter associated with 
the small fluctuation, we will have:
\bea
& & G_{\mu\nu}\to \hat{G}_{\mu\nu} +\epsilon \delta G_{\mu\nu},\;\;
\qquad\qquad
 F_5\to 
\hat{F}_5 +\epsilon\delta F_5,\;\;\nonumber\\
& & C_2\to \epsilon \delta C_2\,,\;\;\qquad
B_2\to \epsilon \delta B_2\,,\;\; \qquad \phi\to 
\epsilon\delta \phi\,,\qquad \chi\to\epsilon\delta \chi\,,\nonumber\\
& & A_\mu^{i}\to \epsilon \delta A_\mu^i,\;\;\; \qquad\varphi^i\to 
\epsilon\delta\varphi^i\,\,,
\label{fluctuation}
\eea
where $(A_\mu^i,\varphi^i)$ denote the fields on the flavor branes with a 
flavor index $i$. So, the procedure should be to expand the action
$$
S=S_{IIB}+ S_{DBI+WZ}\,\,,
$$
up to order $\epsilon^2$, find a subset of 
fluctuations that is closed and diagonalize them. It may 
happen that there is a mixing of glueballs with fields on the branes. 
Clearly the glueballs cannot have a flavor index, so the diagonal 
glueballs should be the same with or without flavors but the mesons can have 
a component of the `closed string fields''. 

The procedure above is quite lengthy, but we believe that  it should be the one to 
apply to have a correct equation for the mesons and glueballs masses.
We can nevertheless approximate this  complicated  computation  by 
neglecting  the mixing between  glueballs and mesons. This approximation is similar to the one made when studying multi-electron atoms, and solving the Schr\"odinger equation of the external electron in the presence of the field created by the nucleus and the  other electrons. Here, the ``electrons'' and ``nucleus'' are the flavor and color branes respectively \footnote{We are grateful to Angel Paredes for suggesting this analogy.}.  This is equivalent to considering a probe brane in 
the background that encodes the dynamics of the field theory with gauge 
group $SU(N_c)$ and flavor group $SU(N_f)$. Doing this probe brane 
computation will indeed take into account the diagrams that scale like 
$N_f/N_c$, because they are encoded in the background where the 
probe fluctuates.

There is another interesting point: in the particular backgrounds we 
studied in the first part of this paper, the $N_f$ quarks we added are 
massive, their mass being related to the radial position $\rho_Q$ where we 
localize the flavor branes. This implies that if we place this probe brane 
at a fixed  position $\rho_p$ above the value of $\rho_Q$ the influence of 
the flavors will be explicit in the differential equations satisfied by 
the fluctuations, while if we place it  below  $\rho_Q$ there 
will be no 
influence whatsoever. This is indeed what happens in this and many other 
theories; one can choose a regularization where the matching
between the high energy and the low energy theory is made exactly at the 
point $\rho_Q$. This seems to be the regularization preferred by our 
set-up.

\subsection{Quenched mesons on the Coulomb branch}
\label{Quenched-mesons}
As argued above,  when $N_f<<N_c$ it is justified to neglect the backreaction and
consider the flavor branes as probes in the background created by the stack of color branes. This defines the so-called quenched approximation, which has been used frequently to determine the spectra of mesons of the theory by analyzing the normalizable fluctuations of the flavor brane probe. In this subsection we shall illustrate how this can be done for our unflavored  background. To begin with, notice that the metric ${\cal G}_{ab}$ induced on the worldvolume of a D3-brane embedded in the ten-dimensional metric (\ref{D3metric}) as  in eqs.  (\ref{D3-wvcoordinates}) and (\ref{Coulomb-flavor-embedding}) is:
\beq
{\cal G}_{ab}\,d\xi^a\,d\xi^b\,=\,
\big[\,H(\rho_Q, \sigma)\,\big]^{-{1\over 2}}\,dx^2_{1,1}\,+\,
{\big[\,H(\rho_Q, \sigma)\,\big]^{{1\over 2}}\over z(\rho_Q, \sigma)}\,
\Big[\, d\sigma^2\,+\,\sigma^2\,(d\psi)^2\,\Big]\,\,.
\label{induced-metric}
\eeq
Let us now study the form of this metric when the worldvolume holographic coordinate $\sigma\to\infty$. For this purpose we can use the asymptotic expressions of $z$ and $H$ obtained in section \ref{44unflavored} (eqs. (\ref{zapprox}) and (\ref{asymp-expansionHg})). One gets that, at leading order, the metric (\ref{induced-metric}) reduces to:
\beq
m^2\sigma^2\,dx^2_{1,1}\,+\,{1\over m^2\sigma^2}\,(d\sigma)^2\,+\,{1\over m^2}\,
(d\psi)^2\,\,,
\label{AdS-wv}
\eeq
where we have redefined the Minkowski coordinates as $x^{0,1}\to \sqrt{z_*}\,x^{0,1}$. Notice that the metric (\ref{AdS-wv}) is of the form $AdS_3\times S^1$, with the $S^1$ factor being given by the $\psi$ coordinate. The radii  of the $AdS_3$ and $S^1$ factors 
in (\ref{AdS-wv}) 
are the same and equal to $1/m$. This UV behavior will be useful to establish the operator/fluctuation dictionary in our case. Notice also that, as $H(\rho_Q, \sigma)$ and 
 $z(\rho_Q, \sigma)$ go to a finite value for $\sigma=0$, the circle parameterized by $\psi$ collapses in the IR at $\sigma=0$, as usually happens in the induced metrics that describe holographic flavor \cite{KK}.

Let us now study the fluctuations of the flavor brane probe around the Coulomb branch configuration (\ref{Coulomb-flavor-embedding}).  To simplify matters we will only consider the fluctuations of the coordinate $\rho$ around the point 
$\rho=\rho_Q$, namely:
\beq
\rho\,=\,\rho_Q\,+\,\hat\rho (x^{\mu}, \psi,\sigma)\,\,.
\label{rho-hatrho}
\eeq
By expanding the DBI+WZ lagrangian density of the flavor brane up to second order in the fluctuation $\hat\rho$ , one gets:
\beq
{\cal L}\,=\,-{T_3\over 2}\,H^{{1\over 2}}\,\sqrt{-\det {\cal G}}\,\,{\cal G}^{ab}\,\,
\partial_a\,\hat\rho\,\partial_b\,\hat\rho\,=\,
-{T_3\over 2}\,\Big[\,{\sigma H\over z}\,\,(\partial_{x^{\mu}}\hat \rho)^2\,+\,
\sigma (\partial_{\sigma}\hat \rho)^2\,+\,{1\over \sigma}\,
(\partial_{\psi}\hat \rho)^2\,\Big]\,\,.
\label{lagrangian-fluct}
\eeq
The equation derived from this lagrangian is given by:
\beq
\partial_{\sigma}\,\big[\,\sigma\partial_{\sigma}\hat \rho\,\big]\,+\,
{H\sigma\over z}\,\partial^2_{x^{\mu}}\,\hat\rho\,+\,{1\over \sigma}\,\partial_{\psi}^2\,
\hat\rho\,=\,0\,\,.
\eeq
In order to find the solutions of this equation, let us separate variables in this 
equation as:
\beq
\hat\rho\,=\,\chi(\sigma)\,e^{ikx}\,e^{il\psi}\,\,,
\label{hatrho-separated}
\eeq
where $l$ is an integer, which  for simplicity we will take to be non-negative, and  $k$ is a momentum vector along the Minkowski $x^{\mu}$ directions. The equation resulting for $\chi(\sigma)$ is:
\beq
\partial_{\sigma}\,\big[\,\sigma\partial_{\sigma}\chi\,\big]\,+\,\Big[\,{\sigma H\over z}\,M^2\,
\,-\,{l^2\over \sigma}\,\Big]\chi\,=\,0\,\,,
\label{separated-fluct}
\eeq
where $M^2\equiv-k^2$. Our purpose is to determine the values of $M$ for which there exist regular solutions of (\ref{separated-fluct}). Notice that $\sigma$ is a non-compact variable that can be arbitrary large and, when $\rho_Q$ is large enough, its minimal value is $\sigma=0$. Moreover, from the behavior of the functions $z(\rho_Q,\sigma)$ and $H(\rho_Q,\sigma)$ for $\sigma=0,\infty$ one can verify that, in both limits, the fluctuation $\chi$ behaves as a linear combination of $\sigma^l$ and $\sigma^{-l}$ for $l\not=0$. When 
$l=0$ the two independent solutions behave as $\chi\sim ({\rm constant}, \log\sigma)$ as 
$\sigma\to 0,\infty$.  Therefore, it is clear that, by fine-tuning $M$, there are  solutions of (\ref{separated-fluct}) that are regular when $\sigma=0,\infty$. The discrete set of values for which this occurs is identified, in the context of the extension of the gauge/gravity correspondence put forward in \cite{KK}, with the mass spectrum of the mesons in the dual field theory.  
Actually, since the induced UV metric is $AdS_3\times S^1$, one can use the full machinery of the AdS/CFT correspondence to determine the conformal dimension of the operator dual to the normalizable fluctuations. From the behavior at large $\sigma$ of the solutions of (\ref{separated-fluct}) one easily concludes that the $\hat\rho$ fluctuations with angular quantum number $l$ are dual to operators with dimension $\Delta=l+1$.

The values of the meson masses can be obtained numerically by means of the shooting technique. In table \ref{Masses} we present the values found by this method for $l=0$.

\begin{table}[ ht]
\begin{center}
\begin{tabular}[b]{|c|c|c|}   
\hline  
\multicolumn{3}{|c|}{Meson spectra for $l=0$}\\
\hline
 $n$  & WKB  & Numerical \\ 
\hline   
\ \ 1 & $4.23$  & $4.22$  \\   
\ \ 2 & $7.32$  & $7.31$  \\   
\ \ 3 & $10.35$  & $10.35$  \\   
\ \ 4 & $13.37$  & $13.36$  \\     
\ \ 5 & $16.37$  & $16.36$  \\
\hline
\end{tabular}
\end{center}
\caption{Numerical and WKB values of $M$ for $l=0$. These values correspond to $\rho_Q=1.5$ and $z_*=5$.}
\label{Masses}
\end{table}

It is also interesting to point out that the fluctuation equation (\ref{separated-fluct}) can be converted into a  Schr\" odinger equation by means of a suitable change of the holographic variable. Indeed, let us define the variable $y$ as follows:
\beq
e^{y}\,=\,\sigma\,\,.
\eeq
Notice that, in this change of variables $\sigma\to\infty$ is mapped into $y\to\infty$, while
$\sigma=0$ corresponds to $y=-\infty$. In terms of $y$, the equation (\ref{separated-fluct}) of the fluctuations can be written as the zero-energy Schr\" odinger equation:
\beq
{d^2\chi\over dy^2}\,-\,V(y)\,\chi\,=\,0\,\,,
\label{Sch}
\eeq
where the potential $V(y)$ is given by:
\beq
V(y)\,=\,l^2\,-\,M^2\,{e^{2y}\, H\over z}\,\,.
\label{potential-Sch}
\eeq
\begin{figure}[ht]
\begin{center}
\includegraphics[width=0.45\textwidth]{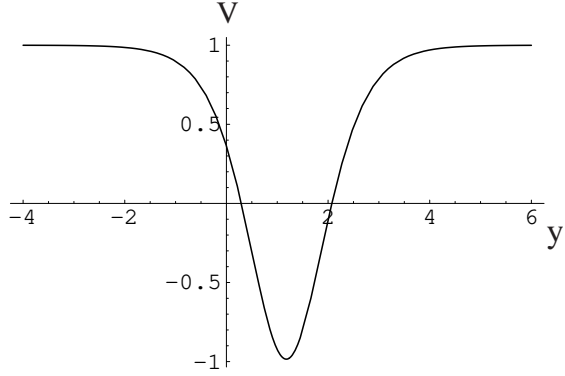}
\end{center}
\caption[potential]{ The Schr\"odinger potential for $l=1$ and $M^2=18.11$, 
which is the mass of the ground state ($n=0$) for this value of $l$. We have set $z_*=5$ and
$\rho_Q=1.5$.} 
\label{potential}
\end{figure}
It is easy to verify by using the behavior of the  functions $H$ and $z$ at $\sigma=0,\infty$ that $V(y)\to l^2$ when $y\to\pm\infty$.  In figure \ref{potential} we have plotted the form of this potential for $l=1$. Notice that $V$ has a unique minimum and that, for $l\not=0$,  the classically allowed region $V\le 0$ of (\ref{Sch}) is of finite size. Then, it is clear that the wave equation for $V$ has a discrete spectrum  of energies and, for certain values of $M$ one of these levels has zero energy as in (\ref{Sch}). A fast, and rather accurate, estimate of the mass spectrum can be obtained by applying the semiclassical WKB method. This method is explained in detail in \cite{Minahan:1998tm,Russo:1998by} (and applied to compute meson masses in \cite{Arean:2005ar,Arean:2006pk}). Here we will limit ourselves to give the final  result for the WKB spectrum, which is:
\beq
M^2_{WKB}\,=\,{\pi^2\over\big[ \zeta(\rho_Q)\big]^2}\,\,(n+1)(n+2l)\,\,,
\label{MWKB}
\eeq
where $n$ is a non-negative integer and  $\zeta(\rho_Q)$ is the following integral:
\beq
\zeta(\rho_Q)\,=\,\int_0^{+\infty}\,\,d\sigma\,\sqrt{{H(\rho_Q,\sigma)\over z(\rho_Q,\sigma)}}\,\,.
\label{zeta-WKB}
\eeq
In (\ref{MWKB}) $n\ge 0$ when  $l\not=0$, whereas  $n\ge 1$ for $l=0$.
Let us  now define the mass gap $M_*$ of the mesonic spectrum as the mass of the lightest meson.  We can get an estimate of this mass gap from
the WKB  formula for the masses. In this case the lightest meson corresponds to taking  $l=0$ and $n=1$ in  (\ref{MWKB}). One has:
\beq
M_*^{WKB}\,=\,{\pi\sqrt{2}\over \zeta(\rho_Q)}\,\,.
\eeq
We can further estimate the value of the integral $\zeta(\rho_Q)$ in (\ref{zeta-WKB}) by using our asymptotic expressions (\ref{zapprox}) and 
(\ref{asymp-expansionHg}):
\beq
\zeta(\rho_Q)\,\approx\,{\sqrt{z_*}\over m^2}\,\,
\int_0^{+\infty}\,
{d\sigma\over \sigma^2\,+\,z_*\,\rho^2_Q}\,\,.
\eeq
This integral can be explicitly evaluated, namely:
\beq
\zeta(\rho_Q)\,\approx\,{\pi\over 2 m^2\,\rho_Q}\,\,.
\label{zeta-approx}
\eeq
Notice that it is independent of $z_*$. Using the value of $ \zeta(\rho_Q)$ obtained above, we find:
\beq
M_*^{WKB}\,\approx\,2\sqrt{2}\,m^2\,\rho_Q\,=\,{2\sqrt{2}\,m_Q\over 
\sqrt{\pi \,g_s\,N_c}}\,\,.
\label{approx-masgap}
\eeq
\begin{figure}[ht]
\begin{center}
\includegraphics[width=0.55\textwidth]{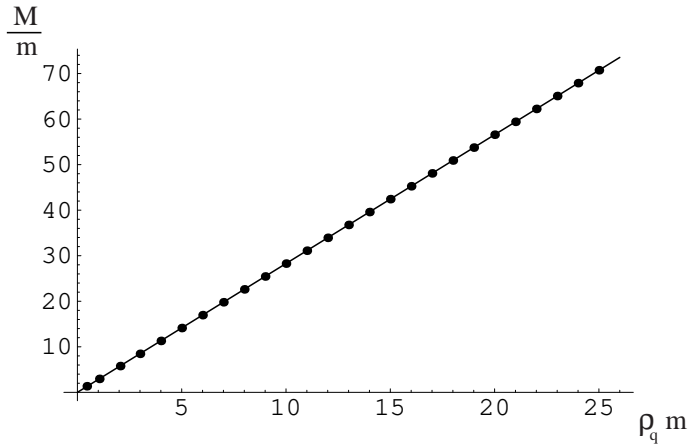}
\end{center}
\caption[massgap]{Numerical mass gap as a function of $\rho_Q$. The straight line corresponds to the approximate WKB prediction. We have set $z_*=5$.  } 
\label{massgap}
\end{figure}
Thus, we predict that the mass gap is linear in the quark mass $m_Q$ and is independent of the IR parameter $z_*$.  Notice that, in principle, the use of the asymptotic expressions (\ref{zapprox}) and 
(\ref{asymp-expansionHg}) is only justified when $\rho_Q$ is large. However, it turns out that (\ref{approx-masgap}) is a rather good approximation to the actual numerical results (see figure \ref{massgap}).  Also, the numerical results show that the mass gap is almost independent of $z_*$, as predicted by our approximate analysis. Furthermore, by using (\ref{zeta-approx}) in (\ref{MWKB}) we also obtain an approximate analytical expression of the WKB meson mass spectra, namely:
\beq
M^{WKB}(n, l)\,\approx\,2\,m^2\,\rho_Q\,
\sqrt{(n+1)(n+2l)}\,\,.
\eeq
This formula works rather well for $l=0,1$.

Up to now we have only considered one particular fluctuation mode of the flavor brane. The remaining modes can be treated similarly and one can find decoupled sets of fluctuations, for which one can obtain the corresponding mass levels. Let us illustrate this fact with a particular example. Let us consider the fluctuation of the coordinates $\theta$ and $\phi$ that determine the position of the flavor brane in the Calabi-Yau manifold, namely:
\beq
\theta\,=\,\theta_0\,+\,\hat\theta (x^{\mu}, \psi, \sigma)\,\,,
\qquad
\phi\,=\,\phi_0\,+\,\hat\phi (x^{\mu}, \psi, \sigma)\,\,,
\eeq
where $\theta_0$ and $\phi_0$ are the constant unperturbed values. By analyzing the behavior of the DBI+WZ action for these perturbations one discovers that these fluctuations are coupled. However, following the same steps as in \cite{Paredes:2006wb}, one can decouple them. Indeed, let us define $\chi_+(\sigma)$ and $\chi_-(\sigma)$ in terms of $\hat\theta$ and $\hat\phi$ by means of the following relation:
\bear
&&\hat\theta\,=\,{1\over 2}\,\Big(\,\chi_+(\sigma)\,+\,\chi_-(\sigma)\,\Big)\,\sin\theta_0\,
e^{ikx}\,\sin (l\psi)\,\,,\rc\rc
&&\hat\phi\,=\,{1\over 2}\,\Big(\,\chi_+(\sigma)\,-\,\chi_-(\sigma)\,\Big)\,
e^{ikx}\,\cos (l\psi)\,\,,
\eear
where $l$ is a non-negative integer. By explicit calculation one can verify that $\chi_+$ and $\chi_-$ satisfy decoupled equations that, actually, can be mapped to (\ref{separated-fluct}) by a simple redefinition. Moreover, from the AdS/CFT dictionary one can associate to $\chi_+$ a field of conformal dimension $\Delta_+=l+3$, while $\chi_-$ is dual to a field with dimension $\Delta_-=l-1$. It is interesting to notice the similarity between these results and the ones found in \cite{Arean:2006pk, CEGK} for the D3-D3 intersection in flat space, which is dual to a 2d defect field theory.

\subsection{Meson spectra on the Higgs branch}
\label{mesons-Higgs}
Let us now consider the fluctuations of the brane probe around a non-trivial holomorphic embedding of the type (\ref{polynomial-embedding}). As in the analysis of the Coulomb branch in subsection  \ref{Quenched-mesons}, we will concentrate on studying the fluctuations of the coordinate $\rho$ around $\rho=\rho_Q$.  If $\hat\rho$ denotes this fluctuation (see eq. (\ref{rho-hatrho})), the corresponding lagrangian up to second order in $\hat\rho$ takes the form:
\beq
{\cal L}\,=\,-{T_3\over 2}\,\,\Big[\,{\cal A}_x\,(\partial_{x^{\mu}}\hat\rho)^2\,+\,
{\cal A}_{\sigma}\,(\partial_{\sigma}\hat\rho)^2\,+\,
{\cal A}_{\psi}\,(\partial_{\psi}\hat\rho)^2\,\Big]\,\,,
\label{lagrangian-Higgs}
\eeq
where the coefficients ${\cal A}_x$, ${\cal A}_{\sigma}$ and ${\cal A}_{\psi}$ are given by:
\bear
&&{\cal A}_x\,=\,{m^2\sigma^2\,(1+p\cos\tilde\theta)^2\,H\,+\,p^2\,z^2\,\sin^2\tilde\theta
\over m^2\,\sigma\,(1+p\cos\tilde\theta)\,z}\,\,,\rc\rc
&&{\cal A}_{\sigma}\,=\,{1\over {\cal A}_{\psi}}\,=\,\sigma\,(1+p\cos\tilde\theta)\,\,,
\label{calA-Higgs}
\eear
with $p=p_2/p_1$ and $\tilde\theta$ being the function of $\sigma$ given in (\ref{theta-sigma}) and $H$ and $z$ are evaluated at $\rho=\rho_Q$. As it should, when $p=0$ the lagrangian (\ref{lagrangian-Higgs}) reduces to (\ref{lagrangian-fluct}).  Moreover, the equation of motion derived from (\ref{lagrangian-Higgs}) is:
\beq
\partial_{\sigma}\,\Big[\,{\cal A}_{\sigma}\,\partial_{\sigma}\,\hat\rho\,\Big]\,+\,
{\cal A}_x\,\partial^2_{x^{\mu}}\hat\rho\,+\,{1\over  {\cal A}_{\sigma}}\,
\partial^2_{\psi}\hat\rho\,=\,0\,\,.
\eeq
From now on we shall restrict ourselves to the analysis of the fluctuations around the embedding with $p=1$ which, as argued above, we think realizes geometrically the brane recombination of the Higgs branch. In this $p=1$ case, the function $\tilde\theta(\sigma)$ is given by:
\beq
\sin^2\Big({\tilde\theta\over 2}\Big)\,=\,{\sigma_*\over \sigma}\,\,,
\label{tilde-theta-p1}
\eeq
where $\sigma_{*}$ is a constant that represents the minimal value of the coordinate 
$\sigma$ (that occurs for $\theta=\pi$). Notice that, when $\sigma_*\not=0$, the probe brane never reaches the point $\sigma=0$. In this case the brane starts at the north pole
$\theta=0$ when $\sigma\to\infty$ and as $\sigma\to \sigma_*$ it approaches the south pole of the $(\theta, \phi)$ two-sphere.  From the point of view of the color brane that is being recombined the fact that $\sigma=0$ is not reached means that a FI term has been switched on. According to our field theory discussion this triggers the Higgs mechanism.

By using (\ref{tilde-theta-p1}) in (\ref{calA-Higgs}) the coefficients  ${\cal A}_x$
and ${\cal A}_{\sigma}$ for this $p=1$ case can be easily written in terms of the coordinate $\sigma$, with the result:
\beq
{\cal A}_x\,=\,{2\,(\sigma-\sigma_*)\,H\over z}\,+\,{2\,\sigma_* \,z\over m^2 \,\sigma^2}\,\,,
\qquad\qquad
{\cal A}_{\sigma}\,=\,2\,(\sigma-\sigma_*)\,\,,
\eeq
and the differential equation for the fluctuations becomes:
\beq
\partial_{\sigma}\,\Big[\,(\sigma-\sigma_*)\,\partial_{\sigma}\hat\rho\,\Big]\,+\,
\Big[\,{(\sigma-\sigma_*)\,H\over z}\,+\,{\sigma_* \,z\over m^2 \,\sigma^2}\,\Big]\,
\partial^2_{x^{\mu}}\hat\rho\,+\,{1\over  4(\sigma-\sigma_*)}\,
\partial^2_{\psi}\hat\rho\,=\,0\,\,.
\eeq
After separating variables as in (\ref{hatrho-separated}), this equation becomes:
\beq
\partial_{\sigma}\,\Big[\,(\sigma-\sigma_*)\,\partial_{\sigma}\chi\,\Big]\,+\,
\Bigg[\,\Big(\,{(\sigma-\sigma_*)\,H\over z}\,+\,{\sigma_* \,z\over m^2 \,\sigma^2}\,\Big)\
M^2\,-\,{l^2\over 4\,(\sigma-\sigma_*)}\,\Bigg]\,\chi\,=\,0\,\,.
\label{separated-fluct-Higgs}
\eeq
Eq. (\ref{separated-fluct-Higgs}) should be compared with the analogous equation   (\ref{separated-fluct}) for the Coulomb branch. Curiously, when $\sigma_*$ vanishes eq. (\ref{separated-fluct-Higgs}) becomes (\ref{separated-fluct}) with $l$ replaced by $l/2$. This makes sense because $\tilde\theta$ is constant (and equal to zero) when $\sigma_*=0$ is taken in (\ref{tilde-theta-p1}). However, the embedding for $\sigma_*=0$ is not the same as the one in the Coulomb branch because the angle $\phi$ is not constant (see (\ref{phi-psi-Higgs}) for $p=1$). Somehow, $\sigma_*=0$ is a singular point in the family of $p=1$ embeddings we are considering. Clearly, for  this singular point,  the mass spectrum of the Higgs branch is related to the one in  the Coulomb branch by means of the following relation:
\beq
M^{Higgs}(n,l)_{\sigma_*= 0}\,=\,M^{Coulomb}\Big(n,{l\over 2}\Big)\,\,,
\qquad\qquad l\in 2\,\mathbb {Z}\,\,.
\label{spectral-flow}
\eeq
\begin{figure}[ht]
\begin{center}
\includegraphics[width=0.45\textwidth]{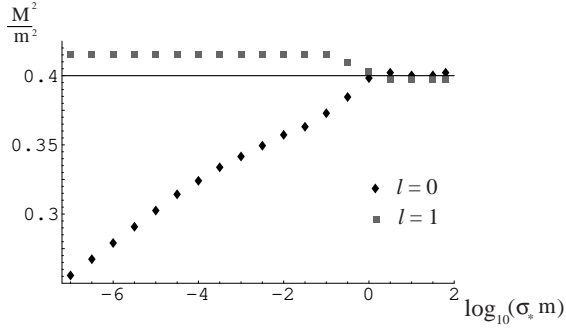}
\end{center}
\caption[MassHiggs]{The mass of the ground state in the Higgs branch as a function of
$\sigma_*$ for $l=0$ and $l=1$. The straight line corresponds to the
prediction of eq. (\ref{Higgs-mgap}). We have set $\rho_Q=1.5$ and $z_*=5$. } 
\label{MassHiggs}
\end{figure}
Notice that the spectral flow relation (\ref{spectral-flow}) only makes sense if $l$ is even.  The numerical calculation of the mass spectra by using the shooting technique shows that the spectrum  for $\sigma_*$ very small differs significantly from the one corresponding to  $\sigma_*=0$ and, therefore, the mass levels are discontinuous at $\sigma_*=0$. Actually, 
when $\sigma_*$ is increased from values very close to zero, the masses vary rapidly until they reach a plateau where they become nearly independent of $\sigma_*$. We have illustrated this fact in figure \ref{MassHiggs}. In order to have a clearer idea on how the masses behave when $\sigma_*$ is not very small, let us apply the WKB method also in this Higgs branch case. With this purpose we first map (\ref{separated-fluct-Higgs}) into a Schr\" odinger equation. First, we define the new variable $y$ as follows:
\beq
e^y\,=\,\sigma\,-\,\sigma_*\,\,,\qquad\qquad
-\infty\,<\,y\,<+\infty\,\,.
\eeq
Let us also define $y_*$ as:
\beq
e^{y_*}\,=\,\sigma_*\,\,.
\eeq
Then, one can demonstrate that (\ref{separated-fluct-Higgs}) can be written as the Schr\" odinger equation (\ref{Sch}), with the potential $V(y)$ being given by:
\beq
V(y)\,=\,{l^2\over 4}\,-\,{M^2\over z}\,\,\Bigg[\,e^{2y}\,H\,+\,{z^2\over m^2}\,\,
{e^{y-y_*}\over \big(1+e^{y-y_*}\big)^2}\,\Bigg]\,\,.
\label{Potential-Higgs}
\eeq
As a check, notice that when $y_*\to-\infty$ (which corresponds to taking $\sigma_*\to 0$) the potential (\ref{Potential-Higgs}) becomes (\ref{potential-Sch}) with $l\to l/2$. Interestingly, the mass spectrum associated with the potential (\ref{Potential-Higgs}) has exactly the same form as in (\ref{MWKB}) with a different value of the prefactor $\zeta$, which is now given by the integral:
\beq
\zeta(\rho_Q, \sigma_*)\,=\,\int_{\sigma_*}^{\infty}\,d\sigma\,\sqrt{
{H(\rho_Q, \sigma)\over z(\rho_Q, \sigma)}\,+\,
{\sigma_*\over m^2\,\sigma^2 \,(\sigma-\sigma_*)}\, z(\rho_Q, \sigma)}\,\,.
\label{zeta-higgs}
\eeq
Let us now estimate the mass gap in the Higgs branch. We first notice that, if $\sigma_*$ is not very small, the second term inside the square root in eq. (\ref{zeta-higgs}) dominates the integral. Thus:
\beq
\zeta (\rho_Q,\sigma_*)\,\approx\,{\sqrt{\sigma_*}\over m}\,\,
\int_{\sigma_*}^{\infty}\,d\sigma {\sqrt{z(\rho_Q,\sigma)}\over \sigma 
\sqrt{ \sigma-\sigma_*}}\,\,.
\eeq
Taking the further approximation in which $z$ is taken to be constant, which will be valid for large $\rho_Q$, we get:
\beq
\zeta (\rho_Q,\sigma_*)\,\approx\,{\sqrt{\sigma_*\,z_*}\over m}\,\,
\int_{\sigma_*}^{\infty}\,d\sigma {1\over \sigma 
\sqrt{ \sigma-\sigma_*}}\,=\,{\pi \sqrt{z_*}\over m}\,\,,
\eeq
which is independent of $\rho_Q$. Using this value we get that the mass gap in the Higgs branch (for large $\rho_Q$) is approximately given by:
\beq
M_*^{Higgs}\,=\,{\sqrt{2}\,\,m\over \sqrt{z_*}}\,\,.
\label{Higgs-mgap}
\eeq
\begin{figure}[ht]
\begin{center}
\includegraphics[width=0.45\textwidth]{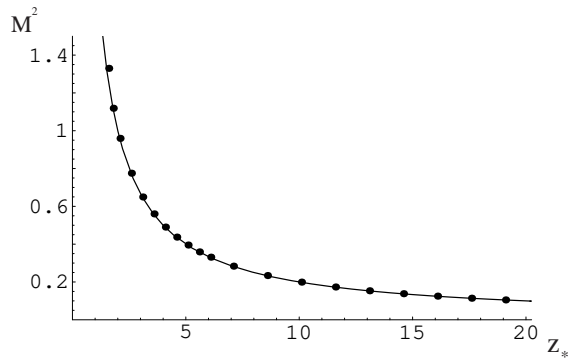}
\end{center}
\caption[MassgapHiggs]{The mass gap in the Higgs branch as a function of $z_*$.  The continuous curve corresponds to the prediction of  eq. (\ref{Higgs-mgap}). We have set $\rho_Q=1.5$  and $\sigma_*=10$.} 
\label{MassgapHiggs}
\end{figure}

Notice that in this case the gap is independent of the quark mass and only depends on the infrared scale $m/\sqrt{z_*}$. Moreover, in the decompactification limit $z_*\to\infty$ the mass gap vanishes, in agreement with the result found for the codimension-two defects in the D3-D3 intersection in flat space (where the discrete spectrum of mesons is lost) \cite{Arean:2007nh}. In figure \ref{MassgapHiggs} we compare the numerical values for the mass gap with those obtained by means of  eq. (\ref{Higgs-mgap}). We see that the agreement between our approximate estimate and the numerical values is excellent. 

\section{Additional supergravity solutions}
\label{EH-section}
Let us now try to find new  simpler solutions of the BPS equations (\ref{BPSsystem}). We shall try an ansatz in which the three functions $g$, $H$ and $z$ depend on the two variables $\rho$ and $\sigma$ as follows:
\beq
g\,=\,g(\sigma)\,\,,\qquad
H\,=\,H(\rho)\,\,,\qquad
z\,=\,z_1(\rho)\,z_2(\sigma)\,\,.
\eeq
Notice that, with this ansatz, the third of the BPS equations  (\ref{BPSsystem}) is automatically satisfied (both sides of the equation vanish). Moreover, by plugging the ansatz on the first equation in  (\ref{BPSsystem}) and by grouping in the same side the terms that depend on the same variable, we get:
\beq
{m^2\,g(\sigma)\over z_2(\sigma)}\,=\,\rho^3\,z_1'(\rho)\,=\,-2c_1\,\,,
\label{first-separation}
\eeq
where we have called the separation constant  $-2c_1$. Eq. (\ref{first-separation}) contains two different equations.  The one  for $z_1(\rho)$ is the following differential equation:
\beq
z_1'(\rho)\,=\,-{2c_1\over \rho^3}\,\,,
\eeq
which can be integrated as:
\beq
z_1(\rho)\,=\,c_2\,+\,{c_1\over \rho^2}\,\,,
\label{z1-rho}
\eeq
with $c_2$ being a new constant.  We also get from (\ref{first-separation}) the form of 
$g(\sigma)$ in terms of $z_2(\rho)$, namely:
\beq
g(\sigma)\,=\,-{2c_1\over m^2}\,z_2(\sigma)\,\,.
\eeq
From the second of the BPS equations (\ref{BPSsystem}), we obtain:
\beq
{m^2 H(\rho)\over z_1^2(\rho)}\,=\,{z_2(\sigma)\dot z_2(\sigma)\over \sigma}\,=\,c_3\,\,,
\label{second-separation}
\eeq
where $c_3$ is another constant. The resulting equation for $z_2(\sigma)$:
\beq
z_2(\sigma)\dot z_2(\sigma)\,=\,c_3\,\sigma\,\,,
\eeq
can be integrated as:
\beq
z_2(\sigma)\,=\,\sqrt{c_4\,+\,c_3\,\,\sigma^2}\,\,.
\label{z2-rho}
\eeq
Moreover, from (\ref{second-separation}) we also  learn that the warp factor $H(\rho)$ is related to $z_1(\rho)$ as follows:
\beq
H(\rho)\,=\,{c_3\over m^2}\,z_1^2(\rho)\,\,.
\eeq
Using the value of $z_1(\rho)$ and $z_2(\rho)$ given in eqs.  (\ref{z1-rho}) and (\ref{z2-rho}), we find that the solution we have is:
\bear
&&g(\sigma)\,=\,-{2c_1\over m^2}\,\sqrt{c_4\,+\,c_3\,\,\sigma^2}\,\,,\rc\rc
&&H(\rho)\,=\,{c_3\over m^2}\,\,\Big[\,c_2\,+\,{c_1\over \rho^2}\,\Big]^2\,\,,\rc\rc
&&z(\rho,\sigma)\,=\,\Big(c_2+{c_1\over \rho^2}\,\Big)\,\sqrt{c_4\,+\,c_3\,\,\sigma^2}\,\,.
\label{gHz-new}
\eear
The ten-dimensional metric corresponding to this solution  can be obtained by plugging the values of $H$ and $z$ given in (\ref{gHz-new}) into the general ansatz (\ref{D3metric}). One gets:
\beq
ds^2_{10}\,=\,ds^2_{6}\,+\,ds^2_{4}\,\,,
\eeq
where the six-dimensional part of the metric $ds^2_{6}$ is independent of $\sigma$ and given by:
\beq
ds^2_6\,=\,H^{-{1\over 2}}\,dx^2_{1,1}\,+\,H^{{1\over 2}}\,
\Big[\,d\rho^2\,+\,\rho^2\,d\Omega_3^2\,\Big]\,\,,
\eeq
and the four-dimensional metric, which does not depends of $\rho$, can be written as:
\beq
ds^2_4\,=\,{\sqrt{c_4\,+\,c_3\,\,\sigma^2}\over m\sqrt{c_3}}\,\,
\Big[\,d\theta^2\,+\,\sin^2\theta\,d\phi^2\,\Big]\,+\,
{\sqrt{c_3}\over m \sqrt{c_4\,+\,c_3\,\,\sigma^2}}\,\,
\Big[\,d\sigma^2\,+\,\sigma^2\,\big(\,d\psi+\cos\theta d\phi\,\big)^2\,\Big]\,\,.\qquad\qquad
\label{ds4}
\eeq
In order to rewrite the metric (\ref{ds4}) in a more familiar form, let us perform a change of variables and define a new radial variable $\zeta$, related to $\sigma$ as follows:
\beq
\zeta^2\,=\,{4\over m\sqrt{c_3}} \,\,\sqrt{c_4\,+\,c_3\,\,\sigma^2}\,\,.
\label{zeta}
\eeq
Actually, if we define the new constant $a$ as follows:
\beq
a^4\,=\,{16 c_4\over m^2 c_3}\,\,,
\eeq
the relation that gives $\sigma$ in terms of $\zeta$ is the following:
\beq
\sigma\,=\,{m\over 4}\,\sqrt{\zeta^4-a^4}\,\,.
\eeq
Clearly, $\zeta\ge a$, which corresponds to the range $\sigma\ge 0$. After this change of variable, one can easily prove that the metric (\ref{ds4}) becomes:
\beq
ds^2_4\,={d\zeta^2\over 1\,-\,\big({a\over \zeta}\big)^4}\,+\,
{\zeta^2\over 4}\,\,\Big[\,d\theta^2\,+\,\sin^2\theta\,d\phi^2\,+\,
\Big(\,1\,-\,\Big({a\over \zeta}\Big)^4\,\Big)\,\big(\,d\psi+\cos\theta d\phi\,\big)^2\,\Big]\,\,,
\label{EH}
\eeq
which is the metric of an Eguchi-Hanson space $EH_4$ with resolution parameter $a$.  When $a=0$ the metric (\ref{EH}) becomes the one corresponding to the $\mathbb{C}^2/\mathbb{Z}_2$ orbifold. Moreover, if we define the new constants $\eta$ and $R$ as:
\beq
\eta\,\equiv\,{\sqrt{c_3}\, c_2\over m}\,\,,\qquad\qquad
R^2\,\equiv\,{\sqrt{c_3}\, c_1\over m}\,\,,
\eeq
then, the warp factor $H$ becomes:
\beq
H\,=\,\Big[\,\eta\,+\,\Big({R\over \rho}\Big)^2\,\Big]^2\,\,.
\label{H-new}
\eeq
Using this result the six-dimensional part of the metric takes the form:\
\beq
ds^2_6\,=\,{dx^2_{1,1}\over \eta\,+\,\Big({R\over \rho}\Big)^2} \,+\,
\Big[\,\eta\,+\,\Big({R\over \rho}\Big)^2\,\Big]\,
\Big[\,d\rho^2\,+\,\rho^2\,d\Omega_3^2\,\Big]\,\,.
\label{ds6}
\eeq
Notice that by rescaling $x^{0,1}$ and $\rho$ in (\ref{ds6}) we can have $\eta=0,1$. The case $\eta=0$ corresponds to a ten-dimensional metric of the form $AdS_3\times S^3\times EH_4$, with $R$ being the common radius of the $AdS_3$ and $S^3$ parts. 
When $\eta\not=0$ the six-dimensional metric flows in the UV region $\rho\to\infty$ to the flat six-dimensional Minkowski space, whereas in the near-horizon region $\rho\to 0$ it becomes  $AdS_3\times S^3$. Notice also that the functions $g$ and $z$ can be written in terms of the  new variables as:
\beq
g(\zeta)\,=\,-{R^2\over 2}\,\,\zeta^2\,\,,\qquad\qquad
{z(\rho,\zeta)\over m^2}\,=\,\Big[\,\eta\,+\,\Big({R\over \rho}\Big)^2\,\Big]\,{\zeta^2\over 4}\,\,.
\eeq
Let us now write the expression of the RR five-form $F_5$ for this solution. In order to do that, let us choose the following  vierbein one-form basis for the $EH_4$ metric (\ref{EH}):
\bear
&&\eta^1\,=\,{\zeta\over 2}\,\,d\theta\,\,,\qquad\qquad\qquad
\eta^2\,=\,{\zeta\over 2}\,\,\sin\theta\,d\phi\,\,,\rc\rc
&&\eta^3\,=\,{\zeta\over 2}\sqrt{\,1\,-\,\Big({a\over \zeta}\Big)^4\,}\,
\big(\,d\psi+\cos\theta\,d\phi\,\big)\,\,,\qquad\qquad
\eta^4\,=\,{d\zeta\over \sqrt{\,1\,-\,\Big({a\over \zeta}\Big)^4}\,}\,\,.
\qquad\qquad\qquad
\eear
In terms of this basis, the  Kahler form ${\cal J}$ of the $EH_4$ space takes the form:
\beq
{\cal J}\,=\,\eta^1\wedge \eta^2\,+\,\eta^3\wedge \eta^4\,\,.
\label{Kahler}
\eeq
To get the expression of $F_5$ we just plug the value of $g$ for our solution, displayed in   (\ref{gHz-new}), in our ansatz of eq. 
 (\ref{CalC4}). After writing the result in terms of the variable
$\zeta$ defined in (\ref{zeta}), one gets:
\beq
F_5\,=\,2R^2\,\Big(\,{1\over \rho^3 H}\,d\rho\wedge dx^0\wedge dx^1\,-\,
\,\omega_3\,\Big)\,\wedge  {\cal J}\,\,,
\eeq
where $ {\cal J}$ is the two-form defined in (\ref{Kahler}). Interestingly, by using the 
explicit value of $H$ written in (\ref{H-new}), one can verify that,
in the $AdS_3\times S^3$ limit $\eta=0$,   $F_5$ takes the form:
\beq
 F_5\,=\,{2\over R}\, \Big[\,{\rm Vol} (AdS_3)\,-\,{\rm Vol} (S^3)\,\Big]\wedge 
 {\cal J}\,\,,
 \eeq
 which seems to be the same type of solution obtained in \cite{Gauntlett:2007ph}.

\subsection{Non-relativistic backgrounds}
According to the papers in \cite{NR}, one can construct a background dual to a non-relativistic theory in $d+1$ dimensions by performing a combination of suitable T-dualities and shifts to a solution of supergravity in $d+3$ dimensions. Here we shall apply this procedure to our unflavored background. The final result will have the isometries of the Galileo group in $d=0$ and, thus, it is somewhat dual to a matrix model. 

As the first step in our procedure, let us introduce the light-cone variables $x^{\pm}$ as:
\beq
x^{\pm}\,=\,x^0\pm x^1\,\,.
\eeq
Next, we will perform a T-duality along the fiber direction $\partial_{\psi}$, followed by a shift in the light-cone coordinate $x^-$ of the form:
 \beq
 dx^-\,\to\,dx^-\,+\,\gamma\,d\psi\,\,.
 \label{shift}
 \eeq
 As the result of these two consecutive operations we obtain a background of type IIA supergravity.  The  corresponding metric and dilaton are given by:
 \bear
&&ds_{IIA}^2\,=\,H^{-{1\over 2}}\,\,\Big[\,-dx^+dx^-\,-\,\gamma dx^+d\psi\,+\,
{z\over m^2}\,
\Big(\,d\theta^2\,+\,\sin^2\theta\,d\phi^2\,\,\Big)\,\Big]\,+\,\rc\rc
&&\qquad\qquad+\,
H^{{1\over 2}}\,\,\,\Big[\,{1\over z}\,
d\sigma^2\,+\,
d\rho^2\,+\,\rho^2\,d\Omega_3^2\,\Big]\,+\,H^{-{1\over 2}}\,{z\over \sigma^2}\,d\psi^2
\,\,,\qquad\qquad\rc\rc
&&e^{2\Phi}\,=\,{z\over \sigma^2 \sqrt{H}}\,e^{2\Phi_0}\,\,,
\label{IIA-metric}
\eear
 while the non-vanishing  NSNS and RR potentials are:
 \bear
 &&B_2\,=\,\cos\theta\,d\phi\wedge d\psi\,\,,\rc\rc
 &&C_3\,=\,g\,\omega_3\,+\,{\sigma\over 2z}\,dx^+\wedge dx^-\wedge d\sigma\,-\,
 \gamma\,{\sigma\over 2z}\,dx^+\wedge d\sigma\wedge d\psi\,\,,\rc\rc
 &&C_5\,=\,-{z\over 2m^2 H}\,dx^+\wedge dx^-\wedge \omega_2\wedge d\psi\,\,.
 \eear
 Let us next perform another T-duality along $\partial_{\psi}$. We get a background of the type IIB theory with constant dilaton and the following metric:
 \bear
&&ds_{IIB}^2\,=\,H^{-{1\over 2}}\,\,\Big[\,-dx^+dx^-\,-\,
{\gamma^2\over 4}\,\,{\sigma^2\over z}\,(dx^+)^2
\,+\,{z\over m^2}\,
\Big(\,d\theta^2\,+\,\sin^2\theta\,d\phi^2\,\,\Big)\,\Big]\,+\,\rc\rc
&&\qquad\qquad+\,
H^{{1\over 2}}\,\,\,\Big[\,{1\over z}\,\,\Big(\,
d\sigma^2\,+\,\sigma^2\,
\Big(\,d\psi+\cos\theta d\phi\,\Big)^2\,\Big)
\,+\,
d\rho^2\,+\,\rho^2\,d\Omega_3^2\,\Big]\,\,,
\qquad\qquad
\label{NR-metric}
\eear
which has  the following potentials:
\bear
 &&B_2\,=\,-{\gamma\over 2}\,{\sigma^2\over z}\,\,dx^+\wedge 
 \big(\,d\psi+\cos\theta d\phi\,\big)\,\,,\rc\rc
 &&C_2\,=\,-{\gamma\over 2}\,{\sigma\over z}\,dx^+\wedge d\sigma\,\,,\rc\rc
 &&C_4\,=\,g\,\omega_3\wedge \big(\,d\psi+\cos\theta d\phi\,\big)\,+\,
 \Big[\,{\sigma\over 2z}\,d\sigma\wedge
(d\psi+\cos\theta d\phi)\,-\,{z \over 2m^2 H}\,\omega_2\,\,\Big]\wedge
dx^+\wedge dx^-\,\,.\qquad\qquad
\eear
Notice that, if we take the shift parameter $\gamma$ to be zero, the two-forms vanish and we recover the original ansatz we started with. 
 From these potentials one can compute the corresponding field strengths as
 $H_3=dB_2$, $F_3=dC_2$ and $F_5=dC_4+C_2\wedge H_3$. By computing directly
 the exterior derivatives and after using the BPS equations, one gets:
 \bear
 &&H_3\,=\,{\gamma\over 2}\,\Bigg[\,{m^2\sigma^2 g\over \rho^3 z^2}\,d\rho\,-\,
 {2\sigma\over z}\,\Big(\,1\,-\,
{m^2\sigma^2 H\over 2 z^2}\,d\sigma\,\Big)\,\Bigg]\wedge dx^+\wedge 
\big(\,d\psi+\cos\theta d\phi\,\big)\,-\,{\gamma\over 2}\,{\sigma^2\over z}\,dx^+\wedge \omega_2\,\,,\rc\rc
&&F_3\,=\,{\gamma\over 2}\,{m^2\sigma g\over \rho^3 z^2}\,
dx^+\wedge d\sigma\wedge d\rho\,\,,\rc\rc
&&F_5\,=\,\Big(1+*\Big)\,\Big[\,dg\wedge \omega_3\wedge 
\big(\,d\psi+\cos\theta d\phi\,\big)\,+\,g\,\omega_2 \wedge \omega_3\,\Big]\,\,.
\label{NRbackground}
\eear
As a check of the correctness of the final result one can verify that the forms satisfy their equations of motion, namely:
\beq
dF_5\,=\,-H_3\wedge F_3\,\,,\qquad\qquad
d{}^*F_3\,=\,H_3\wedge F_5\,\,,\qquad\qquad
d\Big(e^{-2\phi}\,\,{}^*H_3\,\Big)\,=\,F_5\wedge F_3\,\,.
\eeq

Let us interpret the effect of the transformations performed above, on our field theory, following what is explained in the papers \cite{NR}.

As we mentioned above, if we take $\gamma=0$ in (\ref{shift})
we recover the original background, but we  impose that the coordinate $x^{-}$ is periodic and hence it has discrete modding. Indeed, this would reproduce in string theory the ``mass operator'' of the Galileo invariant field theory that has discrete eigenvalues. In consequence, for $\gamma=0$ we would be doing DLCQ of the 2d theory. Now, we can think of the nonzero $\gamma$ case as a particular case of a TsT transformation introduced in \cite{Lunin:2005jy}. In this particular example, it is generating a dipole field theory \cite{Bergman:2001rw}. In a dipole theory we change the product between two generic operators $\Phi_1(x^-,x^+)$ and $\Phi_2(x^-,x^+)$, with charges $(q_1,q_2)$, to the ``star product'':
\beq
\Phi_1(x_{1}^- , x_{1}^+) 
* \Phi_2(x_{2}^-,x_{2}^+)=
e^{i\big(\, q_2\,\partial_{x_1^-} \,-\,q_1\,\partial_{x_2^-}\,\big)}\,\,
\Phi_1(x_{1}^-,x_{1}^+) \,\Phi_2(x_{2}^-,x_{2}^+)\,\,.
\eeq
In our case $(q_1,q_2)$ are R-symmetry charges
(represented by traslations in the angle $\psi$).
As a consequence of this change of product, there will be physical procesess that are affected and some that are not.

In particular, all products of operators that contain fields charged under the R-symmetry will change after the procedure described above. For example, masses of R-charged fields will change. All this will leave us with a non-relativistic field theory, that in the particular case we studied here, will also be non-supersymmetric (fermions, for example, will get a mass different from scalars and gauge fields will remain massless).

As a word of caution, if we want to learn about non-perturbative aspects of this field theory, computing with the background in (\ref{NRbackground}), we should be careful to use the background for values of the coordinates where the curvature is not very large. Note that since the $x^{-}$ circle has zero lenght, strings that wrap the cycle are massless and the supergravity approximation may not be applicable.

\section{Fractional D1/D5  system}
\label{fractional-section}

In this section we will show how to describe the (4,4) two-dimensional gauge theory by means of a system of fractional D1/D5-branes on the orbifold 
$\mathbb{R}^{1,5}\times \mathbb{C}^2/\mathbb{Z}_2$. 
Fractional branes are a kind of brane that are specific to orbifold (and conifold) singular backgrounds. As can be checked by explicit calculations of string scattering amplitudes \cite{Merlatti:2000ne}, these Dp-branes can be viewed as ordinary D($p+2$)-branes wrapped on an integer basis of vanishing two cycles \cite{Douglas:1996xg}. As in the analogous constructions of \cite{Bertolini:2001qa}, this fact will be of fundamental importance to find the supergravity solution corresponding to the D1-D5 system we are investigating. If we parameterize 
$\mathbb{C}^2$ by four cartesian coordinates $x^6, x^7, x^8, x^9$, then the action of $\mathbb{Z}_2$ will be simply given by:
\beq
(x^6, x^7, x^8, x^9)\,\rightarrow \,(-x^6, -x^7, -x^8,- x^9)\,\,.
\eeq
Moreover, if $x^0,\cdots, x^5$ are the coordinates of $\mathbb{R}^{1,5}$, the fractional D1-branes are extended along  $x^0,x^1$ and are stuck at the orbifold fixed point 
$x^6= x^7= x^8=x^9=0$. In this setup the flavor branes are D5-branes extended along $x^0,x^1$, as well as along the four directions of $ \mathbb{C}^2/\mathbb{Z}_2$. Therefore, the full D1/D5 configuration is described by the array:
\begin{center}
\begin{tabular}{|c|c|c|c|c|c|c|c|c|c|c|}
\multicolumn{7}{c}{ }&
\multicolumn{4}{c}{$\overbrace{\phantom{\qquad\qquad\quad}}^{\mathbb{C}^2/\mathbb{Z}_2}$}\\
\hline
&0&1&2&3&4&5&6&7&8&9\\
\hline
$N_c\,\,$ 
D1 &--&--&$\cdot$&$\cdot$&$\cdot$&$\cdot$&$\cdot$&$\cdot$&$\cdot$&$\cdot$\\
\hline
$N_f\,\,$ D5 &--&--&$\cdot$&$\cdot$&$\cdot$&$\cdot$&--&--&--&--\\
\hline
\end{tabular}
\end{center}
To find a supergravity solution for this setup it is interesting to regard the $\mathbb{C}^2/\mathbb{Z}_2$ orbifold as the limit of an ALE space (or, conversely, the ALE space can be thought  of as the blow-up of the orbifold). This ALE space is endowed with an anti-self-dual two-form $\hat\omega_2$, which is the Poincar\'e dual to the exceptional two-cycle $\Sigma_2$ that is obtained by the resolution of the orbifold singularity ($\Sigma_2$ shrinks in the orbifold limit).

The peculiarity of the orbifold backgrounds is the appearance of six-dimensional twisted fields. They are  $p$-form fields that can be seen as coming from higher dimensional $(p+2)$-forms which have components along the non-trivial shrinking two-cycle $\Sigma_2$ of the orbifold at the orbifold fixed point. As such, they are the zero modes of the KK reduction of those $(p+2)$-form fields which have components along the anti-self-dual two-form $\hat\omega_2$ of the orbifold \cite{Aspinwall:1996mn}. In our case the relevant twisted fields are the two-form RR potential $A_{2}$ and the NS-NS scalar $b$.  They are related to the ordinary RR four-form potential $C_{4}$ and to the NS-NS $B_2$ field as:
\be 
C_{4}\, = \, A_{2}\wedge\hat\omega_2\ \ \ , \ \ \  B_{2}\, = \, b\,\hat\omega_2\,\,,
\ee
where the two form $\hat\omega_2$ is normalized as follows:
\be \int_{\mathbb{C}^2/\mathbf{Z}_2}\hat \omega_2\wedge\hat\omega_2\, =\, -1\,\,.
\ee
After performing this KK reduction, the action of the type IIB supergravity theory in the Einstein frame can be consistently truncated to be: 
\bea
\nonumber {S_{IIB}}\,=\, \frac{1}{2\kappa^2_{orb}}\Big\{ \int d^{10}x\, \sqrt{-\det\, G}\, R\,-\,\frac{1}{2}\int d\Phi\wedge^\star d\Phi\, -\, \frac{1}{2}\int e^{\Phi}F_{3}\wedge^\star F_{3}\,\,-\\ -\, \frac{1}{2}\int_{\mathbb{R}^{1,5}} e^{-\Phi}db\wedge^{\star_6} db\, -\, 
\frac{1}{4}\int _{\mathbb{R}^{1,5}}T_{3}\wedge^{\star_6} T_{3}\, +\, 
\frac{1}{2}\int _{\mathbb{R}^{1,5}}A_{2}\wedge db\wedge F_{3}\Big\},
\label{lagrangian}
\eea
where we have not included the contribution of the different sources. In (\ref{lagrangian})
the integrals in the second line are six-dimensional, $\kappa_{orb}=\sqrt{2}\kappa_{10}\,=\,(2\pi)^{{7\over 2}}\,g_s\,(\alpha')^2$,
$F_{3}\,=\,dC_{2}$ is the usual RR three-form field  and $T_{3}$ is the following six-dimensional anti-self-dual three-form:
\be
\label{t3} 
T_{3}\, =\, d A_{2}-C_{2}\wedge db\,\,.
\ee 
The anti-self-duality\footnote{Our conventions are such that, in flat indices, $\epsilon^{0...9}\,=\,\epsilon^{0...6}=1$.} of $T_3$ has  to be imposed by hand. Actually, we will represent $T_3$ as:
\beq
T_3\,=\,{\cal T}_3\,-\,{}^{*6}{\cal T}_3\,\,,
\label{T3-calT3}
\eeq
with ${\cal T}_3$ being given by:
\beq
{\cal T}_{3}\, =\, d {\cal A}_{2}-C_{2}\wedge db\,\,,
\label{calT3}
\eeq
and $d\,{}^{*6}{\cal T}_3=0$. 
From the action (\ref{lagrangian}) one can readily get the following equations of motion for the dilaton and forms:
\bear
d^\star d\Phi\, -\, \frac{1}{2}e^{\Phi}F_{3}\wedge^\star F_{3}\,+\,\frac{1}{2}e^{-\Phi}db\wedge^{\star_6}db\,\wedge \Omega_4=\, 0\,\,,\rc\rc
d(e^{\Phi}\,^\star F_{3})\,=\,db\wedge T_{3}\,\wedge \Omega_4\,\,,\rc\rc
d(e^{-\Phi}\,^{\star_6}db)\,=\,T_{3}\wedge F_{3}\,\,,\rc\rc
d^{\star_6}T_{3}\,=\,-d T_{3}\,=\, F_{3}\wedge db\,\,,
\label{eom-forms}
\eear
where $\Omega_4$ is the following four-form:
\beq
\Omega_4\,=\,\delta(x^6)\cdots \delta(x^9)\,dx^6\wedge\cdots \wedge dx^9\,\,.
\eeq
The Einstein equations derived from (\ref{lagrangian}) read:
\bear
&&R_{MN}\,-\,\frac{1}{2}G_{MN}R\,=\, \frac{1}{2}\left(\partial_M\Phi\partial_N\Phi\,-\frac{1}{2}G_{MN}\partial_P\Phi\partial^P\Phi\right)\,+\,\rc\rc
&&+\,\frac{1}{2}\frac{1}{3!}e^{\Phi}\left(3F_{(3)\,MPQ}F_{(3)\,N}^{PQ}\,-\,\frac{1}{2}G_{MN}F_{(3)}^2\right)\,+\,\rc\rc
&& +\, \delta_4\,\,\frac{\sqrt{-\det\,G_{6}}}{\sqrt{-\det\,G}}\,\,\Bigg\{ \frac{1}{2}e^{-\Phi}\left( \partial_M b\partial_N b\,-\,\frac{1}{2}G_{MN}\partial_Pb\partial^Pb\right)\,
+\frac{1}{8}\,T_{(3)\,MPQ}T_{(3)\,N}^{PQ}\Bigg\}\,\,,\qquad
\label{Einstein-orb}
\eear
where $G_6$ denotes the restriction of the ten-dimensional metric to the six-dimensional subspace $\mathbb{R}^{1,5}$, $\delta_{4}\,=\,\delta(x^6)\cdots \delta(x^9)$ and 
$M,\,N\,=\, 0,\cdots, 9$.  It is understood that in the terms where a $\delta_4$ appears in 
(\ref{Einstein-orb}) (in the last line) these indices run just along the six directions transverse to the orbifold. 

The action (\ref{lagrangian}) must be supplemented with the boundary actions corresponding to the different branes present in the problem, which give rise to source terms that must be added to the equations of motion (\ref{eom-forms}) and (\ref{Einstein-orb}) of the bulk fields. In our particular problem the branes sourcing the bulk fields are $N_c$ fractional D1-branes and $N_f$ D5-branes that wrap the entire orbifold. Therefore, the boundary action is:
\beq
S_{b}\,=\,N_c\,S_{D1}\,+\, N_f\,S_{D5}\,\,.
\eeq
In order to write the expression of $S_{D1}$ and $S_{D5}$, let us recall  \cite{Aspinwall:1996mn} that the NS-NS $b$ field has a non-vanishing constant value for the $\mathbb{C}^2/\mathbb{Z}_2$ orbifold. With our conventions, we can write $b$ as:
\be
b\,=\,-(2\pi^2\alpha'\,+\,\tilde{b})\,\,,
\ee
where $\tilde{b}$ represents the fluctuation around the background value $b=\,-2\pi^2\alpha'$. The action of a fractional D1-brane in the Einstein frame is:
\bear
 \label{fractional}
S_{D1}\,=\,-\frac{T_1}{2}\Bigg\{ \int d^2x\, e^{-{\Phi\over 2}}\sqrt{-det(G_{\alpha\beta}\,+\,e^{-{\Phi\over 2}}\,2\pi\alpha'F_{\alpha\beta}})\left(1\,+\,\frac{\tilde{b}}{2\pi^2\alpha'}\right)\,-\rc\rc
\,- \int C_2 \left(1\,+\,\frac{\tilde{b}}{2\pi^2\alpha'}\right) -\frac{1}{2\pi^2\alpha'}\int  A_2\Bigg\}.
\eear
Similarly, the action for the D5-brane is given by:
\bea
\label{fractional5}
&&S_{D5}\,=\,-\frac{T_5}{2}\left\{ \int d^6x\,  e^{{\Phi\over 2}}\sqrt{-det(G_{\alpha\beta}})\,- \int \,C_6\right\}\,+\,\rc\rc
&& \qquad\qquad\qquad
+\frac{T_1}{2}\frac{1}{4(2\pi^2\alpha')}\left\{ \int d^2x\,  e^{-{\Phi\over 2}}\,\sqrt{-det(G_{\alpha\beta})}\,\, \tilde{b}\,-\,
 \int  A_2\right\},
\eea
where we have not included higher order terms in the twisted sector.
In eqs. (\ref{fractional}) and (\ref{fractional5}) the tensions $T_1$ and $T_5$ are given by:
\beq
T_1\,=\,{1\over 2\pi\alpha' \,g_s}\,\,,\qquad\qquad
T_5\,=\,{1\over (2\pi)^5\,(\alpha')^3\,g_s}\,\,.
\eeq

\subsection{The unflavored solution}
\label{fractionalD1}

First of all, 
let us now  look for a solution corresponding to a stack of $N_c$ fractional D1-branes on the space  $\mathbb{R}^{1,5}\times \mathbb{C}^2/\mathbb{Z}_2$, without D5-branes.
With this purpose, let us make the following ansatz\footnote{$\alpha,\ \beta\,=\,0,1$ label the worldvolume directions of the D1-brane.} for the metric and the dilaton:
\bea
\label{fractionalD1-metric}
&&ds^2= H_1^{-3/4}\eta_{\alpha\beta}dx^{\alpha}dx^{\beta}\,+\,
H_1^{1/4}\Big(\,dr^2\,+\,\frac{r^2}{4}\sum_{i=1}^3\omega^i\,\omega^i\,+\,d\sigma^2\,+\, \frac{\sigma^2}{4}\sum_{j=1}^3\tilde\omega^{j}\,\tilde\omega^{j}\,\Big)\,\,,
\qquad\qquad\rc\rc
&&e^{\Phi}\,=\,H_1^{1/2}\,\,,
\eear
where $r$ is the radial direction transverse to both the D1-branes and the orbifold space, $\omega^{i}$ are the usual left-invariant one forms parametrizing the corresponding $S^3$, $\sigma$ is the radial orbifold direction and $\tilde\omega^{i}$, with the range of angles conveniently chosen, are the left-invariant one-forms parametrizing the orbifolded $S^3$. The left-invariant one-forms are normalized  such that $d\omega^i\,=\, {1\over 2}\,\epsilon_{ijk}\omega^j\wedge\omega^k$ and $d\tilde\omega^{i}\,=\, {1\over 2}\, \epsilon_{ijk}\tilde\omega^{j}\wedge\tilde\omega^{k}$.  We will assume that there is a non-zero $b$ field, that only depends on the coordinate $r$. Moreover,  the warp factor $H_1$ will depend on both radial coordinates. 
Then:
\beq
H_1\,=\,H_1(r,\sigma)\,\,,\qquad\qquad
b=b(r)\,\,.
\eeq
In addition, our background will be endowed with a non-zero RR two-form $C_{2}$, as well as with a two-form potential ${\cal A}_{2}$ in the RR twisted sector:
\beq
C_{2}=(H_1^{-1}-1)\,dx^0\wedge dx^1\,\,,\qquad\qquad
{\cal A}_{2}= a(r)dx^0\wedge dx^1\,\,,
\label{2form-potentials}
\eeq
where $a(r)$ is a function to be determined by the equations of motion.  By using eqs. (\ref{T3-calT3}) and (\ref{calT3}) it is straightforward to find the field strength $T_3$ corresponding to the two-form potentials (\ref{2form-potentials}). One finds:
\beq
T_{3}\,=\, \left[a'-b'(H_1^{-1}-1)\right]dr\wedge dx^0\wedge dx^1\,+\,\frac{H_1\,r^3}{8}\left[a'-b'(H_1^{-1}-1)\right]\omega^1\wedge\omega^2\wedge\omega^3,
\eeq
where the prime denotes the derivative with respect to $r$. Similarly the dot will  denote the derivative with  respect to $\sigma$.

Let us now analyze the different equations of motion for this system.
The equation of motion for $F_{3}$ leads to the following equation for $H_1$ :
\be 
\partial_r(r^3\sigma^3H_1')\,+\,\partial_{\sigma}(r^3\sigma^3\dot{H_1})\,=\,b'H_1\,r^3\left[a'\,-\,b'(H_1^{-1}-1)\right]\delta_4\,-\,N_c\,\kappa^2_{orb}\,T_1\,\delta_8\,\,,
\label{f3}
\ee
where $\delta_8\equiv \delta(x^2)\cdots \delta(x^9)$ and we have included the source term due to the presence of fractional D1-branes.
For the NS-NS twisted scalar $b$ we get the equations:
\bear
&&b'\,=\,-a'\,\,,\label{a}\\ \rc
&&{1\over r^3}\,\,(b'r^3)'\,=\, -\frac{N_c\,\kappa^2_{orb}\,T_1\,}{2\pi^2\alpha'}\,\hat\delta_4\,\,,
\label{b}
\eear
where again sources have been included and $\hat\delta_4\equiv\delta(x^2)\delta(x^3)\delta(x^4)\delta(x^5)$. 
Using this equation we can rewrite (\ref{f3}) as:
\be
\label{laplace}
\partial_r(r^3\sigma^3H_1')\,+\,\partial_{\sigma}(r^3\sigma^3\dot{H_1})\,+\,b'^2r^3\delta_4\,=\,-N_c\,\kappa^2_{orb}\,T_1\,\delta_8,
\ee
One can verify that  (\ref{a}), (\ref{b}) and (\ref{laplace}) imply the fulfillment of the equation of motion for  the dilaton, as well as  the Einstein equations (\ref{Einstein-orb})
(with the aditional source contribution coming from the boundary action). 

The solution of (\ref{b}) is easily determined to be:
\be
\label{bb} 
b\,=\,-2\pi^2\alpha'\,+\,\frac{N_c\,\kappa^2_{orb}\,T_1\,}{2\pi^2\alpha'}\,
\frac{1}{4\,\pi^2\,r^2}\,\,,
\ee
where, to fix the constant term in (\ref{bb}), we have already taken into account that the background value of $b$ is $-2\pi^2\alpha'$. Moreover, 
since $\kappa^2_{orb}\,T_1\,=\,(2\pi)^6\,(\alpha')^3\,g_s$ we can rewrite (\ref{bb}) as:
\beq
b\,=\,-2\pi\alpha'\,\Big[\,1\,-\,{4g_s\alpha'N_c\over r^2}\,\Big]\,\,.
\label{bb-r}
\eeq

To understand how our solution captures the gauge dynamics on the world-volume of the D1-brane we will make a probe computation, similar to the one performed in subsection \ref{probe-analysis}. Accordingly, we shall consider a fractional D1-brane probe in the background just described, in which a worldvolume gauge field $F_{\alpha\beta}$ has been switched on. The D1-brane will be extended along $x^{0,1}$, at fixed values of the remaining coordinates. By expanding the action (\ref{fractional}) in powers of $F_{\alpha\beta}$, we find an expression of the form:
\beq
S_{D1}\,=\,-V\,-\,{1\over 4g_{YM}^2}\,\,\int\,d^2x\,F_{\alpha\beta}\,F^{\alpha\beta}\,+\,\cdots\,\,,
\label{SD1-expanded}
\eeq
where $V$ can be interpreted as the static potential and $g_{YM}$ is the Yang-Mills coupling of the two-dimensional gauge theory. By plugging our ansatz into  (\ref{fractional}) , we get:
\be 
V= \frac{T_1}{2}\Big[ H_1^{-1}\Big(1+\frac{\tilde b}{2\pi^2\alpha'}\Big)-(H_1^{-1}-1)\Big(1+\frac{\tilde b}{2\pi^2\alpha'}\Big)-\frac{a}{2\pi^2\alpha'}\Big]\,=\,-
{T_1\over 4\pi^2\alpha'}\,(a+b)\,\,.
\ee 
Therefore, the expected no-force condition is satisfied if $a$ and $b$ are related as:
\beq
a\,=\,-b\,\,.
\label{a-b}
\eeq
Notice that this relation solves (\ref{a}) and fixes the integration constant of this equation.
From the action (\ref{fractional}) it is also possible to get the value of the two-dimensional Yang-Mills coupling:
\beq
\frac{1}{g^2_{YM}(r)}\,=\,-{b(r)\over 4\pi g_s}\,\,.
\eeq
By using the explicit expression of the function $b(r)$ for our solution (eq. (\ref{bb-r})), and the radius-energy relation $r\,=\,2\pi\alpha'\mu$, we get:
\be 
\frac{1}{g^2_{YM}(r)}\,=\,\frac{\pi\alpha'}{2 g_s}-\frac{N_c}{2\pi\mu^2}\,=\,
\frac{1}{g^2_{YM}}\,\Big[\,1\,-\,{g^2_{YM}\,N_c\over 2\pi^2\mu^2}\,\Big]\,\,,
\ee
where, in the last step, we have introduced  the bare Yang-Mills coupling $g_{YM}^2\,=\,{2g_s\over \pi\alpha'}$.  We notice that, indeed, the field theory result of eq. (\ref{gYM-QFT}) is reproduced by our solution. 

We finish this subsection by recalling that we have found the explicit solution for the functions $a(r)$ and $b(r)$ of our ansatz (eqs. (\ref{bb-r}) and (\ref{a-b})), while the remaining function $H_1(r,\sigma)$ is given by the solution of (\ref{laplace}) (which we have not been able to solve in terms of elementary functions, although we think that it is doable).

\subsection{The flavored solution}
\label{fractionalD1D5}

We now look for a solution corresponding to the complete fractional D1-D5 brane system, the inclusion of the D5-branes accounting for the presence of flavors on the gauge theory side. First of all, we make an ansatz of the standard form for the metric and dilaton in terms of two warp factors $H_1$ and $H_5$, namely:
\bea
\label{fractionalD1D5-metric}
&&ds^2= H_1^{-3/4}H_5^{-1/4}\eta_{\alpha\beta}dx^{\alpha}dx^{\beta}\,+\,H_1^{1/4}H_5^{3/4}\Big(dr^2\,+\,\frac{r^2}{4}\sum_{i=1}^3\omega^i\omega^i\Big)\,+\\ 
&&\qquad\qquad\qquad\qquad\qquad\qquad\qquad
+H_1^{1/4}H_5^{-1/4}\Big(d\sigma^2\,+\, \frac{\sigma^2}{4}\sum_{j=1}^3\tilde\omega^{i}\,\tilde\omega^{j}\Big)\,\,,\rc\rc
&&e^{\Phi}\,=\,H_1^{1/2}H_5^{-1/2}\,\,,
\eear
where $H_1$ and $H_5$ depend on the radial variables $r$ and $\sigma$ as follows:
\beq
H_1=H_1(r,\sigma)\,\,,\qquad\qquad
H_5\,=\,H_5(r)\,\,.
\eeq
As in the unflavored case of section (\ref{fractionalD1}), our background will contain a non-vanishing $b=b(r)$ field. Actually, it is convenient to factor out in $b$ the warp factor 
$H_5(r)$ and to represent it in terms of a new function $Z=Z(r)$ as:
\beq
b(r)\equiv -Z(r)\Big[\,H_5(r)\,\Big]^{-1}\,\,.
\label{b-Z}
\eeq
The two-forms $C_{2}$ and $ {\cal  A}_{2}$ will have now two types of components, due to the presence of the flavor branes. We will represent them as:
\beq
C_{2}=(H_1^{-1}-1)dx^0\wedge dx^1\,+\,{\cal C}_f\,\Omega_2\,\,,
\qquad\qquad
 {\cal  A}_{2}= a(r)\,dx^0\wedge dx^1\,-\,{\cal C}_f\,{Z\over H_5}\,\Omega_2,
\eeq
where ${\cal C}_f$ is a constant accounting for the presence of the D5-branes that wrap the orbifolded space and  $\Omega_2$ is a two-form defined by the condition $d\Omega_2\,=\,\frac{1}{8}\omega_1\wedge\omega_2\wedge\omega_3$. The constant ${\cal C}_f$  is proportional to the number of flavors $N_f$. Actually, by imposing  the quantization of the RR field strength $F_{3}$ we can write it as:
\beq
{\cal C}_f\,=\, -2g_s\alpha' N_f\,\,.
 \eeq
The value of the three-form field strength $T_{3}$ for this flavored ansatz is:
\beq
T_{3}= \Big[a'+(ZH_5^{-1})'(H_1^{-1}-1)\,
-\,{\cal C}_f\,\frac{Z}{r^3\,H_1\,H_5}\Big]
\big(dr\wedge dx^0\wedge dx^1+\frac{r^3\,H_1H_5}{8}\,\omega^1\wedge\omega^2\wedge\omega^3\big)\,\,.\qquad
\eeq
From the action (\ref{lagrangian}) it is easy to find the equations of motion. After including the D1- and D5- brane sources (determined by the boundary actions (\ref{fractional}) and (\ref{fractional5})), we get the following two equations involving the NS-NS twisted scalar $b$:
\bea
\label{af}
a'\,=\,-b'\,+\, \frac{Z}{r^3H_5^{2}}\,\,
{{\cal C}_f\over H_1(r,\sigma=0)}\,\,,\rc\rc
 \label{bf}
{(H_5b'r^3)'\,+\,{\cal C}_f\,b'\over r^3}\,=\, -\frac{\kappa_{orb}^2\,T_1}{2\pi^2\alpha'}\,\left(N_c\,-\frac{N_f}{4}\right)\,\,\hat\delta_4\,\,,
\eea 
which are the flavored  generalization of  (\ref{a}) and (\ref{b}). Using this result, we get from the equation of $F_{3}$:
\be
 \partial_r(r^3\sigma^3H_1')'\,+\, \partial_{\sigma}\,(r^3\sigma^3H_5\,\dot{H}_1)\,=\,-b'^2H_5r^3\delta_4\,-\,N_c\,\kappa_{orb}^2\,T_1\,\delta_8\,\,.
\label{f3f}
\ee
For any function $f(r, \sigma)$, let us  define the Laplacian operator $\Delta$ as:
 \be
 \Delta f(r,\sigma)\,\equiv\,\frac{1}{r^3} \partial_r(f' r^3)\,+\,
 \frac{H_5}{\sigma^3}\partial_{\sigma}\,(\dot{f}\sigma^3)\,\,.
 \ee
Then,  we can rewrite (\ref{f3f}) as:
\be 
r^3\sigma^3\Delta H_1=\,-b'^2H_5r^3\delta_4\,-\,N_c\,\kappa_{orb}^2\,T_1\,\delta_8\,\,.\label{f3fl}
\ee
Using the relations (\ref{bf})-(\ref{f3fl}) one can demonstrate also that the dilaton and the Einstein equations reduce to:
\bear
\label{H5}
\Delta H_5\,=\,-(2\pi)^2g_s\alpha' N_f\hat\delta_4\,\,,\\ \rc
\Delta H_1\,+\,\frac{H_5(ZH_5^{-1})'^2}{\sigma^3}\,\delta_4\,=\,-{(2\pi)^6(\alpha')^3 \,g_s\, N_c\over r^3\sigma^3}\,\delta_8
\label{H1f}
.\eea
By combining all the equations, one can also prove that the function $Z(r)$  defined in (\ref{b-Z}) satisfies the equation:
\beq
{1\over r^3}\,(r^3\,Z')'\,=\,32\,\pi^4\,(\alpha')^2\,g_s\,\Big(\,N_c\,-\,{N_f\over 2}\,\Big)\,
\hat\delta_4\,\,.
\label{Z-eq}
\eeq
The solution to eqs.  (\ref{H5}) and (\ref{Z-eq}) is easily found to be:
\bea
\label{bbf} 
Z&=& 2\pi^2\alpha'\left(1-2g_s\alpha'\,\frac{2N_c-N_f}{r^2}\right)\,\,,\rc\rc
H_5&=&1\,+\,g_s\alpha'\frac{N_f}{r^2}\,\,.
\eea

The twisted RR potential which is relevant for the boundary action is $ A_2$, appearing in (\ref{t3}). One can show that for our solution it can be taken as:
\be
 A_2\,=\,-b\,dx^0\wedge dx^1\,-\,\left[2\alpha'g_sN_f\,b\,-\,4\pi^2g_s\alpha'^2\,(4N_c-N_f)\right]\Omega_2.
 \label{A2}
\ee
Let us now perform a probe analysis similar to the one carried out for the unflavored case. Accordingly, let us consider a fractional D1-brane extended along the Minkowski directions and let us expand its action (\ref{fractional})  as in  (\ref{SD1-expanded}).  By using the value of $A_2$  displayed in (\ref{A2}) it is possible to show that the potential $V$ vanishes. Moreover, one can also obtain the value of the Yang-Mills coupling constant, namely:
\beq
 {1\over g^2_{YM}(r)}\,=\,{Z(r)\over 4\pi g_s}\,\,.
 \label{gYM-flavored-orbifold}
 \eeq
 By substituting in the right-hand side of (\ref{gYM-flavored-orbifold}) the value of the function $Z(r)$ given in (\ref{bbf}) one gets:
\be 
\frac{1}{g^2_{YM}(r)}\,=\,\frac{\pi\alpha'}{2 g_s}-\frac{2N_c-N_f}{4\pi\mu^2}\,=\,
\frac{1}{g^2_{YM}}\,\,\Big[\,1\,-{g^2_{YM}\over 2\pi\mu^2}\,\Big(N_c\,-\,{N_f\over 2}\Big)
\Big]\,\,,
\ee
where again $r\,=\,2\pi\alpha'\mu$ and the bare Yang-Mills coupling is $g_{YM}^2\,=\,{2g_s\over \pi\alpha'}$ . Then, the field theory result of eq. (\ref{gYM-QFT}) is  also reproduced by our solution when flavor branes are added. Again, to find the complete solution we would have to solve for the warp factor $H_1$ in (\ref{H1f}), something we will not attempt to do here.

\section{Summary and conclusions}
\label{conclusions}

In this paper we have found string duals to gauge theories  in two dimensions with ${\cal N}=(4,4)$ supersymmetry and a large number of colors. These duals are engineered by considering D3-branes wrapping a two-cycle of a Calabi-Yau cone of complex dimension two. We also obtained the dual of the ${\cal N}=(4,4)$ theories with flavors both  in the quenched ($N_f/N_c \to 0$) and unquenched limit ($N_f/N_c $ finite). 
We checked that our solutions correctly capture the perturbative running of gauge couplings,  despite the fact that the resulting geometries develop an infrared singularity. Actually, this phenomenon  is typical for the gravity duals of field theories with such an amount of supersymmetry.

 We have also explored the holographic realization of the Higgs mechanism for our setup and the meson spectrum in both the Coulomb and Higgs branches of the theory. As an alternative string dual we have found supergravity solutions representing a system of fractional D1/D5-branes on an orbifold and we have verified that these solutions also capture the one loop beta functions of the gauge theory (see ref. \cite{Divecchia} for a discussion of the relation between the wrapped and fractional brane approaches).

The geometries we have found  present a naked singularity in the IR. As usual, such singularity signals some interesting infrared physics.   We know  that, in cases with such an amount of supersymmetry, the singularity can be consistently screened by an enhan\c{c}on (it is common in this case to talk about the excision of the singularity). The enhan\c{c}on is the locus where the sources of the background become actually tensionless and the geometry of the supergravity solution seems to end there: it is not possible to determine the geometry inside the enhan\c{c}on shell. On the field theory side, this corresponds to the inability to get instanton corrections. These euclidean non-perturbative configurations contribute to physical correlators with a factor proportional to the exponential of minus their action: $\exp[-\frac{8\pi^2}{g^2}]$. It is easy to see that at the enhan\c{c}on locus $g^2 \to \infty$ and thus these contributions become important (they are of ${\cal O}(1)$). The fact that the gravity solution ends there corresponds to the fact that it does not include them. This lack of information of the gravity dual is due to the fact that these configurations are suppressed in the large $N_c$  't Hooft limit, according to which  the 't Hooft coupling $N_c \,g^2 $ is kept fixed. Instanton corrections are thus exponentially suppressed in this limit.
This is reflected also in the computation of the beta function. The gravity solution gives the exact perturbative answer but all the non-perturbative instanton corrections are missing: they are suppressed in the 't Hooft limit.

This is different from the case with less supersymmetry (four real supercharges) where the singularity can be actually removed and the geometry is determined up to the deep IR. The difference is due to the fact that in this case the relevant non-perturbative configurations are the so-called fractional instantons, whose action goes as $\frac{8\pi^2}{N_c \,g^2}$: their contribution is of ${\cal O}(1)$ in the large $N_c$  't Hooft limit and it can thus be captured by the gravity duals. In the two dimensional cases we are considering here, this would correspond to finding a completely regular solution describing the dual of the (2,2) supersymmetric field theory. There are at least two possible scenarios to engineer such a dual using wrapped branes: D3-branes wrapped on a two-cycle or D5-branes wrapped on a four-cycle of a three (complex)-dimensional Calabi-Yau.
It should be possible to explicitly determine this geometry in the corresponding gauged supergravity approach. The D3-brane case looks more promising for the de-singularization of the background: a non-abelianization of the gauge field responsible for the twist along the cycle where the relevant D3-brane is wrapped has to be implemented  (this is analogous to what happens in the four dimensional case studied in \cite{CVMN}). We are working in this direction \cite{22flavors}.

The three-dimensional analogue of our unflavored background was analyzed in \cite{Divecchia}. This background is created by D4-branes wrapping a two-cycle of a Calabi-Yau  manifold in the type IIA theory. By using the same techniques employed here, one can add flavor to this solution and check that the field theory results are matched \cite{3dflavors}.

Another interesting point to discuss is the validity, for our flavored system, 
of the supergravity approximation which, as is well-known, requires the Ricci tensor to be small in string units. As already discussed above, the geometry is singular in the IR.  Let us now explore its behavior outside the enhan{\c{c}}on region. Instead of looking directly at the Ricci tensor, it is more useful to analyze the right-hand side of the Einstein equation (\ref{Einstein}). Clearly, the energy-momentum tensor of the brane (\ref{TMN}) gives a singular  contribution  localized at the position $\rho=\rho_Q$ of the flavor brane, which corresponds to the wedge shape of the function $z$ (see figure \ref{Flavored}).  On the contrary, in the  RR five-form the flavors contribute with a Heaviside function (see eq. (\ref{calF5-flavored})), which is non-vanishing outside the location of the flavor source. By inspecting (\ref{calF5-flavored}) one readily concludes that the contribution of $N_f$ to $F_5$ (and thus to the Ricci tensor) is just an additive constant to $g$.  
Recall that $g$ is proportional to $N_c$. Actually, one can show that  in many observables  $N_c$ and $N_f$ appear in the combination $N_c-{N_f\over 2}$, which scales as $N_c$ in the Veneziano limit $N_c, N_f\to\infty$ with $N_f/N_c$ fixed.  Therefore, outside the enhan{\c{c}}on and the location of the flavor brane source, the curvature in the flavored model is small in the Veneziano limit, as it was in the unflavored model  in the  't Hooft limit.  Notice that this is in contrast to what happens to the flavored Klebanov-Witten model of ref. \cite{Benini:2006hh} (see \cite{Bigazzi:2008zt} for a clear discussion) and is similar to the behavior of the case studied in \cite{Casero:2006pt}.

Finally, let us now comment on the validity of using the DBI+WZ action in our setup. This was already discussed, for example, in \cite{HoyosBadajoz:2008fw} (see section 7 of that paper). We have $N_f$ flavor branes and, in principle, we would like to find a solution where they are all coincident. This solution would be dual to the QFT with $SU(N_f)$ global symmetry (when we smear, we are breaking this global flavor group to $U(1)^{N_f}$). Notice nevertheless, that the localized solution mentioned above may need the DBI+WZ action to be corrected. Indeed, since the string coupling $g_s\sim N_c^{-1}$ in the usual scaling for D3-branes, we find that open string diagrams, correcting the brane action, are weighted by factors of $g_s N_f\sim \frac{N_f}{N_c}\sim 1$. This would imply that (unless some cancellation happens) we should not trust the tree-level DBI+WZ action. Now, it is clear that the smearing avoids this potential problem. Indeed, when `separating'  the flavor branes, the correction to the tree-level action will be suppressed by the mass of the open strings between flavor branes. Of course, there is another possibility one should not discard, that is the fact that even in the localized case, the corrections aluded above do indeed cancel due to SUSY. If this is the case, the smearing should be seen as just a technical trick to get simpler equations.

\section*{Acknowledgments}

We are grateful to F. Bigazzi,  E. Conde,  A. Cotrone, I. Kirsch, A. Paredes, D. Rodriguez-Gomez, J. Shock and  D. Zoakos for discussions. 
This  work was supported in part by MEC and  FEDER  under grant
FPA2005-00188,  by the Spanish Consolider-Ingenio 2010 Programme CPAN (CSD2007-00042), by Xunta de Galicia (Conselleria de Educacion and grant PGIDIT06PXIB206185PR) and by  the EC Commission under  grant MRTN-CT-2004-005104.

\vskip 1cm
\renewcommand{\theequation}{\rm{A}.\arabic{equation}}
\setcounter{equation}{0}
\appendix
\section{BPS equations}
\label{BPSequations}
We will determine the functions $z(\rho,\sigma)$, $H(\rho,\sigma)$ and $g(\rho,\sigma)$ of our ansatz by imposing that the background preserves eight supersymmetries. This requirement will lead us to a set of first-order BPS equations whose fulfillment will imply the equations of motion of type IIB supergravity. In order to obtain these  BPS equations let us consider the supersymmetric variations of the dilatino $\lambda$ and of  the gravitino $\psi_M$ for the type of background  that we  are analyzing.  These  variations of the dilatino and gravitino are:
\bear
&&\delta\lambda\,=\,{1\over 2}\,\,\Gamma^{M}\,\partial_ M\,\Phi\,\epsilon\,\,,\rc\rc
&&\delta\psi_ M\,=\,\Big[\,D_ M\,+\,{e^{\Phi}\over 8}\,\,{1\over 2\cdot  5!}\,\,
F_{N_1\cdots N_5}^{(5)}\,\Gamma^{N_1\cdots N_5}\,\Gamma_{M}\,\,
(i\tau_2)\,\,\Big]\,\epsilon\,\,,
\label{susyvariationsD3}
\eear
where $\Phi$ is the dilaton, $\epsilon$ is a doublet of Majorana-Weyl spinors of fixed ten-dimensional chirality and $\tau_2$ is the second Pauli matrix (which acts on the doublet $\epsilon$). The Killing spinors of the background are the $\epsilon$'s for which $\delta\lambda=\delta\psi_M=0$. These Killing spinors generate the supersymmetries preserved by the background. Their existence imposes some non-trivial constraints on the different fields of type IIB supergravity. For example, it is clear from the first equation in (\ref{susyvariationsD3}) that the dilatino equation $\delta\lambda=0$ requires that the dilaton $\Phi$ should be constant. Accordingly, in what follows we will take $\Phi=0$. Moreover, the different components of the gravitino equation $\delta\psi_M=0$ will lead us to the desired system of BPS equations when a set of projections are imposed on the spinors $\epsilon$. In order to obtain them, let us choose the following vielbein basis for the metric  (\ref{D3metric}):
\bear
&&e^{0,1}\,=\,H^{-{1\over 4}}\,dx^{0,1}\,\,,\qquad
e^2\,=\,H^{-{1\over 4}}\,{\sqrt{z}\over m}\,d\theta\,\,,\qquad
e^{3}\,=\,H^{-{1\over 4}}\,{\sqrt{z}\over m}\,\sin\theta\,d\phi\,\,,\rc\rc
&&e^{4}\,=\,{H^{{1\over 4}}\over \sqrt{z}}\,d\sigma\,\,,\qquad
e^{5}\,=\,{H^{{1\over 4}}\over \sqrt{z}}\,\sigma\,\,
(\,d\psi\,+\,\cos\theta\,d\phi\,)\,\,,\qquad
e^{6}\,=\,H^{{1\over 4}}\,d\rho\,\,,\rc\rc
&&e^{7}\,=\,\rho \,H^{{1\over 4}}\,d\alpha_1\,\,,\qquad
e^{8}\,=\,\rho \,H^{{1\over 4}}\,\sin\alpha_1\, d\alpha_2\,\,,\qquad
e^{9}\,=\,\rho \,H^{{1\over 4}}\,\sin\alpha_1\,\sin\alpha_2\,d\alpha_3\,\,.\qquad\qquad
\label{D3frame}
\eear
The ansatz for $F_5$ in this unflavored case has been written in (\ref{F5}) in terms of the magnetic component ${\cal F}_5=d{\cal C}_4$ and of its Hodge dual. From the expression of ${\cal C}_4$ in (\ref{CalC4}) we can readily compute the value of 
${\cal F}_5$. In terms of the vielbein one-forms (\ref{D3frame}) the five-form ${\cal F}_5$ is given by:
\beq
{\cal F}_5\,=\,{H^{-{5\over 4}}\sqrt{z}\over \rho^3\sigma}\,\,
\Big(\,\dot g\,\sqrt{z}\,e^4\,+\,g'\,e^6\,\Big)
\wedge e^7\wedge e^8\wedge e^9\wedge e^5\,+\,
{m^2\,gH^{-{1\over 4}}\over  \rho^3 z}\,e^2\wedge e^3\wedge  e^7\wedge e^8\wedge e^9\,\,,
\label{calF5-unflavored}
\eeq
while its Hodge dual can be written as:
\beq
{}^*{\cal F}_5\,=\,{H^{-{5\over 4}}\sqrt{z}\over \rho^3\sigma}\,\,
\Big(\,-\dot g\,\sqrt{z}\,e^6\,+\,g'\,e^4\,\Big)
\wedge e^0\wedge e^1\wedge e^2\wedge e^3\,+\,
{m^2\,gH^{-{1\over 4}}\over  \rho^3 z}\,e^0\wedge e^1\wedge  e^4\wedge e^5\wedge e^6\,\,.
\label{Hodge-calF5-unflavored}
\eeq
We can now evaluate the right-hand side of the gravitino variation in (\ref{susyvariationsD3}). In order to do so, we shall impose to the spinor $\epsilon$ the following set of projections:
\beq
\Gamma_{2345}\,\epsilon\,=\,\epsilon\,\,,\qquad\qquad
\Gamma_{0123}\,(i\tau_2)\,\epsilon\,=\, \epsilon\,\,,
\label{projections}
\eeq
where $\Gamma_{\bar M_1\, \bar M_2\cdots}$ denotes the antisymmetrized product of constant Dirac matrices and the indices in (\ref{projections}) correspond to the frame (\ref{D3frame}). After imposing these projections, one can show that the different components of the gravitino equation $\delta\psi_M=0$ are satisfied if 
the system (\ref{BPSsystem})  of first-order BPS equations  for $z$, $H$ and $g$ holds. Moreover,
when the equations  in (\ref{BPSsystem}) are satisfied one can verify that one has the following expression for the Killing spinors:
\beq
\epsilon\,=\,H^{-{1\over 8}}\,e^{-{\psi\over 2}\Gamma_{23}}\,
e^{{\alpha_1\over 2}\Gamma_{67}}\,e^{{\alpha_2\over 2}\Gamma_{78}}\,
e^{{\alpha_3\over 2}\Gamma_{89}}\,\,\eta\,\,,
\label{epsilon-eta}
\eeq
where $\eta$ is a constant spinor that satisfies the same projections as in
(\ref{projections}), namely:
\beq
\Gamma_{2345}\,\eta\,=\,\eta\,\,,\qquad\qquad
\Gamma_{0123}\,(i\tau_2)\,\eta\,=\, \eta\,\,.
\label{eta-projections}
\eeq
Notice that, after imposing these conditions to $\eta$, we are left with eight preserved supersymmetries, as it should for a (4,4) two-dimensional gauge theory.

Notice that, according to its Bianchi identity,  $F_5$ should be a closed five-form, namely one should have:
\beq
dF_5\,=\,0\,\,.
\label{dF5}
\eeq
The fulfillment of (\ref{dF5}) is not automatic for our ansatz. Indeed, one can check that 
(\ref{dF5}) is equivalent to the following partial differential equation (PDE):
\beq
\partial_{\rho}\,\Big[\,{z g'\over \sigma \rho^3 H^2}\,\Big]\,+\,
\partial_{\sigma}\,\Big[\,{z^2 \dot g\over \sigma \rho^3 H^2}\,\Big]\,=\,
{m^4\,g\,\sigma \over \rho^3 z^2}\,\,.
\label{PDEforF5}
\eeq
One can verify that the equation (\ref{PDEforF5})  is satisfied as a consequence of the system (\ref{BPSsystem}). Let us check this fact by explicit calculation. First of all, by using the third and fourth equations in the system (\ref{BPSsystem}) we can rewrite the two terms of the left-hand side of (\ref{PDEforF5})
as:
\bear
&&{z g'\over \sigma \rho^3 H^2}\,=\,z\,\partial_{\sigma}\big(\,H^{-1}\,\big)\,\,,
\rc\rc
&&{z^2 \dot g\over \sigma \rho^3 H^2}\,=\,-\,z\,\partial_{\rho}\big(\,H^{-1}\,\big)\,-\,{m^2\,g\over \rho^3 H}\,\,.
\eear
Then, by explicit calculation one can easily prove that:
\beq
\partial_{\rho}\,\Big[\,{z g'\over \sigma \rho^3 H^2}\,\Big]\,+\,
\partial_{\sigma}\,\Big[\,{z^2 \dot g\over \sigma \rho^3 H^2}\,\Big]\,=\,
\Big[\, z'\,-\,{m^2\,g\over \rho^3}\,\Big]\,\partial_{\sigma}\big(\,H^{-1}\,\big)\,+\,
 {H'\over H^2}\,\dot z\,-\,{m^2\,\dot g\over\rho^3 H}\,\,.
\eeq
After using the first and second equation in (\ref{BPSsystem}), we can rewrite the previous equation as:
\beq
\partial_{\rho}\,\Big[\,{z g'\over \sigma \rho^3 H^2}\,\Big]\,+\,
\partial_{\sigma}\,\Big[\,{z^2 \dot g\over \sigma \rho^3 H^2}\,\Big]\,=\,-
{m^2\over \rho^3 H}\,\Big[\,\dot g\,-\,{\sigma\rho^3\over z}\,H'\,\Big]\,\,,
\eeq
and (\ref{PDEforF5}) can be proved by using again the fourth equation in (\ref{BPSsystem}). Thus, $F_5$ should be represented, at least locally, as $dC_4$. A possible value of this four-form potential has been written in eq. (\ref{C4}). 

Finally, one can also verify that the Einstein equations are satisfied as a consequence of the first-order system (\ref{BPSsystem}) (see subsection \ref{BPSequations-flavored}, where this fact is checked for the backreacted metric).

\subsection{Flavored BPS equations}
\label{BPSequations-flavored}
For the backreacted background, our ansatz for the metric is just the same as in the unflavored case (eq. (\ref{D3metric})). Moreover, the five-form $F_5$ contains an extra term (see eq. (\ref{calF5-flavored})), which takes care of the fact that the Bianchi identity is violated as in (\ref{newBianchi}). The actual form of ${\cal F}_5$ in the present case is:
\beq
{\cal F}_5\,=\,[\,\dot g\,d\sigma\,+\,g' d\rho\,]\,\wedge \omega_{3}\wedge (d\psi+\cos\theta d\phi)\,+\,[\,g\,-\,2\pi\,g_s\,(\alpha')^2\,N_f\,\Theta(\rho-\rho_Q)\,]
\,\omega_3\,\wedge\,\omega_2\,\,.
\label{calF5-flavored2}
\eeq
By computing the Hodge dual of ${\cal F}_5$, we get:
\bear
&&{}^*{\cal F}_5\,=\, {z\over m^2\sigma\rho^3 H^2}\,(\,g'\,d\sigma\,-\,z\dot g\,d\rho)\,\wedge dx^0\wedge dx^1\,\wedge \omega_2\,+\,\rc\rc
&&\qquad+\,
{\sigma m^2 \over \rho^3 z^2}\,
\big[\,g\,-\,2\pi\,g_s\,(\alpha')^2\,N_f\,\Theta(\rho-\rho_Q)\,\big]\,
d\rho\wedge d\sigma \wedge dx^0\wedge dx^1\wedge
(d\psi+\cos\theta d\phi)\,\,.\qquad\qquad
\label{Hodge-calF5-flavored}
\eear
We can now repeat the analysis performed in the unflavored case to arrive at the system (\ref{BPSsystem}). We will still impose the projection conditions (\ref{projections}). Actually, it is clear by comparing (\ref{calF5-flavored2}) and (\ref{Hodge-calF5-flavored}) with their unflavored counterparts (\ref{calF5-unflavored}) and (\ref{Hodge-calF5-unflavored}) that the resulting BPS equations in this backreacted case can be obtained from (\ref{BPSsystem}) by substituting:
\beq
g\,\to\,g\,-2\pi\,g_s\,(\alpha')^2\,N_f\,\Theta(\,\rho-\rho_Q\,)\,\,,
\eeq
in the terms in which $g$ is not differentiated. This is, indeed, the difference between the flavored and unflavored BPS systems written in  eqs. (\ref{BPSsystem}) and (\ref{flavored-BPSsystem}), respectively. Moreover, the Killing spinors are still given by (\ref{epsilon-eta}) and (\ref{eta-projections}) and, thus, the number of unbroken supersymmetries is also eight for the backreacted backgrounds.

The modified Bianchi identity (\ref{newBianchi}) for this flavored case is satisfied if the following PDE for the functions of the ansatz holds:
\beq
\partial_{\rho}\,\Big[\,{z g'\over \sigma \rho^3 H^2}\,\Big]\,+\,
\partial_{\sigma}\,\Big[\,{z^2 \dot g\over \sigma \rho^3 H^2}\,\Big]\,=\,
{\sigma m^4 \over \rho^3 z^2}\,\big[\,g\,-\,2\pi\,g_s\,(\alpha')^2\,N_f\,\Theta(\rho-\rho_Q)\,\big]\,\,.
\label{flavored-PDEforF5}
\eeq
Eq. (\ref{flavored-PDEforF5}) can be verified by direct calculation, following the same strategy used to prove (\ref{PDEforF5}) from the BPS system. However it is simpler to check that, after subtracting an appropriate piece to $F_5$, the result can be represented in terms of a four-form potential. Indeed, by using the equations in (\ref{flavored-BPSsystem}) one can verify that:
\beq
F_5\,+\,2\pi\,g_s\,(\alpha')^2\,N_f\,\Theta(\rho-\rho_Q)\,\omega_2\wedge \omega_3\,=\,dC_4\,\,,
\label{F5dC4-flavored}
\eeq
where  $F_5$ is given by the ansatz (\ref{calF5-flavored2})-(\ref{Hodge-calF5-flavored}) and $C_4$ has the same expression as the one written in (\ref{C4}) for the unflavored case. It is now obvious  from (\ref{F5dC4-flavored}) that $F_5$ satisfies (\ref{newBianchi}).

The Einstein equations in the Einstein frame are:
\beq
R_{MN}\,-\,{1\over 2}\,G_{MN}\,R\,=\,{1\over 96}\,
F_{MP_1\cdots P_4}^{(5)}\,\,F_{N}^{(5)\,\,P_1\cdots P_4}\,+\,T_{MN}\,\,,
\label{Einstein}
\eeq
where $T_{MN}$ is the energy-momentum tensor of the smeared flavor brane, defined as:
\beq
T_{MN}\,=\,-{2\kappa_{10}^2\over \sqrt{-G}}\,{\delta S_{DBI}\over \delta G^{MN}}\,\,.
\eeq
Taking into account the form (\ref{smearedDBI}) of  $S_{DBI}$ we get the following expression for $T_{MN}$ in flat components:
\beq
T_{\bar M \bar N}\,=\,-{(2\pi)^4\,g_s\,(\alpha')^2\over 2}\,\,
\Big[\,\eta_{\bar M \bar N}\,\big|\,\Omega\,\big|\,-\,{1\over 5!}\,{1\over \big|\,\Omega\,\big|}
\,\,\Omega_{\bar M\bar P_1\cdots \bar  P_5}\,\Omega_{\bar N\bar R_1\cdots \bar R_5}\,
\eta^{\bar P_1 \bar R_1}\cdots \eta^{\bar P_5 \bar R_5}\,\Big]\,\,,
\label{TMN}
\eeq
where $\Omega$ is the smearing form defined in (\ref{Omega}) and $\big|\,\Omega\,\big|$ is written in (\ref{modulusOmega}). 
In order to evaluate the different components of the tensor (\ref{TMN}) it is convenient to write $\Omega$ in terms of the frame forms. We get:
\beq
\Omega\,=\,-{N_f\over 8\pi^3}\,{m^2\over z\rho^3 \sqrt{H}}\,\,
\delta(\rho-\rho_Q)\,\,e^2\wedge e^3\wedge e^6\wedge e^7\wedge e^8\wedge e^9\,\,.
\eeq
From this expression we obtain the modulus of $\Omega$, namely:
\beq
\big|\,\Omega\,\big|\,=\,{N_f\over 8\pi^3}\,{m^2\over |z|\rho^3 \sqrt{H}}\,
\delta(\rho-\rho_Q)\,\,.
\eeq
Then, the explicit values of the different components of $T_{\bar M\bar N}$ are:
\bear
&&-T_{00}\,=\,T_{11}\,=\,T_{44}\,=\,T_{55}\,=\,\pi\,g_s\,(\alpha')^2\,N_f\,
{m^2\over |z|\rho^3 \sqrt{H}}\,\delta(\rho-\rho_Q)\,\,,\rc\rc
&&T_{22}\,=\,T_{33}\,=\,T_{66}\,=\,T_{77}\,=\,T_{88}\,=\,T_{99}\,=\,0\,\,.
\eear
By using these values, it is straightforward (but tedious) to verify that the Einstein   equations  (\ref{Einstein}) are satisfied as a consequence of the first-order equations (\ref{flavored-BPSsystem}).

\vskip 1cm
\renewcommand{\theequation}{\rm{B}.\arabic{equation}}
\setcounter{equation}{0}
\section{The dual of the (4,4) theory from gauged supergravity}
\label{gaugedsugra44}
Let us consider the  five-dimensional gauged supergravity of ref. \cite{Cvetic:1999xp} which, apart from the five-dimensional metric $g_5$, contains three $U(1)$ gauge fields $A_{\mu}^i$, $ (i=1,2,3)$ and two scalars $\varphi_1$ and $\varphi_2$, which we arrange as a  two-component vector  $\vec \varphi\equiv (\varphi_1, \varphi_2)$. We also consider three two-component vectors $\vec a_1$, $ \vec a_2$ and $\vec a_3$ whose inner products are given by:
\beq
\vec a_i\cdot\vec a_j\,=\,4\delta_{ij}-{4\over 3}\,\,.
\eeq
According to \cite{Maldacena:2000mw} we will represent these vectors as:
\beq
\vec a_1\,=\,({2\over \sqrt{6}}, \sqrt{2})\,\,,\qquad\qquad
\vec a_2\,=\,({2\over \sqrt{6}}, -\sqrt{2})\,\,,\qquad\qquad
\vec a_3\,=\,(-{4\over \sqrt{6}}, 0)\,\,.
\eeq
Then, the bosonic part of the  lagrangian  density of the system is given by:
\bear
&&{\cal L}\,=\,\sqrt{-\det g_5}\,\Big[\,R\,-\,{1\over 2}\,(\partial_{\mu}\varphi_1)^2\,-\,
{1\over 2}\,(\partial_{\mu}\varphi_2)^2\,+\,4 \hat g^2\,\sum_{i}\,
e^{{1\over 2} \vec a_i\cdot \vec \varphi}\,-\,\rc\rc
&&\qquad\qquad\qquad
-\,{1\over 4}\,\sum_i 
e^{ \vec a_i\cdot \vec \varphi}\,\Big(F_{\mu\nu}^i\Big)^2\,+\,
{1\over 4}\,\epsilon^{\mu\nu\rho\sigma\lambda}\,
F_{\mu\nu}^1\,F_{\rho\sigma}^2\,A_{\lambda}^3\,\,\Big]\,\,,
\label{5d-lagrangian}
\eear
where $R$ is the Ricci scalar for the  five-dimensional metric $g_5$, $F_{\mu\nu}^i$ are the field strengths of the abelian gauge fields $A^i_{\mu}$ and $ \hat g$ is a coupling constant.  Following \cite{Cvetic:1999xp, Maldacena:2000mw}, let us introduce the quantities $X^i$ and $X_i$ as:
\beq
X^i\,\equiv\,e^{-{1\over 2} \vec a_i\cdot \vec \varphi}\,\,,\qquad\qquad
X_i\,\equiv\,{1\over 3}\,e^{{1\over 2} \vec a_i\cdot \vec \varphi}\,\,.
\label{Xis}
\eeq
This theory contains two dilatinos $\lambda_i$ ($i=1,2)$  and a gravitino $\psi_{\mu}$, whose supersymmetric variations are given by:
\bear
&&\delta\lambda_i\,=\,\Big[\,{3\over 8}\,\partial_{\varphi_i} X_j\, \Gamma^{\mu\nu}\,
F_{\mu\nu}^{j}\,-\,{i\over 4}\Gamma^{\mu}\,\partial_{\mu} \varphi^i\,+\,{3 i\hat g\over 2}\,
V_j\,\partial_{\varphi_i} X^j\,\Big]\,\epsilon\,\,,\rc\rc
&&\delta\psi_{\mu}\,=\,\Big[\,D_{\mu}\,+\,{i\over 8}\,X_i\,\big(\,
\Gamma_{\mu}^{\,\,\nu\rho}\,-\,4\delta_{\mu}^{\nu}\,\Gamma^{\rho}\,\big)\,
F_{\nu\rho}^{i}\,+\,{\hat g\over 2}\,\Gamma_{\mu}\,X^i\,V_i\,-\,
{3i\over 2}\hat g\,V_i\,A_{\mu}^i\,\Big]\,\epsilon\,\,,
\label{SUSY-5d}
\eear
where the quantity  $V_i$ is defined as:
\beq
V_i\,\equiv\,{1\over 3}(1,1,1)\,\,.
\eeq

\subsection{Background with (4,4) SUSY}
Five-dimensional gauged supergravity is the right theory to describe the supergravity solution corresponding to wrapped D3-branes. As first pointed out in \cite{ Maldacena:2000mw}, the D3-brane can be regarded as a domain wall of the five-dimensional space and supersymmetry is realized by switching on appropriate gauge fields, which implement the corresponding topological twisting. Accordingly, let us adopt the following ansatz for the five-dimensional metric:
\beq
ds_5^2\,=\,e^{2f}\,\big[\,dx^2_{1,1}\,+\,dr^2\,\big]\,+\,{e^{2h}\over m^2}\,
\big[\,(d\theta)^2\,+\,\sin^2\theta\,(d\phi)^2\,\big]\,\,,
\label{5dmetric}
\eeq
with $f=f(r)$ and $h=h(r)$ being functions of the radial variable $r$ to be determined. We will consider a truncated version of the theory just described, in which the scalar field $\varphi_2=0$. Moreover, we will redefine the remaining scalar field  $\varphi_1$ as follows:
\beq
\varphi\,\equiv\,{\varphi_1\over \sqrt{6}}\,\,,
\label{5dscalar}
\eeq
and we will assume that $\varphi$ only depends on the radial coordinate $r$. 
Notice that in this truncated version of the  theory the quantities $X^i$ and $X_i$ defined  in (\ref{Xis}) reduce to:
\beq
X_{i}\,=\,{1\over 3}\,\,\big(\,e^{\varphi}, e^{\varphi}, e^{-2\varphi})\,\,,
\qquad\qquad
X^{i}\,=\,\big(\,e^{-\varphi}, e^{-\varphi}, e^{2\varphi})\,\,.
\label{truncatedXis}
\eeq
 In addition we will assume the following values of the $U(1)$ gauge fields:
\beq
A^1\,=\,0\,\,,\qquad\qquad
A^2\,=\,0\,\,,\qquad\qquad
A^3\,=\,{1\over m}\,\cos\theta d\phi\,\,,
\label{5dgauge}
\eeq
where $m$ is a constant and,  for simplicity, we will fix  the coupling constant $\hat g$ as $\hat g=m$. 

We shall require that the solution given by our ansatz  is supersymmetric, which is equivalent to demanding that $\delta \lambda_i=\delta\psi_{\mu}=0$, \ie\ that the right-hand side of (\ref{SUSY-5d}) vanishes for some Killing spinors $\epsilon$. Actually, we shall impose the following projections on $\epsilon$:
\beq
\Gamma_{12}\,\epsilon\,=\,-i\epsilon\,\,,
\qquad\qquad
\Gamma_{r}\,\epsilon\,=\,-\epsilon\,\,.
\label{5dprojections}
\eeq
After plugging our ansatz (\ref{5dmetric}), (\ref{5dscalar}) and (\ref{5dgauge}) for the metric and fields into the supersymmetry variations (\ref{SUSY-5d}), and by  imposing the projections (\ref{5dprojections}), we arrive at the following system of first-order BPS equations \cite{ Maldacena:2000mw}:
\bear
&&h'\,=\,{m\over 3}\,e^{f}\,\Big[\,e^{-2\varphi-2h}\,+\,2e^{-\varphi}\,+\,e^{2\varphi}\,\Big]
\,\,,\rc\rc
&&f'\,=\,{m\over 6}\,e^{f}\,\Big[\,-e^{-2\varphi-2h}\,+\,2\,\Big(\,
2e^{-\varphi}\,+\,e^{2\varphi}\,\Big)\,\Big]\,\,,
\rc\rc
&&\varphi'\,=\,{m\over 3}\,e^{f}\,\Big[\,e^{-2\varphi-2h}\,-\,2\,\Big(\,
e^{2\varphi}\,-\,e^{-\varphi}\,\Big)\,\Big]\,\,,
\label{5dBPSsystem}
\eear
where the prime denotes derivatives with respect to the radial variable $r$. 
In order to integrate this system, let us notice that we can get from it the following  equations satisfied by combinations  of the functions of the ansatz:
\beq
2h'+\varphi'\,=\,m\,e^{f-\varphi}\,\Big[\,e^{-2h-\varphi}\,+\,2\,\Big]\,\,,
\qquad\qquad
2f'+\varphi'\,=\,2m\,e^{f-\varphi}\,\,.
\eeq
Moreover, let us define a new radial variable $\tau$, related to $r$ by means of the equation:
\beq
{d\over dr}\,\equiv\,m\,e^{f-\varphi}\,{d\over d\tau}\,\,,
\eeq
and the  two new functions $\Lambda_1$ and $\Lambda_2$ as:
\beq
\Lambda_1\,\equiv\,2h+\varphi\,\,,\qquad\qquad \Lambda_2\,=\,2f+\varphi\,\,.
\eeq
Then,  one can easily verify  from (\ref{5dBPSsystem}) that the differential equations for $\Lambda_{1,2}$  are:
\beq
{d\Lambda_1\over d \tau}\,-\,e^{-\Lambda_1}\,=\,2\,\,,\qquad\qquad
{d\Lambda_2\over d \tau}\,=\,2\,\,,
\eeq
whose integral is just:
\beq
e^{\Lambda_1(\tau)}\,=\,\alpha\,e^{2\tau}\,-{1\over 2}\,\,,
\qquad\qquad
e^{\Lambda_2(\tau)}\,=\,\beta\,e^{2\tau}\,\,,
\eeq
with $\alpha$ and $\beta$ being constants of integration. Moreover, the equation for $\varphi$ can be written in terms of $\Lambda_1$ as follows:
\beq
{d\varphi\over d\tau}\,+\,{2\over 3}\,e^{3\varphi}\,=\,{1\over 3}\,e^{-\Lambda_1}\,
+\,{2\over 3}\,\,,
\eeq
which can also be easily integrated. In terms of the original functions $f$, $h$ and $\varphi$  of our ansatz, the solution of the BPS system is thus:
\bear
&&e^{2h+\varphi}\,=\,\alpha\,e^{2\tau}\,-{1\over 2}\,\,,\rc\rc
&&e^{2f+\varphi}\,=\,\beta\,e^{2\tau}\,\,,\rc\rc
&&e^{-3\varphi}\,=\,{\alpha e^{2\tau}\,-\,\tau\,-\gamma\over 
\alpha\,e^{2\tau}\,-{1\over 2}}\,\,,
\label{gsugrasolution}
\eear
where $\gamma$ is a new constant of integration.

\subsection{The uplifted metric}
By using the equations written in \cite{Cvetic:1999xp} one can get the expression of the metric and five-form of the ten-dimensional background corresponding to the solution of five-dimensional gauged supergravity just found.  Actually, the uplifting formula for the metric is \cite{Cvetic:1999xp}:
\beq
ds_{10}^2\,=\,\sqrt{\Delta}\,\, ds_5^2\,+\,{3\over \hat g^2 \sqrt{\Delta}}\,
\sum_{i=1}^{3}
X_{i}\,\Big(d\mu_i^2\,+\,\mu_i^2\,\big(\,d\phi^i\,+\,\hat g\,A^i\,\big)^2\,\Big)\,\,,
\label{upliftedmetric}
\eeq
where $\hat g$ is the same coupling constant  as in (\ref{5d-lagrangian}) and, as before, we  will take $\hat g=m$.  In (\ref{upliftedmetric}) the $\phi^i$ are three angles varying between $0$ and $2\pi$ and 
the quantities $\mu_i$ satisfy the relation:
\beq
\sum_{i=1}^{3}\mu_i^2\,=\,1\,\,,
\eeq
while $X_{i}$ are defined in terms of the scalar field $\varphi$ as in
(\ref{truncatedXis}). The quantity $\Delta$ appearing in (\ref{upliftedmetric}) is defined as:
\beq
\Delta\,=\, \sum_{i=1}^{3}\,X^{i}\,\mu_i^2\,\,.
\eeq
Let us  parameterize the $\mu_i$'s  in terms of two angles $(\tilde\theta, \tilde\psi)$, $0\le \tilde\theta\le \pi/2$,  as:
\beq
\mu_1\,=\,\cos\tilde\theta\sin\tilde\psi
\,\,,\qquad\qquad
\mu_2\,=\,\cos\tilde\theta\cos\tilde\psi\,\,,\qquad\qquad
\mu_3\,=\,\sin\tilde\theta\,\,.
\eeq
When only a scalar field is non-zero as in (\ref{truncatedXis}),  $\Delta$ reduces to  the following expression:
\beq
\Delta\,=\,e^{-\varphi}\,\cos^2\tilde\theta\,+\,e^{2\varphi}\,\sin^2\tilde\theta\,\,.
\eeq
Taking into account that the following combination:
\beq
d\Omega_3^2\,=\,(d\tilde\psi)^2\,+\,\sin^2\tilde\psi\,(d\phi^1)^2\,+\,
\cos^2\tilde\psi\,( d\phi^2)^2\,\,,
\eeq
represents the line element of a three sphere, one  can check that :
\beq
e^{\varphi}\,\sum_{i=1,2}\,(d\mu_i^2\,+\,\mu_i^2\,d\phi_i^2\,)\,+\,e^{-2\varphi}\,
(d\mu_3)^2\,=\,e^{-\varphi}\,\Delta\, (d\tilde\theta)^2\,+\,e^{\varphi}\,
\cos^2\tilde\theta\,d\Omega_3^2\,\,.
\eeq
Taking this into account,  the uplifted ten-dimensional metric (\ref{upliftedmetric}) when the gauge fields are taken as in (\ref{5dgauge}),   can be written as:
\beq
ds_{10}^2\,=\,\sqrt{\Delta}\,ds_{5}^2\,+\,{1\over m^2\sqrt{\Delta}}\,
\Big[\, e^{-\varphi}\,\Delta\, (d\tilde\theta)^2\,+\,e^{\varphi}\,
\cos^2\tilde\theta\,d\Omega_3^2\,+\,\sin^2\tilde\theta\,e^{-2\varphi}\,
\big(d\phi^3+\cos\theta d\phi\,\big)^2\,\Big]\,\,.
\eeq
Let us rewrite this metric by using the explicit form (\ref{5dmetric}) of the five-dimensional metric. If we denote $\psi\equiv \phi^3$, we obtain:
\bear
&&ds_{10}^2\,=\,\sqrt{\Delta}\,\Big[\,
e^{2f}\,dx^2_{1,1}\,+\,{e^{2h}\over m^2}\,
\big[\,(d\theta)^2\,+\,\sin^2\theta\,(d\phi)^2\,\big]\,+\,e^{2f}\,d r^2\,\Big]\,\,+\rc\rc
&&+\,{1\over m^2\sqrt{\Delta}}\,
\Big[\, e^{-\varphi}\,\Delta\, (d\tilde\theta)^2\,
+\,\sin^2\tilde\theta\,e^{-2\varphi}\,
\big(d\psi+\cos\theta d\phi\,\big)^2\,+\,
e^{\varphi}\,\cos^2\tilde\theta\,d\Omega_3^2\,\,\Big]\,\,.
\label{gaugesugraD3metric}
\eear

\subsection{The five-form}
By using the formulae written in \cite{Cvetic:1999xp} we can  also get the explicit expression of the RR five-form $F_5$.  It turns out that $F_5$ can be written in terms of a magnetic 
part ${\cal F}_5$ and its Hodge dual, and that the ${\cal F}_5$  component can be obtained from a potential four-form ${\cal C}_4$, whose explicit expression takes the form (\ref{CalC4}) with:
\beq
g\,=\,{e^{-\varphi}\cos^4\tilde\theta\over m^4\,\Delta}\,\,.
\label{g-sugra}
\eeq
Actually, the explicit expression  of ${\cal F}_5$ is given by:
\bear
&&{\cal F}_5\,=\,-{2\over m^4}\,\,{e^{\varphi}+e^{-\varphi}\Delta\over \Delta^2}\,\,
\sin\tilde\theta\,\cos^3\tilde\theta\, d\tilde\theta\wedge \omega_3\wedge 
(d\phi_3+\cos\theta d\phi)\,-\,\rc\rc
&&\qquad
-\,{3\over m^4}\,\,{e^{\varphi}\sin^2\tilde\theta\,\cos^4\tilde\theta\over \Delta^2}\,\varphi'\,
dr\wedge \omega_3\wedge (d\phi_3+\cos\theta d\phi)\,+\,
{e^{-\varphi}\over m^4}\,\,{\cos^4\tilde\theta\over \Delta}\,\omega_3\wedge\omega_2
\,\,,\qquad\qquad
\label{CalF5-gsugra}
\eear
where $\omega_3$ and $\omega_2$ are the same as in (\ref{omega3}) and (\ref{omega2}). The quantization condition of the RR five-form $F_5$ is:
\beq
{1\over 2\kappa_{10}^2}\,\,\int_{{\cal M}_5}\,F_5\,=\,N_c\,T_3.
\label{fluxquantization}
\eeq
Taking into account that:
\beq
 2\kappa_{10}^2\,=\,(2\pi)^7\,g_s^2\,(\alpha')^4\,\,,\qquad\qquad
 T_3\,=\,{1\over (2\pi)^3\,g_s\,(\alpha')^2}\,\,,
 \eeq
 and thus:
 \beq
 2\kappa_{10}^2\,T_3\,=\,(2\pi)^4\,g_s\,(\alpha')^2\,\,,
 \eeq
 we can convert (\ref{fluxquantization}) into:
 \beq
 \int_{{\cal M}_5}\,F_5\,=\,(2\pi)^4\,g_s\,(\alpha')^2\,N_c\,\,.
 \eeq

We can use the form of ${\cal F}_5$ written in (\ref{CalF5-gsugra})  in the  flux quantization condition (\ref{fluxquantization}) to determine the value of $m$. To do this we must integrate ${\cal F}_5$ at $\tau\to\infty$ along the transverse five-sphere parameterized by $\tilde\theta$, $\phi_3$ and the transverse three-sphere  whose volume element is $\omega_3$.  It is clear that the only contribution comes from the first term in (\ref{CalF5-gsugra}). Taking into account that $\varphi\to 0$ as $\tau\to\infty$, one has that $\Delta\to 1$ and, thus:
\beq
{{\cal F}_5}_{\big|_{S^5}}\,=\,{4\over m^4}\,\sin\tilde\theta\,\cos^3\tilde\theta\,
d\tilde\theta\wedge d\phi_3\wedge\omega_3\,\,,
\eeq
and, therefore:
\beq
\int_{S^5}\,{\cal F}_5\,=\,{4\over m^4}\,
\int_0^{{\pi\over 2}}d\tilde\theta\sin\tilde\theta \cos^3\tilde\theta\,
\int_0^{2\pi} d\phi_3\,\int_{S^3}\omega_3\,=\,{4\pi^3\over m^4}\,\,.
\eeq
By using this value in the quantization condition (\ref{fluxquantization}), one gets that $m$ is given by:
\beq
{1\over m^4}\,=\,4\pi\,(\alpha')^2\,g_s\,N_c\,\,.
\label{m-gsugra}
\eeq
Notice that $m$ is the same as in (\ref{m}) and that $m^{-4}$ is just equal to what we called $g_0$ in the ten-dimensional approach (see eq. (\ref{g0})).

\subsection{Identification with the 10d variables}

Let us now try to identify the metric (\ref{gaugesugraD3metric}) with the one  in (\ref{D3metric}) obtained in the ten-dimensional approach of section \ref{44unflavored}. By identifying the parts corresponding to the Minkowski space and the cycle one gets:
\beq
H^{-{1\over 2}}\,=\,\sqrt{\Delta}\, \,e^{2f}\,\,,\qquad\qquad
H^{-{1\over 2}}\,z\,=\,\sqrt{\Delta}\, \,e^{2h}\,\,,
\label{H-z}
\eeq
which leads to the following identification of the function $z$:
\beq
z\,=\,e^{2(h-f)}\,\,.
\label{z-h-f}
\eeq
Moreover, by comparing the term of the fiber $\psi$ and that of the three-sphere $d\Omega_3^2$ in (\ref{gaugesugraD3metric}) and (\ref{D3metric}), we get:
\beq
{H^{{1\over 2}}\,\sigma^2\over z}\,=\,{\sin^2\tilde\theta\,e^{-2\varphi}\over m^2\sqrt{\Delta}}\,\,,\qquad\qquad
H^{{1\over 2}}\,\rho^2\,=\,{\cos^2\tilde\theta\,e^{\varphi}\over m^2\sqrt{\Delta}}\,\,.
\label{fiber-sphere}
\eeq
By using the expressions of $H$ and $z$ obtained in (\ref{H-z}) and (\ref{z-h-f})  in (\ref{fiber-sphere}), we can obtain the expression of the two radial coordinates $\sigma$ and $\rho$ of the ten-dimensional approach in terms of the variables of gauged sugra, namely:
\beq
\sigma\,=\,{\sin\tilde\theta\,e^{h-\varphi}\over m}\,\,,\qquad\qquad
\rho\,=\,{\cos\tilde\theta\,e^{f+{\varphi\over 2}}\over m}\,\,.
\label{sigma-rho}
\eeq
In order to completely identify the two metrics (\ref{gaugesugraD3metric}) and (\ref{D3metric}), let us calculate $d\rho$ and $d\sigma$ from (\ref{sigma-rho}). One gets:
\bear
&&d\rho\,=\,{e^{f+{\varphi\over 2}}\over m}\,\Big[\,
\Big(\,f'+{\varphi'\over 2}\,\Big)\,\cos\tilde\theta\,dr\,-\,\sin\tilde\theta\,d\tilde\theta\,\Big]
\,\,,\rc\rc
&&d\sigma\,=\,{e^{h-\varphi}\over m}\,\Big[\,\big(\,h'-\varphi'\,\big)\,\sin\tilde\theta\,dr\,+\,
\cos\tilde\theta\,d\tilde\theta\,\Big]\,\,.
\label{drho-dsigma}
\eear
Moreover, from the the BPS system (\ref{5dBPSsystem}) , one can easily demonstrate that:
\beq
f'+{\varphi'\over 2}\,=\,m\,e^{f-\varphi}\,\,,\qquad\qquad
h'-\varphi'\,=\,m\,e^{f+2\varphi}\,\,.
\eeq
Plugging this result into (\ref{drho-dsigma}) we obtain:
\bear
&&d\rho\,=\,e^{2f-{\varphi\over 2}}\,\cos\tilde\theta\,dr\,-\,{e^{f+{\varphi\over 2}}\over m}\,
\sin\tilde\theta\,d\tilde\theta\,\,,\rc\rc
&&d\sigma\,=\,e^{f+h+\varphi}\,\sin\tilde\theta\,dr\,+\,
{e^{h-\varphi}\over m}\,\cos\tilde\theta\,d\tilde\theta\,\,.
\eear
It is now easy to show that:
\beq
H^{{1\over 2}}\,(d\rho)^2\,+\,{H^{{1\over 2}}\over z}\,(d\sigma)^2\,=\,
{\sqrt{\Delta}\over m^2}\,e^{-\varphi}\,(d\tilde\theta)^2\,+\,
\sqrt{\Delta}\,\,e^{2f}\,(dr)^2\,\,,
\eeq
from which it follows that  (\ref{gaugesugraD3metric}) and (\ref{D3metric}) are equivalent.

\subsection{The solution in the 10d variables}
Let us now rewrite the solution (\ref{gsugrasolution}) found by integrating the first-order system of gauged supergravity equations in terms of the variables used in the ten-dimensional approach. We will verify that the result is just that written in eqs. (\ref{implicit-sol}), (\ref{g-rho-sigma}) and
(\ref{H-rho-sigma}). Actually, from the relation (\ref{z-h-f}) between $z$ and the metric functions $h$ and $f$, it is simple to relate $\tau$ and $z$. One gets:
\beq
e^{2\tau}\,=\,{1\over 2(\alpha-\beta z)}\,\,.
\eeq
By using this result one can easily write the right-hand side of (\ref{gsugrasolution}) in terms of $z$. It is convenient to write this result in terms of  a new constant $z_*$,  defined as:
\beq
z_*\equiv {\alpha\over \beta}\,\,.
\eeq
One obtains:
\bear
&&e^{2h+\varphi}\,=\,{1\over 2}\,{z\over z_*-z}\,\,,\rc\rc
&&e^{2f+\varphi}\,=\,{1\over 2}\,{1\over z_*-z}\,\,,\rc\rc
&&e^{-3\varphi}\,=\,{z_*\,+\,(z_*\,-\,z)\,
\big[\,\log(z_*\,-\,z)\,+\,\kappa\,\big]\over z}\,\,,
\label{sugrasolutioninz}
\eear
where $\kappa$ is the following combination of the integration constants $\beta$ and $\gamma$:
\beq
\kappa\equiv \,\log(2\beta)\,-\,2\gamma\,\,.
\eeq
By using the solution (\ref{sugrasolutioninz}) in (\ref{sigma-rho}), one can also get the values of $\cos\tilde\theta$ and $\sin\tilde\theta$ in terms of $\rho$, $\sigma$ and $z$. One obtains:
\beq
\cos^2\tilde\theta\,=\,2 m^2\,(z_*-z)\,\rho^2\,\,,
\qquad\qquad
\sin^2\tilde\theta\,= \,{2m^2 (z_*-z)\over \Gamma(z)}\,\,\sigma^2\,\,,
\eeq
where the function $\Gamma(z)$ has been defined  in (\ref{Gamma(z)}). 
By using now the fact that $\cos^2\tilde\theta+\sin^2\tilde\theta=1$, one can immediately obtain the equation that implicitly determines $z$ as a function of $(\sigma, \rho)$, namely (\ref{implicit-sol}).  It is also easy to get $\Delta$ in the new variables:
\beq
\Delta\,=\,2m^2 (z_*-z)\,\Big[\,\rho^2+{z\over \Gamma^2(z)}\,\sigma^2\,\Big]\,
\Big[\,{\Gamma(z)\over z}\,\Big]^{{1\over 3}}\,\,.
\eeq
Using this result one can obtain the warp factor $H$ from the identification (\ref{H-z})
and, similarly, one can get the five-form function $g$ from (\ref{g-sugra}). The result is just the one written in eqs. (\ref{g-rho-sigma}) and
(\ref{H-rho-sigma}) of section \ref{44unflavored}.

Notice that, by differentiating the implicit relation  (\ref{implicit-sol}), one can obtain the partial derivatives of $z(\rho,\sigma)$. For the first derivatives one gets:
\beq
z'\,=\,{2\rho (z_*-z)\over \rho^2+{z\over \Gamma^2(z)}\,\sigma^2}\,\,,
\qquad\qquad
\dot z\,=\,{2\sigma\, (z_*-z)\over 
\Big[\,\rho^2+{z\over \Gamma^2(z)}\,\sigma^2\,\Big]\,\Gamma(z)}\,\,.
\label{first-z-derivatives}
\eeq
One can also obtain the second partial derivatives of $z(\rho,\sigma)$ by differentiating (\ref{first-z-derivatives}). Using this result one can verify that, indeed, the function $z(\rho,\sigma)$ defined in (\ref{implicit-sol}) solves the PDE written in (\ref{PDE-z-unflavored}).  Moreover, by plugging the values of $z'$ and $\dot z$ written in (\ref{first-z-derivatives})  in the first and second equations of the system (\ref{BPSsystem}) one can obtain the functions $g(\rho,\sigma)$ and the warp factor $H(\rho,\sigma)$.  The result is, again, the one written in eqs. (\ref{g-rho-sigma}) and (\ref{H-rho-sigma}). As a final check of these expressions one can verify by computing the derivatives of $H$ and $g$ (using again (\ref{first-z-derivatives})) that the last two equations in the system (\ref{BPSsystem}) are also satisfied. Therefore, one concludes that (\ref{implicit-sol}),  (\ref{g-rho-sigma}) and  (\ref{H-rho-sigma}) provide the sought after solution of  the first-order BPS system (\ref{BPSsystem}).

 \renewcommand{\theequation}{\rm{C}.\arabic{equation}}
\setcounter{equation}{0}
\section{SUSY embeddings}
\label{Higgs-embeddings}

The embeddings of the D3-brane that preserve the supersymmetry of the unflavored background are those which satisfy the condition \cite{swedes}:
\beq
\Gamma_{\kappa}\,\epsilon\,=\,\epsilon\,\,,
\label{kappa-condition}
\eeq
where $\Gamma_{\kappa}$ is a matrix which depends on the embedding of the D3-brane and $\epsilon$ is a Killing spinor of the background. In order to specify the precise  form of $\Gamma_{\kappa}$, let us define the induced Dirac matrices on the D3-brane worldvolume as:
\beq
\gamma_a\,=\,\partial_a\, X^{M}\, E_{M}^{\bar M}\,\Gamma_{\bar M}\,\,,
\label{induced-matrices}
\eeq
where $E_{M}^{\bar M}$ denotes the vielbein coefficients, which are the ones needed to express the frame one-forms $e^{\bar M}$ of the ten-dimensional geometry, written explicitly in (\ref{D3frame}),  in terms of the differentials of the coordinates, namely:
\beq
e^{\bar M}\,=\,E_{M}^{\bar M}\,dX^{M}\,\,.
\eeq
Then, when the worldvolume gauge field $F$ is zero, the matrix $\Gamma_{\kappa}$ for the D3-brane takes the form \cite{bbs}:
\beq
\Gamma_{\kappa}\,=\,{1\over 4!}\,\,{1\over \sqrt{-\det \hat G_4}}\,\,\,
\epsilon^{a_1\cdots a_4}\,\,\gamma_{a_1\cdots a_4}\,\,(i\tau_2)\,\,,
\label{Gamma-kappa-D3}
\eeq
where $\gamma_{a_1\cdots a_4}$ denotes the antisymmetrized product of the induced 
matrices (\ref{induced-matrices}) and $\hat G_4$ is the induced metric on the D3-brane worldvolume. Let us now assume that we choose, as in (\ref{D3-wvcoordinates}), $x^0$, $x^1$, $\psi$ and $\sigma$ as worldvolume coordinates. The kappa symmetry matrix (\ref{Gamma-kappa-D3}) for this choice of coordinates becomes:
\beq
\Gamma_{\kappa}\,=\,{1\over \sqrt{-{\rm \det}\,\hat G_4}}\,\,
\gamma_{x^0 x^1\psi\sigma}\,\,(i\tau_2)\,\,.
\label{Gamma-kappa-D3-explicit}
\eeq
We will now restrict ourselves to embeddings which are of the form (\ref{embedding-ansatz}), {\it i.e.}
in which the only non-trivial scalars are $\theta$ and $\phi$, which could depend on the other two $CY_2$ coordinates $\psi$ and $\sigma$. The induced gamma matrices for such embeddings can be obtained from (\ref{induced-matrices}) and are given by:
\bear
&&\gamma_{x^{0,1}}\,=\,H^{-{1\over 4}}\,\,\Gamma_{0,1}\,\,,\rc\rc
&&\gamma_{\psi}\,=\,{H^{-{1\over 4}}\sqrt{z}\over m}\,\,
\Big[\,\partial_{\psi}\theta\,\Gamma_2\,+\,\sin\theta\,\partial_{\psi}\phi\,\Gamma_3\,\Big]\,+\,{H^{{1\over 4}}\sigma\over \sqrt{z}}\,\Big[\,1\,+\,\cos\theta\,\partial_{\psi}\phi\,\Big]\,\Gamma_5\,\,,\rc\rc
&&\gamma_{\sigma}\,=\,{H^{-{1\over 4}}\sqrt{z}\over m}\,\,
\Big[\,\partial_{\sigma}\theta\,\Gamma_2\,+\,\sin\theta\,\partial_{\sigma}\phi\,\Gamma_3\,\Big]\,+\,{H^{{1\over 4}}\over \sqrt{z}}\,\Big[\,\Gamma_4\,+\,\sigma\cos\theta\,
\partial_{\sigma}\phi\,\Gamma_5\,\Big]\,\,.
\label{induced-gamma-Higgs}
\eear
 In order to find the embeddings of the type (\ref{embedding-ansatz}) that are kappa symmetric and preserve the same supersymmetries as the background, we should compute the action of the antisymmetrized product $\gamma_{x^0x^1\psi\sigma}\,(i\tau_2)$ on the Killing spinors $\epsilon$ (see eq. (\ref{Gamma-kappa-D3-explicit})). To perform this calculation it is interesting to realize that, from the projections satisfied by $\epsilon$ (eq. (\ref{projections})), one has:
\bear
&&\Gamma_{0123}\,(i\tau_2)\,\epsilon\,=\,-\Gamma_{0145}\,(i\tau_2)\,\epsilon\,=\,\epsilon
\,\,,\rc\rc
&&\Gamma_{0125}\,(i\tau_2)\,\epsilon\,=\,-\Gamma_{0134}\,(i\tau_2)\,\epsilon\,=\,-\Gamma_{35}\,\epsilon\,\,,\rc\rc
&&\Gamma_{0135}\,(i\tau_2)\,\epsilon\,=\,\Gamma_{0124}\,(i\tau_2)\,\epsilon\,=\,\Gamma_{25}\,\epsilon\,\,.
\eear
By using these conditions, one can easily verify that:
\beq
H^{{1\over 2}}\,
\gamma_{x^0x^1\psi\sigma}\,(i\tau_2)\,
\epsilon\,=\,\big[\,c_I\,+\,c_{35}\,\Gamma_{35}\,+\,c_{25}\,\Gamma_{25}\,\big]\,\epsilon\,\,,
\label{gamma-epsilon-Higgs}
\eeq
where the $c$ coefficients appearing on the right-hand side of (\ref{gamma-epsilon-Higgs}) are given by:
\bear
&&c_I\,=\,{\sigma H^{{1\over 2}}\, \over z}\,\big[\,1\,+\,\cos\theta\,\partial_{\psi}\phi\,\big]\,+\, {z\over m^2 H^{{1\over 2}}\,}\,
\sin\theta\,\Big[\,\partial_{\psi}\theta\,\partial_{\sigma}\phi\,-\,
\partial_{\psi}\phi\,\partial_{\sigma}\theta\,\Big]\,\,,\rc\rc
&&c_{35}\,=\,{\sin\theta\over m}\,\,\partial_{\psi}\,\phi\,+\,{\sigma\over m}\,\Big[\,
\partial_{\sigma}\theta\,+\,\cos\theta\,\big(\,\partial_{\psi}\phi\,\partial_{\sigma}\theta\,-\,
\partial_{\psi}\theta\,\partial_{\sigma}\phi\,\big)\,\Big]\,\,,\rc\rc
&&c_{25}\,=\,{1\over m}\,\Big[\,\partial_{\psi}\,\theta\,-\,\sigma\,\sin\theta\,\partial_{\sigma}\phi\,\Big]\,\,.
\eear
To realize the kappa symmetry condition without imposing any extra projection on the spinor $\epsilon$, the right-hand side of (\ref{gamma-epsilon-Higgs}) should contain only the term proportional to the unit matrix. Therefore, we must impose the following
 BPS conditions:
\beq
c_{35}\,=\,c_{25}\,=\,0\,\,,
\eeq
which reduce to the following system of  equations:
\bear
&&\sin\theta\,\partial_{\psi}\,\phi\,+\,\sigma\,\Big[\,
\partial_{\sigma}\theta\,+\,\cos\theta\,\big(\,\partial_{\psi}\phi\,\partial_{\sigma}\theta\,-\,
\partial_{\psi}\theta\,\partial_{\sigma}\phi\,\big)\,\Big]\,=\,0\,\,,\rc\rc
&&\partial_{\psi}\,\theta\,-\,\sigma\,\sin\theta\,\partial_{\sigma}\phi\,=\,0\,\,.
\label{Higgs-system}
\eear
Notice that, if $\theta$ and $\phi$ are constant, the system (\ref{Higgs-system}) is automatically solved. This is precisely the embedding of flavor branes  in the Coulomb branch used in section \ref{44flavored} (see eq.  (\ref{Coulomb-flavor-embedding})). Moreover,  one can check that, if the BPS system (\ref{Higgs-system}) holds, one has:
\beq
\sqrt{-\det \hat G_4}_{\,\,| BPS}\,=\,H^{-{1\over 2}}\,{c_I}_{\,\,| BPS}\,\,,
\eeq
and, indeed, the kappa symmetry condition $\Gamma_{\kappa}\,\epsilon\,=\,\epsilon$ is fulfilled when the embedding  (\ref{embedding-ansatz}) solves the BPS system (\ref{Higgs-system}). Before finding its more general solution, let us try to solve this system by means of a more restrictive ansatz of the form:
\beq
\theta\,=\,\theta(\sigma)\,\,,
\qquad\qquad
\phi\,=\,\phi(\psi)\,\,.
\label{restricted-ansatz-Higgs}
\eeq
Notice that the second equation of the BPS system (\ref{Higgs-system}) is automatically satisfied by our ansatz (\ref{restricted-ansatz-Higgs}), while the first equation reduces to:
\beq
\partial_{\psi}\phi\,=\,-{\sigma\,\partial_{\sigma}\theta\over \sin\theta\,+\,\sigma\cos\theta \,\partial_{\sigma}\theta}\,\,.
\eeq
Consistency of the above equation with the assumed dependences of $\theta$ and $\phi$ in (\ref{restricted-ansatz-Higgs}) implies that both sides of the equation must be independent of both $\sigma$ and $\psi$. Accordingly, let us write:
\beq
\partial_{\psi}\phi\,=\,p\,\,,
\eeq
with $p$ being constant. This equation can be integrated as:
\beq
\phi\,=\,p\,\psi\,+\,\phi_0\,\,,
\label{phi-psi-Higgs}
\eeq
where $\phi_0$ is a new constant. Moreover, the equation for $\theta(\sigma)$ becomes:
\beq
\partial_{\sigma}\theta\,=\,-{p\sin\theta \over \sigma (1+p\cos\theta)}\,\,,
\eeq
which can be straightforwardly integrated, namely:
\beq
\sigma\,=\,{C\over \Big[\,\sin\big( {\theta\over 2}\big)\,\Big]^{1+{1\over p}}\,
\Big[\,\cos\big( {\theta\over 2}\big)\,\Big]^{1-{1\over p}}}\,\,,
\label{theta-sigma}
\eeq
where $C$ is another constant of integration. Interestingly, the result we have found  in (\ref{phi-psi-Higgs}) and (\ref{theta-sigma}) can be compactly written in terms of the two complex variables $\zeta_1$ and $\zeta_2$ defined in (\ref{def-zetas}). In fact one can prove that these two equations are equivalent to the following complex equation:
\beq
\zeta_1\,\zeta_2^p\,=\,{\rm constant}\,\,.
\label{z1z2-restricted}
\eeq
Actually, it can be proved easily that any holomorphic relation between $\zeta_1$ and $\zeta_2$ solves the system (\ref{Higgs-system}). Indeed, let us assume that $\zeta_1$ and $\zeta_2$ are related by:
\beq
\zeta_1\,=\,f(\zeta_2)\,\,,
\label{zeta1-fzeta2}
\eeq
where $f(\zeta_2)$ is an arbitrary holomorphic function of $\zeta_2$.  In order to check that (\ref{zeta1-fzeta2}) solves (\ref{Higgs-system}) let us take, as in (\ref{embedding-ansatz}), the coordinates $\psi$ and $\sigma$ as independent variables to parameterize the embedding. Then, by computing the derivatives with respect to them of the relation 
(\ref{zeta1-fzeta2}), we get:
\beq
\partial_{\psi}\,\zeta_1\,=\,f'(\zeta_2)\,\partial_{\psi}\,\zeta_2\,\,,\qquad\qquad
\partial_{\sigma}\,\zeta_1\,=\,f'(\zeta_2)\,\partial_{\sigma}\,\zeta_2\,\,.
\eeq
By eliminating $f'(\zeta_2)$ from the above equations, we arrive at the following relation:
\beq
\partial_{\psi}\,\log\zeta_1\,\,\partial_{\sigma}\,\log\zeta_2\,=\,
\partial_{\psi}\,\log\zeta_2\,\,\partial_{\sigma}\,\log\zeta_1\,\,,
\eeq
which, after using the definition of $\zeta_1$ and $\zeta_2$ in  (\ref{def-zetas}),  can be shown to be equivalent to the system (\ref{Higgs-system}). It is now clear that (\ref{zeta1-fzeta2}) is the generalization of the solution (\ref{z1z2-restricted}).


\end{document}